\documentclass[11pt, openright, twoside]{report}
\usepackage[backend=biber,style=numeric,sorting=none, maxnames=99, minnames = 99]{biblatex}
\usepackage{subfiles}
\usepackage[nohyperlinks]{acronym}
\usepackage{graphicx}
\usepackage[hidelinks]{hyperref}
\usepackage[T1]{fontenc}
\usepackage{geometry}
\usepackage{chngcntr}
\counterwithin{table}{section}
\counterwithin{figure}{section}
\usepackage{afterpage}
\usepackage[export]{adjustbox}
\usepackage{datetime}
\usepackage{lipsum}
\usepackage{setspace}
\usepackage{booktabs}
\usepackage{multicol}
\usepackage{multirow}
\usepackage{fontspec}
\usepackage{tocbibind}
\usepackage{nomencl}
\usepackage{etoolbox}
\usepackage{xcolor}
\usepackage{mathrsfs}
\usepackage{tabularx}
\usepackage{caption}
\usepackage{tikz}
\usepackage{subcaption}
\usepackage{amsmath}
\usepackage{placeins}
\usepackage{float}
\usepackage{cleveref}
\usepackage{amssymb}
\usetikzlibrary{positioning}
\usepackage{graphicx}
\usepackage{tikz}
\usepackage{chngcntr}

\newcounter{subfig}

\graphicspath{{./chapters/chapter4/figures}}
\graphicspath{{./chapters/chapter2/figures}}
\makenomenclature

\newdateformat{monthyeardate}{%
  \monthname[\THEMONTH], \THEYEAR}
\newcommand\myemptypage{
    \null
    \thispagestyle{empty}
    \newpage
    }

\newgeometry{
	paper=a4paper, 
        total={210mm,297mm },
	inner=3.23cm, 
	outer=1.5cm, 
	bindingoffset=.5cm, 
	top=2.5cm, 
	bottom=2.5cm, 
}
\usepackage{setspace}
\usepackage[explicit]{titlesec}  

\titleformat{\chapter}[display]
{\fontsize{27}{27}\sffamily\bfseries\filleft}{\hfil \thechapter}{0pt}{{#1}}  

\titleformat{\section}{\Large\bfseries\sffamily}{\thesection}{0.5em}{\MakeUppercase{#1}}
\titleformat{\subsection}{\large\bfseries\sffamily}{\thesubsection}{0.5em}{\MakeUppercase{#1}}
\titleformat{\subsubsection}{\sffamily}{\thesubsubsection}{0.5em}{\MakeUppercase{#1}}

\newcommand{\ttitle}{Satellite-Based Quantum Communication: Performance Evaluation of Discrete-Variable Quantum Key Distribution Protocols}
\newcommand{\supname}{Prof. Subhashish Banerjee} 
\newcommand{\degree}{Doctor of Philosophy (Ph.D.)}
\newcommand{\authorname}{Muskan}
\newcommand{\rn}{P19PH204}
 
\newcommand{\deptname}{Department of Physics}

\begin{document}
 
\setstretch{1}
\begin{titlepage}
\begin{center}
\begin{flushright}
\Huge{\textbf{{\ttitle}}}
\end{flushright}
\vfill
\begin{flushright}
\Large{\textit{A thesis submitted by}}\\
\huge{\textbf{\authorname}}
\end{flushright}
\vfill
\begin{flushright}
\Large{\textit{in partial fulfillment of the requirements for the award of the degree of}}\\
\huge{\textbf{\degree}}
\end{flushright}
\vfill
\includegraphics[width=0.25\textwidth,right]{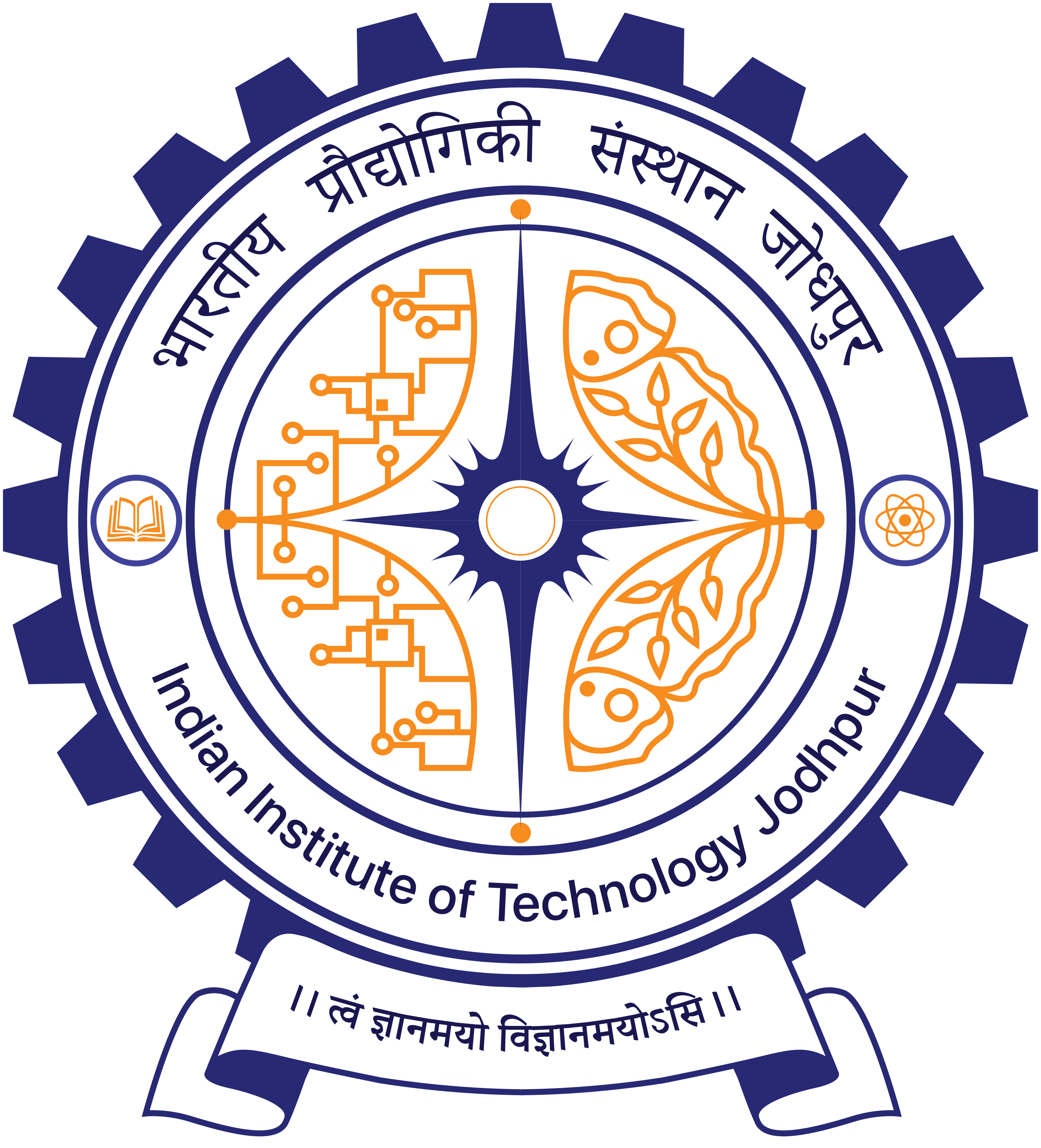}
\begin{flushright}
\Large{\textbf{Indian Institute of Technology Jodhpur}}\\
\Large{\textbf{\deptname}}\\ 
\Large{\emph{\monthyeardate\today}}
\end{flushright}

 
\end{center}
\end{titlepage}

\myemptypage
\pagenumbering{roman}
\setcounter{page}{3}


\begin{flushright}
\huge{ \textbf{Declaration}}
\end{flushright}
{ I hereby declare that the work presented in this thesis titled ``\emph{\ttitle{}}", submitted to the Indian Institute of Technology Jodhpur in partial fulfilment of the requirements for the award of the degree of~ \degree, is a bonafide record of the research work carried out under the supervision of \supname. The contents of this thesis, in full or in parts, have not been submitted to, and will not be submitted by me to, any other Institute or University in India or abroad for the award of any degree or diploma.}

\vspace{2cm}

\begin{flushright}
\raggedleft
\includegraphics[width=0.15\linewidth]{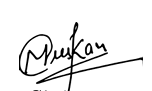}\\

\large{\textbf{Signature}}\\
\emph{\authorname}\\
\rn
\end{flushright}

\myemptypage

\begin{flushright}
\huge{\textbf{Certificate}}
\end{flushright}
  This is to certify that the thesis titled ``\emph{\ttitle{}}", submitted by \authorname{} (\rn{}) to the Indian Institute of Technology Jodhpur for the award of the degree of \degree{}, is a bonafide record of the research work done by her under my supervision. To the best of my knowledge, the contents of this report, in full or in parts, have not been submitted to any other Institute or University for the award of any degree or diploma.
\vspace{2cm}

\begin{flushright}
\raggedleft
\includegraphics[width=0.15\linewidth]{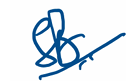}\\
\large{\textbf{Signature}}\\
\supname\\
Ph.D Thesis Supervisor 
\end{flushright}
\myemptypage
\begin{flushright}
\huge{\textbf{Abstract}}
\end{flushright}
\addcontentsline{toc}{chapter}{Abstract}

Quantum Key Distribution (QKD) has emerged as a fundamentally secure approach to communication in the era of quantum computing, offering protection against threats posed to classical cryptographic schemes such as RSA and Diffie-Hellman. This thesis presents a comprehensive performance analysis of satellite-based QKD protocols, focusing on both prepare-and-measure and entanglement-based schemes under realistic atmospheric and operational conditions. The study begins by introducing the theoretical foundations of quantum communication, including qubits, entanglement, and quantum entropy, and motivates the need for satellite-based QKD to overcome the distance limitations of fiber-based systems. Subsequently, the thesis evaluates four prominent QKD protocols-BB84, B92, BBM92, and E91-using a circular beam propagation model that incorporates atmospheric effects such as diffraction, turbulence, attenuation, and pointing errors, along with environmental noise contributions for uplink and downlink. Comparative numerical simulations reveal that protocol performance is strongly influenced by channel asymmetries, beam propagation characteristics, and noise, providing guidance on optimal protocol selection for low Earth orbit (LEO) satellite links. The research further investigates high-dimensional (HD) QKD protocols, specifically HD-BB84 and HD-Extended B92, using the elliptic-beam approximation to account for turbulence-induced distortions for both uplink and downlink. Simulations under varying system dimensions, weather conditions, and zenith angles demonstrate that HD-BB84 achieves higher key rates, superior noise tolerance, and more favorable probability distributions of the key rate compared to HD-Extended B92, highlighting the advantages of high-dimensional encoding for robust satellite-based QKD. Finally, the thesis presents a finite-key and asymptotic-key analysis of the efficient and standard BB84 protocols for CubeSat downlink channels. Atmospheric modeling, statistical fluctuations, and decoy-state methods are incorporated to evaluate realistic key-rate performance under different weather conditions. Results indicate that the efficient BB84 protocol outperforms the standard variant in both finite and asymptotic regimes, offering higher key rates, improved stability, and broader operational range.
The findings of this thesis provide a detailed understanding of the practical performance limits and operational considerations for satellite-based QKD, while identifying avenues for future research, including alternative noise models, different beam propagation schemes, other orbital configurations, and broader protocol comparisons.

\myemptypage

\addcontentsline{toc}{chapter}{Acknowledgements}

\begin{flushright}
\huge{\textbf{Acknowledgements}}

\end{flushright}

I would like to begin by expressing my heartfelt gratitude to everyone who has supported and guided me during my Ph.D. journey. First and foremost, I extend my sincere appreciation to my supervisor, \textbf{Prof. Subhashish Banerjee}, for giving me the opportunity to pursue my Ph.D. under his mentorship. His constant motivation and steadfast support have been truly instrumental in shaping my doctoral journey. During moments when I felt emotionally drained, he encouraged me to take the time I needed to recover and return with renewed clarity. One of the most invaluable aspects of his guidance has been the academic freedom he offered-freedom that not only strengthened my research capabilities but also enabled meaningful introspection, helping me recognize my shortcomings and grow continuously on both professional and personal fronts. For his unwavering faith in me, I remain profoundly grateful. I would also like to extend my sincere thanks to \textbf{Prof. Anirban Pathak}, who played a pivotal role in the early stages of my research career. My interactions and visits to him provided clarity and direction when I was just beginning this journey. Throughout my Ph.D., his insightful discussions, thoughtful questions, and innovative ideas inspired me to explore new dimensions in research. His intellectual curiosity and ability to challenge conventional thinking have been a true source of inspiration; one that I will carry with me throughout my career. I also extend my gratitude to \textbf{Prof. V. Narayanan} for his helpful discussions and valuable insights which were always encouraging and thought-provoking. \\

A special note of appreciation goes to one of the key people in my journey-my collaborator and, above all, my best friend, \textbf{Dr. Ramniwas Meena} (Meenu). From the very first day of my Ph.D. journey, he has been a constant source of support. His unwavering support-both professionally and personally-has been an unfailing source of strength for me. His presence, guidance, and friendship have made this journey not just meaningful, but truly memorable. I would now like to express my heartfelt appreciation to \textbf{Dr. Arindam Dutta}, whose presence in my Ph.D. journey was nothing short of transformative. He entered my academic life at a particularly significant stage, and his influence has been profound. His clarity of thought, innovative approach to problem-solving, and unwavering dedication to research deeply shaped my understanding of what it means to pursue meaningful work. With him, every discussion became an opportunity to learn, every idea a space to grow. He motivated me a lot, challenged my perspectives, and helped refine my thinking with remarkable patience and insight. His mentorship not only strengthened my research but also helped me evolve into a more confident and thoughtful researcher. For the depth, direction, and inspiration he brought into my work, I remain truly indebted.\\

I express my deepest gratitude to my \textbf{parents}, who have always believed in me, trusted my choices, and given me the freedom to pursue the path of research. Their unwavering support has been the bedrock of my progress. My heartfelt thanks also go to my sister, \textbf{Surbhi Di} and my brother-in-law, \textbf{Jagpal Jeeju}, whose constant encouragement and faith in me-often expressed through their reassuring words, \textbf{“Tu kar sakti hai aur tu kar legi”}-gave me immense strength during challenging times.
I would also like to extend my warm thanks to all my labmates and my circle of friends-Jai Lalita, Baibhav, Monika, Devvrat, Nihar, Mahima, Satish, Neha Pathania, Rashid Malik Ansari, Dourgabatee Rout, Taniya, Sachin, and Aman-as well as my school-time best friend Abhishek, for their companionship, support, and constant encouragement throughout this journey. \\

Lastly, but certainly not least, I would like to extend my gratitude to \textbf{Myself}. This may seem unusual to some, but I believe it is important to acknowledge one’s own efforts, resilience, and perseverance. When I began my Ph.D. journey, everything felt uncertain and unclear. I did not know where to start or how to move forward. Yet, step by step, I kept trying with patience and determination. Even when results did not come, even when every effort seemed to go in the wrong direction, I still went to the lab each day with the hope that I would learn something new and make even the smallest amount of progress. There were many days when nothing changed, but I continued-driven by consistency, dedication, and the quiet faith that one day things would fall into place. With belief in myself and trust in God, I kept moving forward. Today, as I write the final pages of my thesis, I am grateful to the version of myself who refused to give up. I know this is not the end-there are many milestones yet to achieve-but I am proud to have reached this important stage of completing my thesis. Acknowledging my own strength is a reminder that perseverance and hope can truly shape one’s journey.

\myemptypage
\newpage
\thispagestyle{empty}
\chapter*{\centerline{}  }

     \vspace*{1in}

\textit{
This thesis is dedicated to my family for their endless support and encouragement, and to myself for the patience and determination that made this work possible.}\\

\begin{flushright}
\tableofcontents
\clearpage
\listoffigures
\listoftables
\clearpage
\null

\renewcommand{\acsfont}[1]{\textsc{#1}} 

\setlength{\columnsep}{4pc}
\setstretch{0.5}
\begin{flushright}
\Huge{\textbf{List of Abbreviations}}
\end{flushright}
\addcontentsline{toc}{chapter}{List of Abbreviations}



\begin{acronym}[TDMA] 
\setstretch{0.2}
 \acro{QKD}{Quantum Key Distribution}
 \acro{LEO}{Low Earth Orbit}
 \acro{HD}{High Dimensional}
 \acro{RSA}{Rivest-Shamir-
Adleman }
 \acro{DH}{Diffie-Hellman}
 \acro{QBER}{Quantum Bit Error Rate}
 \acro{PDT}{Probability Distribution Of Transmittance}
 \acro{PDR}{Probability Distribution of Key Rate}
 \acro{WCP}{Weak Coherent Pulse }
 \acro{DV}{Discrete- Variable}
 \acro{CV}{Continuous-Variable}
 \acro{SDI}{Source-Device-Independent}
 \acro{MDI}{Measurement-Device-Independent}
 \acro{DPR}{Distributed-Phase-Reference}
 \acro{DPS}{Differential Phase Shift}
 \acro{COW}{Coherent
One-Way}
\acro{PNS}{Photon Number Splitting}
\acro{QND}{Quantum Non-Demolition}
\acro{SDP}{SemiDefinite Programming}
\acro{MEO}{Medium Earth Orbit}
\acro{GEO}{Geostationary Equatorial Orbit}
\acro{OAM}{Orbital Angular Momentum}
 \end{acronym}
 \clearpage
\null
\end{flushright}

\clearpage
\pagenumbering{arabic}
\setcounter{page}{1} 


\chapter{INTRODUCTION}

\section{Introductory Outline}
Quantum physics, which provides a fundamental description of matter and energy at atomic and subatomic scales, has profoundly shaped modern science. Its core principles, including superposition, entanglement, and the uncertainty principle, underpin a range of specialized fields such as quantum communication, quantum information, quantum computing, and quantum optics. Among these, the secure transmission of information has always been of paramount importance, giving rise to the field of cryptography, which evolved from simple methods of concealing messages to sophisticated techniques ensuring confidentiality, integrity, and authenticity. Classical cryptographic schemes, including public-key systems such as  Rivest-Shamir-Adleman (RSA) scheme \cite{RSA78} and Diffie-Hellman  (DH) scheme, rely on the computational difficulty of specific mathematical problems \cite{DH22}. However, the development of quantum computing, exemplified by Shor’s algorithm \cite{S94}, poses a serious threat to these methods by enabling efficient solutions to problems like integer factorization and discrete logarithms. This potential vulnerability underscores the need for fundamentally secure alternatives. Quantum Key Distribution (QKD) addresses this challenge by exploiting the principles of quantum mechanics to achieve inherently secure key exchange, independent of computational assumptions. The present thesis focuses on secure quantum communication, motivated by its critical role in enabling unconditionally secure information transfer in the emerging era of quantum technologies.
\subsection{Brief Introduction Of Quantum Communication and Motivation of This Thesis}
From foundational experiments to revolutionary advancements, quantum physics and mechanics have profoundly influenced both our understanding of the universe and our technological capabilities. As the second quantum revolution continues to unfold, quantum technologies are emerging as transformative tools across computing, communication, and sensing, placing them at the forefront of modern scientific and technological innovation. The advent of quantum mechanics not only reshaped the theoretical framework of physics but also opened avenues for novel technological developments, one of the most significant being quantum communication.
Quantum communication is an emerging discipline that leverages the principles of quantum mechanics to address the limitations of classical communication systems. With the rapid progress in quantum computing, there is a growing need to develop technologies capable of interconnecting multiple quantum processors to form large-scale quantum networks, often referred to as the quantum internet \cite{S02, GT07}. Beyond enabling communication between quantum devices, quantum communication offers distinct advantages over its classical counterpart, including enhanced security, higher data integrity, and improved information transfer efficiency. The foundational principles of quantum communication are rooted in quantum mechanics, which extends and refines the classical laws of physics through novel concepts and physical interpretations. These principles provide a robust theoretical basis for understanding and manipulating quantum systems for communication purposes. Building upon these ideas, researchers have demonstrated that the intrinsic properties of quantum systems-such as superposition and entanglement-can be harnessed to achieve unprecedented levels of communication security. Consequently, a variety of QKD protocols and cryptographic techniques have been developed, establishing quantum communication as a cornerstone of secure information exchange in the emerging quantum era.\\
QKD is a fundamental aspect of quantum communication that enables two parties to establish a shared secret key whose security is ensured by the fundamental laws of quantum mechanics. The first QKD
 protocol (BB84) was proposed a decade before Shor’s findings highlighted the vulnerability of classical cryptography. This protocol leverages fundamental quantum properties, such as the no-cloning theorem, the measurement postulate, and Heisenberg’s uncertainty principle, to ensure security, allowing it to be implemented with polarization-encoded single photons or other forms of photonic qubits. Over the past few decades, QKD has developed into a reliable framework for achieving information-theoretic security, with successful demonstrations across both optical fiber and free-space communication channels. Satellite-based QKD has emerged as a promising solution to overcome the distance limitations inherent in fiber-based quantum communication systems. Building on this motivation, this thesis focuses on evaluating the performance of various QKD protocols-namely BB84, B92, E91 and BBM92-through the computation of Quantum Bit Error Rate (QBER) and secure key rate using the circular-beam transmittance approximation for satellite-based links. The analysis incorporates key atmospheric and system parameters to enable a detailed comparison of protocol performance under diverse link conditions. To further enhance the efficiency and robustness of satellite-based QKD, this work extends to HD protocols, which offer higher key rates and improved noise tolerance. In this context, the Probability Distribution of Transmittance  (PDT) and the Probability Distribution of Key Rate (PDR) are studied under varying weather conditions using the elliptical-beam approximation, providing a more accurate representation of beam deformation caused by atmospheric turbulence. As satellite-based QKD becomes increasingly viable, it is also crucial to accurately model the security of key generation under realistic constraints, particularly in finite-key scenarios. Recent advancements in finite-key analysis have significantly improved secure key rate estimation by addressing statistical fluctuations and optimizing security parameters. Building upon these developments, this thesis investigates the performance of tight statistical techniques for parameter estimation and error correction to compute the finite-key block size, alongside the asymptotic key rate for Weak Coherent Pulse (WCP)-based efficient BB84 and standard BB84 protocols. The analysis is performed using the elliptical-beam model in a CubeSat-based QKD downlink scenario, providing a realistic assessment of protocol performance under practical space-to-ground conditions.
This study aims to provide a comprehensive understanding of satellite-based QKD performance, guiding the design of more efficient and resilient quantum communication systems.
\\
This thesis aims to advance the field of quantum communication and cryptography by addressing key challenges, and proposing innovative solutions. The central contribution of this work lies in the rigorous analysis of key rates for various QKD protocols implemented in satellite-based QKD systems. To ensure that the thesis remains comprehensive and self-contained, the fundamental tools of quantum communication are discussed in Section \ref{Fundamental Tools of Quantum Communication}, with emphasis on concepts directly relevant to the research presented herein. An initial overview of QKD, including its underlying protocols, security considerations, and the decoy-state method for practical implementations, is provided in Section \ref{QKD}. Subsequently, Section \ref{Satellite-Based QKD} elaborates on the challenges and limitations associated with fiber-based QKD systems and highlights the motivation for transitioning toward satellite-based QKD. Furthermore, concise introductions and motivations for the research presented in Chapters 2, 3, and 4 are provided in Sections \ref{ch2int}, \ref{ch3int}, and \ref{ch4int}, respectively, focusing on key rate analysis of satellite-based QKD protocols, HD satellite QKD, and finite key rate analysis for different quantum communication protocols. Finally, the chapter concludes with a summary and an overview of the thesis structure in Section \ref{summary}.

\section{Fundamental Tools of Quantum Communication}
\label{Fundamental Tools of Quantum Communication}
\subsection{Quantum States and Basis}
A quantum system is mathematically represented within a Hilbert space \(\mathcal{H}\). The state of the system is characterized by a density operator, which is a linear operator acting on \(\mathcal{H}\) that is Hermitian, positive semi-definite, and has a unit trace. Quantum states constitute the fundamental units of quantum information processing, serving as the essential framework upon which all quantum communication protocols are built. Analogous to the classical binary states \(0\) and \(1\), a quantum system can also exist in distinct states, conventionally represented in Dirac notation as \(|0\rangle\) and \(|1\rangle\). These states form an orthonormal basis of a two-dimensional Hilbert space. Unlike a classical system, however, a quantum system can exist in a superposition of these basis states-an inherently quantum phenomenon with no classical analogue. Consequently, the state of the simplest quantum system, known as a qubit, can be expressed as \cite{NC10}:
\[
|\psi_{qubit}\rangle = \alpha |0\rangle + \beta |1\rangle,
\]
where \(\alpha\) and \(\beta\) are complex probability amplitudes satisfying the normalization condition \(|\alpha|^2 + |\beta|^2 = 1\). In a classical framework, it is always possible to determine definitively whether a system is in state \(0\) or \(1\). However, this determinism does not hold for quantum systems. When a measurement is performed on a qubit, the outcome collapses probabilistically to either \(|0\rangle\) or \(|1\rangle\), yielding the result ``0'' with probability \(|\alpha|^2\) and ``1'' with probability \(|\beta|^2\). 
By parametrizing the complex coefficients \(\alpha\) and \(\beta\) in terms of two real parameters \(\theta\) and \(\omega\), the qubit state can be equivalently represented as:
\[
|\psi\rangle = \cos\frac{\theta}{2}\,|0\rangle + e^{i\omega}\sin\frac{\theta}{2}\,|1\rangle,
\]
where parameters \(\omega\) and \(\theta\) define a unique point on the surface of a unit three-dimensional sphere, known as the Bloch sphere. The Bloch sphere provides an intuitive geometric representation of a single-qubit state and serves as a powerful visualization tool in quantum computation and quantum information theory. However, this geometric intuition becomes increasingly limited when extended to multi-qubit systems.
The concept of a qubit is defined within a two-dimensional Hilbert space spanned by the basis states \(|0\rangle\) and \(|1\rangle\). This framework can be generalized to higher-dimensional quantum systems. For instance, a single qutrit, which resides in a three-dimensional Hilbert space spanned by \(|0\rangle\), \(|1\rangle\), and \(|2\rangle\), can be represented as
\[
|\psi_{qutrit}\rangle = \alpha |0\rangle + \beta |1\rangle + \gamma |2\rangle,
\]
where the complex coefficients \(\alpha\), \(\beta\), and \(\gamma\) satisfy the normalization condition 
\(|\alpha|^2 + |\beta|^2 + |\gamma|^2 = 1\).
\paragraph{Basis}
The two fundamental states of a qubit are denoted by \(|0\rangle\) and \(|1\rangle\). These states can, for example, represent two orthogonal polarization directions of a photon. In vector form, the computational basis states are written as
\[
|0\rangle = 
\begin{pmatrix} 
1 \\ 
0 
\end{pmatrix}, 
\quad 
|1\rangle = 
\begin{pmatrix} 
0 \\ 
1 
\end{pmatrix}.
\]

These states form an orthonormal basis for the two-dimensional Hilbert space of a qubit, commonly referred to as the computational basis. Any arbitrary qubit state \(|\psi\rangle\) in this basis can be expressed as \cite{W21}
\[
|\psi\rangle = 
\begin{pmatrix} 
\alpha \\ 
\beta 
\end{pmatrix},
\]
where \(\alpha\) and \(\beta\) are complex probability amplitudes satisfying the normalization condition \(|\alpha|^2 + |\beta|^2 = 1\).

Another important basis in quantum information theory is the Hadamard basis, defined by the states
\[
|+\rangle = \frac{1}{\sqrt{2}}
\begin{pmatrix} 
1 \\ 
1 
\end{pmatrix}, 
\quad 
|-\rangle = \frac{1}{\sqrt{2}}
\begin{pmatrix} 
1 \\ 
-1 
\end{pmatrix}.
\]

The computational and Hadamard bases are related as
\[
|+\rangle = \frac{|0\rangle + |1\rangle}{\sqrt{2}}, 
\quad 
|-\rangle = \frac{|0\rangle - |1\rangle}{\sqrt{2}}.
\]

The Hadamard transformation \(H\) provides a mapping between these two bases:
\[
H|0\rangle = |+\rangle, \quad H|1\rangle = |-\rangle,
\]
with the matrix representation
\[
H = \frac{1}{\sqrt{2}}
\begin{pmatrix} 
1 & 1 \\ 
1 & -1 
\end{pmatrix}.
\]

It can be verified that \(H^2 = I\), implying \(H^{-1} = H\). This property confirms that the Hadamard transformation is its own inverse, allowing bi-directional mapping between the computational and Hadamard bases.

\subsection{Density Matrix}
The density matrix formalism provides a convenient and comprehensive framework for describing the state of a quantum system \cite{NC10}. If a quantum system is found in the state \(|\psi_i\rangle\) with probability \(p_i\), the density matrix \(\rho\) of the system is defined as
\[
\rho = \sum_i p_i |\psi_i\rangle \langle \psi_i|.
\]
The density operator \(\rho\) is Hermitian, positive semi-definite, and satisfies the normalization condition \(\mathrm{Tr}(\rho) = 1\). A quantum state is said to be \textit{pure} if \(\mathrm{Tr}(\rho^2) = 1\), and \textit{mixed} if \(\mathrm{Tr}(\rho^2) < 1\). Since the density operator $\rho$ acts on a state vector $|\psi\rangle \in \mathcal{H}$, it belongs to the space $\mathcal{B}(\mathcal{H})$, which represents the set of all bounded linear operators defined on the Hilbert space $\mathcal{H}$. Notably, $\mathcal{B}(\mathcal{H})$ itself constitutes a Hilbert space.

\subsection{Composite Systems}
A composite quantum system consisting of two subsystems associated with Hilbert spaces $\mathcal{H}_A$ and $\mathcal{H}_B$ is described by the tensor product space $\mathcal{H}_A \otimes \mathcal{H}_B$ \cite{W21}. States in this composite space are linear combinations of tensor product states $|\phi\rangle \otimes |\psi\rangle$, where $|\phi\rangle \in \mathcal{H}_A$ and $|\psi\rangle \in \mathcal{H}_B$. 

If $\{|e_i\rangle\}_i$ and $\{|f_j\rangle\}_j$ denote orthonormal bases for $\mathcal{H}_A$ and $\mathcal{H}_B$, respectively, then the set $\{|e_i\rangle \otimes |f_j\rangle\}_{i,j}$ forms an orthonormal basis for the composite Hilbert space $\mathcal{H}_A \otimes \mathcal{H}_B$. Accordingly, the dimension of the composite Hilbert space is the product of the dimensions of the individual subsystems:
\[
\dim(\mathcal{H}_A \otimes \mathcal{H}_B) = \dim(\mathcal{H}_A) \, \dim(\mathcal{H}_B).
\]
The tensor product of vectors in Hilbert spaces satisfies the following properties for any scalar $\lambda \in \mathbb{C}$ and vectors $|\phi\rangle, |\phi_1\rangle, |\phi_2\rangle \in \mathcal{H}_A$ and $|\psi\rangle, |\psi_1\rangle, |\psi_2\rangle \in \mathcal{H}_B$:

\begin{enumerate}
    \item $\lambda (|\phi\rangle \otimes |\psi\rangle) = (\lambda |\phi\rangle) \otimes |\psi\rangle = |\phi\rangle \otimes (\lambda |\psi\rangle),$
    
    \item $(|\phi_1\rangle + |\phi_2\rangle) \otimes |\psi\rangle = |\phi_1\rangle \otimes |\psi\rangle + |\phi_2\rangle \otimes |\psi\rangle,$
    
    \item $|\phi\rangle \otimes (|\psi_1\rangle + |\psi_2\rangle) = |\phi\rangle \otimes |\psi_1\rangle + |\phi\rangle \otimes |\psi_2\rangle.$
\end{enumerate}
To obtain a vector representation of states in a composite system, we employ the definition of the tensor product from linear algebra.  Suppose we have the two dimensional vectors

\[
\mathbf{u} = 
\begin{pmatrix} u_1 \\ u_2 \end{pmatrix}, 
\quad 
\mathbf{v} = 
\begin{pmatrix} v_1 \\ v_2 \end{pmatrix}.
\]

The tensor product of these vectors is defined as
\[
\mathbf{u} \otimes \mathbf{v} =
\begin{pmatrix} u_1 \\ u_2 \end{pmatrix} \otimes 
\begin{pmatrix} v_1 \\ v_2 \end{pmatrix} =
\begin{pmatrix} u_1 v_1 \\ u_1 v_2 \\ u_2 v_1 \\ u_2 v_2 \end{pmatrix}.
\]
\subsection{Entanglement}
Consider a quantum state $|\phi_A\rangle \in \mathcal{H}_A$ and another state $|\phi_B\rangle \in \mathcal{H}_B$, where $\mathcal{H}_A$ and $\mathcal{H}_B$ are the Hilbert spaces of the respective subsystems. The combined state of the composite system is denoted by $|\phi_{AB}\rangle \in \mathcal{H}_A \otimes \mathcal{H}_B$ \cite{P13}.

The combined state may or may not be expressible as a tensor product of the subsystems. If it is possible to write the combined state as
\[
|\phi_{AB}\rangle = |\phi_A\rangle \otimes |\phi_B\rangle,
\]
then the state is called a separable state; otherwise, it is entangled and the condition for the entangled state is

\[
|\phi_{AB}\rangle  \neq  |\phi_A\rangle \otimes |\phi_B\rangle.
\]

For example, the states
\[
|0\rangle \otimes |0\rangle, \quad \frac{1}{\sqrt{2}}(|0\rangle \pm |1\rangle) \otimes |1\rangle
,\]
are separable states, whereas
\[
\frac{1}{\sqrt{2}}(|00\rangle \pm |11\rangle), \quad \frac{1}{\sqrt{2}}(|01\rangle \pm |10\rangle)
,\]
are examples of entangled states.

For mixed states, the condition for entanglement can be expressed in terms of the density matrix as
\[
\rho \neq \sum_{j=1}^{n} p_j \, \alpha_j \otimes \beta_j,
\]
where $\alpha_1, \alpha_2, \dots, \alpha_n$ and $\beta_1, \ldots, \beta_n$ are the states of the first and second systems respectively, and \(\sum_{j=1}^{n} p_j = 1\). 

\subsection{Quantum Entropies}
Entropy serves as a quantitative measure of the information content or uncertainty associated with a signal. It can also be interpreted as a measure of the randomness or disorder present within a system. For a discrete random variable \( X \) with possible outcomes \( i = 1, 2, \dots, n \), each occurring with probability \( p_i \), the Shannon entropy \( H(X) \) is defined as  
\[
H(X) = -\sum_{i=1}^{n} p_i \log_2 p_i
.\]
It is also referred to as the classical Shannon entropy \cite{M08}. Consider two random variables \( X \) and \( Y \) with a joint probability mass function \( p(x, y) \). The joint Shannon entropy quantifies the total uncertainty associated with the pair \((X, Y)\) and is expressed as  
\[
H(X, Y) = -\sum_{x,y} p(x, y) \log_2 p(x, y)
.\]
The conditional entropy of \( X \) given \( Y \) represents the average uncertainty remaining about \( X \) when \( Y \) is known. It is mathematically defined as  
\[
H(X|Y) = H(X, Y) - H(Y)
.\]
The quantum entropy, also known as the von Neumann entropy \cite{MMF+13}, is an extension of the classical Shannon entropy, introduced by John von Neumann. In a quantum mechanical system, the concept of a random variable is replaced by a density matrix \( \rho \). Accordingly, for a given density matrix \( \rho \), the von Neumann entropy is defined as  
\[
S(\rho) = -\mathrm{Tr}(\rho \log \rho)
.\]
Here, the logarithms are taken with base 2. If the density matrix \( \rho \) can be expressed in terms of its eigenvalues \( \lambda_k \) and corresponding eigenvectors \( |k\rangle \) as  
\[
\rho = \sum_{k} \lambda_k |k\rangle \langle k|,
\]
then the von Neumann entropy of \( \rho \) is given by  
\[
S(\rho) = -\sum_{k} \lambda_k \log_2 \lambda_k.
\]
The von Neumann entropy quantifies the information content in a quantum system.
We now discuss the most generalized form of quantum entropy, known as the Rényi entropy. The Rényi entropy is a generalization of the von Neumann entropy and is widely used in quantum information theory to measure the uncertainty or information content of a quantum state. It extends the concept of the classical Rényi entropy to the quantum regime. This measure was first introduced by the Hungarian mathematician Alfréd Rényi in 1961. The Rényi entropy of order \( \alpha \) for a quantum state represented by a density matrix \( \rho \) is defined as \cite{MMF+13} 
\[
S_{\alpha}(\rho) = \frac{1}{1 - \alpha} \log_2 \left( \mathrm{Tr}\,\rho^{\alpha} \right),
\]
where \( \alpha \in (0, 1) \cup (1, \infty) \), and the logarithm is taken with base 2. In the limit \( \alpha \to 1 \), the Rényi entropy converges to the von Neumann entropy.

\section{Quantum Key Distribution}
\label{QKD}
QKD enables two communicating parties to establish a shared secret key through the transmission of quantum signals \cite{GRT+02, SBC+09}. The primary objective of QKD is to provide information-theoretic security, ensuring that the generated key remains secure even against adversaries with unlimited computational resources. In a typical setup, Alice (the sender) and Bob (the receiver) distribute keys over an insecure quantum channel, such as free space or optical fiber, while classical communication is employed for post-processing. Importantly, only the key distribution phase relies on quantum mechanics; subsequent steps are entirely classical. The classical channel must be authenticated, allowing Alice and Bob to verify each other’s identities. Although an eavesdropper (Eve) may monitor the classical communication, active participation is prevented. Conversely, the quantum channel is fully accessible to adversarial manipulation. The security task of QKD is therefore to ensure robustness against such eavesdropping attempts while enabling the detection of any intrusion in real time. Owing to these capabilities, QKD offers not only secure key establishment but also the potential to play a crucial role in the development of secure communication infrastructures in the modern era.

\subsection{Fundamentals of QKD}

QKD builds on the principles of quantum mechanics to enable secure generation of symmetric keys between two distant legitimate parties. Unlike classical cryptographic systems, where security depends on computational complexity assumptions, QKD achieves information-theoretic security by leveraging the unique properties of quantum states, such as no-cloning and the inevitable disturbance introduced by measurement. The fundamental mechanisms that underpin QKD include the no-cloning theorem and the Heisenberg uncertainty principle, which collectively guarantee that any eavesdropping attempt can be detected through an increase in the error rate.

\subsubsection{No-Cloning Theorem as a Foundation for Quantum Security}
One of the fundamental principles that ensures the security of quantum communication protocols is the no-cloning theorem \cite{WZ82,SIG+05,NMA+26}. This theorem establishes that it is impossible to design a universal procedure capable of producing an exact copy of an arbitrary, unknown quantum state. The proof follows directly from the linearity of quantum mechanics \cite{NC10}.
Let an arbitrary pure state $|\psi\rangle$ be stored in register $A$, and let register $B$ be initialized in a standard state $|s\rangle$. The initial state of the composite system is given by  
\begin{equation}
|\psi\rangle \otimes |s\rangle.
\end{equation}
If perfect cloning were possible, there would exist a unitary operator $U$ such that  
\begin{equation}
U \big(|\psi\rangle \otimes |s\rangle\big) = |\psi\rangle \otimes |\psi\rangle,
\end{equation}
for all $|\psi\rangle$.
Consider two distinct pure states $|\psi\rangle$ and $|\phi\rangle$. For a universal cloner, the following conditions must simultaneously hold:
\begin{align}
U \big(|\psi\rangle \otimes |s\rangle\big) &= |\psi\rangle \otimes |\psi\rangle, \\
U \big(|\phi\rangle \otimes |s\rangle\big) &= |\phi\rangle \otimes |\phi\rangle.
\end{align}
Taking the inner product of these relations yields
\begin{equation}
\langle \psi | \phi \rangle = \big(\langle \psi | \phi \rangle \big)^2.
\end{equation}
This condition is satisfied only if $\langle \psi | \phi \rangle = 0$ or $\langle \psi | \phi \rangle = 1$, i.e., the states are either mutually orthogonal or identical. Therefore, no unitary operation can perfectly clone arbitrary non-orthogonal quantum states.
As a consequence, an unknown quantum state cannot be copied without error. Although approximate and state-dependent cloning transformations have been proposed, they necessarily involve a reduction in fidelity and cannot achieve exact replication. This impossibility result is a cornerstone of QKD, since it guarantees that an eavesdropper cannot duplicate transmitted quantum states without introducing detectable disturbances.\\
In addition to the no-cloning theorem, the security of quantum communication protocols fundamentally relies on the impossibility of perfectly distinguishing nonorthogonal quantum states \cite{NC10}. Since nonorthogonal states cannot be discriminated without error, any measurement performed by an eavesdropper inevitably introduces disturbances that can be detected by the legitimate parties. This principle forms a key basis for the security of various QKD protocols.
\subsubsection{The Heisenberg Uncertainty Principle and Its Significance in Quantum Communication}
Heisenberg Uncertainty Principle is a fundamental concept in quantum mechanics that states that two non-commuting, or canonically conjugate, observables cannot be simultaneously measured with arbitrary precision \cite{H27}. Examples of such pairs include position and momentum ($\hat{X}, \hat{P}$), rectilinear and diagonal polarizations of photons ($\sigma_Z, \sigma_X$), and angular position and orbital angular momentum ($\theta, l$) of photons. For polarization observables, this principle is mathematically expressed as  

\[
\Delta \sigma_X \, \Delta \sigma_Z \geq \frac{\hbar}{2},
\]  

where $\Delta \sigma_X$ and $\Delta \sigma_Z$ denote the uncertainties in diagonal and rectilinear polarizations, respectively. The Heisenberg Uncertainty Principle underpins the security of QKD protocols, as it ensures that measurement in one basis inherently randomizes the outcome in its conjugate basis \cite{W83}. For instance, a photon prepared in the horizontal polarization state $|H\rangle$ produces a deterministic result when measured in the rectilinear basis $\{|H\rangle, |V\rangle\}$, but measurement in the diagonal basis $\{|D\rangle, |A\rangle\}$, where 

\[
|H\rangle = \frac{1}{\sqrt{2}}(|D\rangle + |A\rangle),
\]

yields a completely random outcome.  

Consequently, any attempt by an eavesdropper to extract information from the quantum channel necessarily disturbs the quantum states, generating detectable errors. This unavoidable disturbance provides a fundamental, physics-based guarantee of security, forming a cornerstone of unconditional security in QKD protocols \cite{NC10}. The information-disturbance trade-off implied by the uncertainty principle ensures that any eavesdropping attempt can be reliably detected, thereby preventing covert interception of the key.  

\subsection{QKD Protocols: Classification and Evolution of Quantum Cryptographic Schemes}\label{QKD Protocols: Classification and Evolution of Quantum Cryptographic Schemes}

\subsubsection{DISCRETE-VARIABLE (DV) QKD Protocols}
In DV QKD protocols, the encoding is done in discrete variables of a quantum state like the polarization of single photons with examples being BB84 \cite{BB84}, B92 \cite{B92}, SARG04 protocol \cite{SAR03}.
\paragraph{Prepare-and-Measure Protocols} 
The term prepare-and-measure arises from the operational structure of this class of QKD protocols, where the sender (Alice) prepares quantum states-commonly realized as polarized photons-and the receiver (Bob) subsequently measures these transmitted states. The security of such protocols is fundamentally rooted in the Heisenberg uncertainty principle, which asserts that it is impossible to measure a quantum state without disturbing it \cite{NS18, PAB+20}. In practice, this implies that any attempt by an adversary to gain information unavoidably introduces detectable disturbances in the transmitted states. Furthermore, the no-cloning theorem reinforces this security, as it establishes that an arbitrary and unknown quantum state (qubit) cannot be perfectly copied or amplified without altering its properties. Together, these principles ensure that any eavesdropping attempt manifests as an increase in the error rate observed during transmission. By monitoring these error parameters, Alice and Bob can reliably detect the presence of an eavesdropper and decide whether the communication channel remains secure.  
\paragraph{BB84}
In 1984, Charles H. Bennett and Gilles Brassard introduced the first quantum cryptographic protocol, known as BB84~\cite{BB84}, which laid the foundation for secure quantum communication. The protocol establishes a method for two legitimate parties-commonly referred to as Alice and Bob-to generate a shared secret key over an insecure communication channel, a task that is fundamentally impossible using classical means without relying on computational assumptions. In the BB84 framework, Alice and Bob are assumed to operate within secure laboratories and are connected by two types of channels: a quantum channel, through which qubits are transmitted, and an authenticated classical channel used for public communication. While the classical channel is assumed to be immune to tampering, the quantum channel is considered fully accessible to an eavesdropper, Eve, who can attempt to intercept and measure the transmitted quantum information but cannot modify the classical communication without detection. The protocol relies on the unique properties of quantum mechanics, particularly the no-cloning theorem and the behavior of measurements in non-orthogonal bases. Alice encodes each bit of her raw key into the polarization state of a single photon, randomly selecting one of two mutually unbiased bases: the computational (or rectilinear) basis $\{|0\rangle, |1\rangle\}$, and the diagonal (or Hadamard) basis $\{|+\rangle, |-\rangle\}$, where $|\pm\rangle = (|0\rangle \pm |1\rangle)/\sqrt{2}$. In this encoding scheme, the states $|0\rangle$ and $|+\rangle$ represent the logical bit 0, while the states $|1\rangle$ and $|-\rangle$ represent bit 1. Bob, independently and without prior knowledge of Alice’s choice of basis, measures each incoming photon using a randomly selected basis. When Bob’s measurement basis coincides with Alice’s encoding basis, the measurement outcome perfectly matches Alice’s transmitted bit. Conversely, if the bases are mismatched, Bob obtains an outcome that is completely uncorrelated with Alice’s bit. After a sufficiently large number of qubits have been exchanged, Alice and Bob publicly announce the bases used for each transmission over the classical channel, retaining only those bits for which their bases match. This process, known as sifting, produces a sifted key shared between the two parties, which is identical in the ideal case of noiseless transmission and no eavesdropping. To ensure the security of the key, Alice and Bob perform a parameter estimation phase, in which a subset of the sifted key is revealed to estimate the QBER. The QBER quantifies the fraction of bits that disagree between Alice and Bob and serves as a direct indicator of disturbances in the quantum channel, which may arise due to environmental noise, imperfections in the transmission and detection systems, or the presence of an eavesdropper. If the observed QBER exceeds a predetermined threshold, the protocol is considered insecure, and the key generation process is aborted. Otherwise, Alice and Bob proceed to information reconciliation, where error correction techniques are employed to reconcile any discrepancies and ensure that both parties hold identical keys. Following this, privacy amplification is performed to reduce any partial information that Eve may have gained about the key, thereby generating a final, secure cryptographic key suitable for use in conventional encryption schemes. The BB84 protocol represents a paradigm shift in cryptography, as its security is based on the fundamental laws of quantum mechanics rather than computational complexity. By leveraging the principles of superposition, measurement disturbance, and quantum uncertainty, BB84 guarantees that any attempt at eavesdropping inevitably introduces detectable errors, allowing Alice and Bob to quantify and mitigate potential information leakage. This pioneering work has inspired a vast range of QKD protocols and continues to serve as the cornerstone of both theoretical and experimental developments in quantum cryptography, particularly in the context of long-distance and satellite-based quantum communication networks.
\paragraph{B92}
The B92 protocol, proposed by Bennett in 1992 \cite{B92}, is a simplified variant of the BB84 protocol. Unlike BB84, which employs four quantum states, B92 uses only two non-orthogonal states, typically chosen as $|0\rangle$ and $|+\rangle$, to encode the logical bits 0 and 1, respectively. In the B92 scheme, Alice randomly prepares each qubit in either of the two states and transmits them to Bob. Upon reception, Bob randomly chooses a measurement basis for each photon, either the computational basis $\{|0\rangle, |1\rangle\}$ or the diagonal basis $\{|+\rangle, |-\rangle\}$. The key feature of B92 is that only certain measurement outcomes allow Bob to unambiguously infer Alice’s transmitted state. Specifically, outcomes $|1\rangle$ or $|-\rangle$ provide definitive information about the bit sent, while outcomes $|0\rangle$ or $|+\rangle$ are inconclusive and discarded. Bob then communicates which qubits yielded conclusive results so that Alice can retain the corresponding entries, forming the raw key for further processing. The security of the B92 protocol arises directly from the non-orthogonality of the transmitted states, which prevents an eavesdropper from perfectly distinguishing them without introducing detectable errors. Any attempt at measurement by an eavesdropper inevitably disturbs the qubits, allowing Alice and Bob to detect potential interception through the observed error rate. By using only two states, B92 achieves a simpler implementation than BB84, making it particularly suitable for practical experimental setups and systems with limited resources, while still providing a robust framework for secure QKD. \\
 The BB84 and B92 protocols are the most prominent examples of prepare and measure QKD protocols. Other variants, such as the SARG04 and six-state protocols, further extend this concept by using different sets of quantum states or measurement strategies to enhance robustness against channel noise and eavesdropping. 
\subsubsection{Entanglement-Based Protocols} 
\paragraph{E91}
The E91 protocol is an entanglement-based QKD scheme \cite{E91} in which a source emits pairs of maximally entangled spin-\(\tfrac{1}{2}\) particles (or polarization-entangled photons) in the singlet state. One particle from each pair is sent to Alice and the other to Bob. For each received particle, Alice and Bob independently and randomly choose one of three analyzer orientations lying in the transverse (\(x\!-\!y\)) plane. The azimuthal angles defining Alice’s settings are
\[
\phi_1^{(a)}=0,\quad \phi_2^{(a)}=\tfrac{\pi}{4},\quad \phi_3^{(a)}=\tfrac{\pi}{2},
\qquad
\phi_1^{(b)}=\tfrac{\pi}{4},\quad \phi_2^{(b)}=\tfrac{\pi}{2},\quad \phi_3^{(b)}=\tfrac{3\pi}{4}.
\]
Each local measurement yields a dichotomic outcome \(x,y\in\{+1,-1\}\). When the two parties happen to select the same orientation, the singlet correlations produce perfect anticorrelation, enabling those outcomes to be mapped to raw key bits after sifting.

Outcomes obtained with different analyzer settings are used to estimate nonlocal correlations. Let \(P_{\alpha\beta}(a_i,b_j)\) denote the joint probability of outcomes \(\alpha,\beta\in\{+1,-1\}\) given settings \(a_i\) and \(b_j\). The correlation function is
\begin{equation}
E(a_i,b_j)= P_{++}(a_i,b_j)+P_{--}(a_i,b_j)-P_{+-}(a_i,b_j)-P_{-+}(a_i,b_j),
\label{eq:Eij}
\end{equation}
which, for the singlet, is equivalent to the quantum prediction \(E(a_i,b_j)=-\mathbf{a}_i\!\cdot\!\mathbf{b}_j\).
To certify security, Alice and Bob evaluate the CHSH parameter using a subset of setting pairs:
\begin{equation}
S=E(a_1,b_1)-E(a_1,b_3)+E(a_3,b_1)+E(a_3,b_3).
\label{eq:CHSH}
\end{equation}
Local-hidden-variable models satisfy \(|S|\le 2\), whereas quantum mechanics requires \(|S|=2\sqrt{2}\) for appropriately chosen angles. Any eavesdropping attempt that gains information necessarily alters the two-party statistics and diminishes the Bell violation, driving the observed \(S\) toward the classical bound.

After parameter estimation (Bell test) confirms a statistically significant violation of Eq. \eqref{eq:CHSH}, the parties keep only the measurement results obtained with identical analyzer orientations to form the sifted key. In this way, E91 integrates key generation with an inherent security test grounded in Bell’s theorem.

\paragraph{BBM92}
The BBM92 protocol, introduced by Bennett, Brassard, and Mermin, is an another entanglement-based QKD scheme \cite{BBM92} that avoids reliance on Bell's theorem. In this protocol, an entangled source distributes photon pairs to Alice and Bob, who each randomly choose to measure along the $Z$ or $X$ basis. After measurement, they announce only their basis choices and retain the data corresponding to matching bases, which are perfectly correlated (or anti-correlated) and thus form the raw key. To detect eavesdropping, a subset of outcomes is publicly compared; any adversarial disturbance would break the correlations and produce detectable errors. Unlike Ekert's E91 protocol, BBM92 does not require testing Bell inequalities-its security follows directly from the impossibility of faking entanglement correlations without introducing observable errors. Conceptually, BBM92 can be seen as the entanglement-based counterpart of BB84: in BB84, Alice actively prepares and sends single-qubit states, while in BBM92, the randomness arises naturally from entanglement measurements. This intrinsic symmetry makes BBM92 a foundational protocol for entanglement-based QKD systems and a bridge between prepare-and-measure and entanglement-assisted approaches.

\subsubsection{Continuous-Variable (CV) QKD Protocols}
  The development of continuous-variable QKD (CV-QKD) has evolved from theoretical concepts to practical and secure implementations. In CV QKD, the message
is encoded in continuous variables like quadratures of coherent or squeezed state \cite{PMB26}. The earliest CV-QKD scheme, proposed in 1999, employed squeezed states for secret key generation \cite{R99}, but experimental challenges in producing such states led to the introduction of a more practical approach in 2002 using coherent states and homodyne detection, known as the GG02 protocol \cite{GG02}. This protocol, relying on easily generated laser light, sparked extensive research interest. Later, heterodyne-based no-switching protocols \cite{WLB04} were introduced to enhance performance. Despite the success of Gaussian-modulated schemes, achieving ideal Gaussian modulation in practice remained difficult, motivating the proposal of discrete-modulated CV-QKD protocols \cite{LG09, MB25}. Variants like unidimensional \cite{UG15} and phase-sensitive multimode schemes \cite{MKB25} have further improved noise tolerance and key rates. Alongside these, two-way, Source-Device-Independent (SDI), and Measurement-Device-Independent (MDI) CV-QKD protocols \cite{GVK+22} have been developed to address device trust and security loopholes. 
\subsubsection{Distributed-Phase-Reference (DPR) QKD Protocols}
  Compared to other QKD schemes, DPR QKD protocols offer relatively simple experimental implementation and high communication efficiency. These protocols are primarily categorized into two main types: the Differential Phase Shift (DPS) QKD protocol and the Coherent One-Way (COW) QKD protocol. The DPS protocol was first proposed in 2002~\cite{IWY02}, while the COW protocol was introduced in 2005~\cite{SBG+05}. Since their introduction, significant progress has been made in improving their experimental realizations and extending their achievable communication distances.

In addition to the well-known prepare-and-measure, entanglement-based, CV, and DPS protocols, several other QKD schemes have been developed to enhance security, efficiency, and practical applicability. Two-way QKD protocols, such as the Ping-Pong protocol \cite{BF02}, utilize a bidirectional quantum channel to improve resilience against certain attacks. MDI QKD protocols \cite{LCQ12} eliminate vulnerabilities associated with detection devices, enabling secure key distribution even with untrusted measurement units. Twin-field QKD protocols \cite{LYD+18} further extend communication distances by exploiting single-photon interference, surpassing the rate-distance limit of conventional schemes. HD QKD protocols encode information in higher-dimensional quantum states, increasing the key rate per photon and enhancing robustness against noise. These diverse approaches collectively expand the landscape of QKD, addressing practical challenges and paving the way for secure quantum communication over long distances.
\subsection{Weak Coherent Pulses and the Decoy State Method}
Deterministic single-photon sources are still in the early stages of development. Consequently, alternative approaches such as probabilistic single-photon sources and WCPs have been introduced. WCPs are essentially laser pulses that are strongly attenuated to approximate single-photon behavior. However, this substitution must be carefully considered in the security analysis of QKD protocols, as it may introduce potential vulnerabilities that an eavesdropper could exploit. A strongly attenuated laser pulse can be modeled as a coherent state characterized by a low mean photon number. These coherent states represent superpositions of different photon-number states. Regardless of how small the mean photon number becomes, there always exists a finite probability that a pulse contains multiple photons. The existence of multiphoton components renders the system susceptible to a Photon Number Splitting (PNS) attack \cite{L00}. In this attack, the eavesdropper (Eve) employs a Quantum Non-Demolition (QND) measurement to ascertain the number of photons in Alice’s transmitted pulse without disturbing the encoded quantum state. If the pulse contains only a single photon, Eve blocks it, preventing its detection by Bob. However, if the pulse comprises multiple photons, Eve retains one photon for her measurement and transmits the remaining photons to Bob through an ideal lossless channel, thereby gaining information without introducing detectable errors. Eve stores the intercepted photons in quantum memory until Bob publicly reveals his measurement bases. She then measures the stored photons using the same bases as Bob. Consequently, Eve obtains identical measurement outcomes to those of Bob, rendering the key generation process insecure. Therefore, only single-photon pulses contribute to secure key generation, while multiphoton pulses are entirely vulnerable to interception. Choosing a very low average photon number effectively suppresses the occurrence of multi-photon pulses, but this simultaneously increases the probability of transmitting vacuum states, which reduces the efficiency of the communication. As a result, the secure key rate exhibits a quadratic dependence on the channel transmittance, in contrast to the linear scaling observed when employing a true single-photon source.\\
A significant advancement in QKD was achieved with the introduction of the decoy-state method \cite{H03}. The PNS attack exploits the ability of an eavesdropper to identify and block single-photon pulses. Consequently, the effective transmittance of the quantum channel between Alice and Bob-including the actions of Eve-becomes dependent on the photon number of the transmitted signals, rendering the channel non-passive.
 By testing the quantum channel during a QKD session, the presence of a PNS attack can be detected, allowing the protocol to be aborted if necessary. In decoy-state QKD, the characteristics of the otherwise unknown channel are probed by analyzing its response to input signals of varying photon numbers. Alice and Bob implement the standard QKD operations using weak laser pulses with different intensities-typically one ``signal state'' and one or more ``decoy states''-and separately evaluate the transmittance and QBER for each intensity. If Eve attempts a PNS attack, she inevitably introduces differing losses between the signal and decoy states, which can be detected. It can thus be shown that the secure key rate regains linear scaling with channel transmittance when the decoy-state method is employed.
The initial security proofs for the decoy-state method were based on the assumption of an infinite number of decoy states. Subsequent studies demonstrated that the simpler vacuum+weak decoy-state approach can achieve a key rate nearly equivalent to that obtained in the infinite decoy-state case \cite{MQZ+05}. A key assumption underlying decoy-state protocols is that the signal and decoy states are indistinguishable in every aspect except their average photon number. This ensures that Eve cannot determine whether a received pulse originates from a signal or a decoy state.

\section{Satellite-Based QKD}
\label{Satellite-Based QKD}
QKD, and more broadly quantum communication, possess the potential to fundamentally revolutionize the secure transmission of information over the internet. Photons, which serve as the natural carriers of quantum information, are already widely utilized in classical optical fiber networks to achieve high data transmission rates. However, despite remarkable advancements in recent years~\cite{MPR+19, BBR+18}, realizing long-distance quantum communication remains a formidable challenge due to the intrinsic transmission losses encountered during propagation through optical fibers. To overcome these limitations, several schemes for the implementation of quantum repeaters have been proposed in recent years, providing a feasible pathway to extend quantum communication over large distances and seamlessly integrate them into quantum networks~\cite{MAK+15, ATL15, ZPD+18, SGL18, SSD+11}. Nevertheless, considering the substantial technological hurdles that must be addressed before quantum repeaters become practically realizable, satellite-based free-space optical links have emerged as the most promising near-term solution for achieving long-distance QKD~\cite{BBM+15}. These systems can capitalize on the mature technologies developed in satellite engineering and optical communication over the past few decades. Over the last twenty years, numerous feasibility studies have explored the potential of satellite-based quantum communication~\cite{BBM+15, BAL17, BMH+13, BTD+09}, and several experimental demonstrations have successfully validated the readiness of the underlying technology for real-world deployment~\cite{LCL+17, YCL+17, RXY+17}. Optical links established between satellites and ground stations facilitate quantum communication over distances exceeding 1000~km~\cite{LCL+17}. A typical free-space optical link consists of two primary subsystems: a transmitting telescope and a receiving telescope. The transmitter emits a well-collimated optical beam directed toward the receiving station, where the receiver telescope collects the incoming photons and guides them to the detection and analysis modules for further processing. \\
The most fundamental satellite-based quantum communication configurations are categorized into downlinks and uplinks. In a downlink configuration, the optical signal is transmitted from the satellite to an optical ground station, where the receiver is located. Conversely, in an uplink configuration, the ground station acts as the transmitter, directing the optical beam toward the satellite, which houses the receiver module. Downlink channels generally exhibit significantly higher transmittance compared to uplink channels. This advantage arises because, in downlinks, atmospheric turbulence and absorption affect the optical beam only during the final portion of its propagation, thereby introducing relatively minor losses. In contrast, in uplink scenarios, the beam encounters atmospheric disturbances at the beginning of its propagation path, resulting in increased beam divergence and, consequently, greater overall transmission losses.

\section{Key rate analysis of QKD protocols in satellite-based QKD}
\label{ch2int}
The first QKD protocol, BB84, was introduced by Bennett and Brassard in 1984 \cite{BB84}. Since then, both theoretical and experimental studies of QKD have advanced significantly \cite{SBC+09, SPR17, PAB+20}. However, practical deployment remains constrained by the challenges of generating, maintaining, and manipulating quantum resources with current technology. This motivated the development of simpler protocols requiring fewer quantum resources. For instance, BB84 employs four quantum states and two measurement bases, whereas the B92 protocol, proposed by Bennett in 1992, uses only two nonorthogonal states and two bases \cite{B92}, though it is more sensitive to noise. The E91 protocol, introduced by Ekert in 1991, utilizes entangled photon pairs and ensures security via Bell inequality violations \cite{E91}, providing device-independent security but at the cost of reduced key generation efficiency due to additional measurements for Bell tests. To address these limitations, the BBM92 protocol was proposed in 1992 by Bennett, Brassard, and Mermin \cite{BBM92}, which combines entangled states with the BB84 measurement and sifting procedure. This approach eliminates the need for Bell tests, simplifies experimental implementation, and increases the fraction of detected events contributing to the sifted key, making it more practical for high-rate QKD.\\

In Chapter 2, we analyze the performance of four QKD protocols-BB84, B92, BBM92, and E91-encompassing both prepare-and-measure and entanglement-based schemes, for asymptotic key rate computation in satellite-based QKD using a circular beam transmittance model \cite{BMH+13}. The study considers both uplink and downlink scenarios, incorporating atmospheric transmittance from MODTRAN 6 \cite{BCK+14} and background noise, including stray counts, under day and night conditions for LEO satellite at 500 km altitude. Compared to medium and geostationary orbits, the shorter LEO propagation distance reduces free-space losses, increases photon detection probability, and enhances secure key rates. The circular beam model accounts for diffraction and Gaussian-distributed pointing errors in downlink links, while uplink configurations additionally include turbulence-induced beam broadening based on the Hufnagel-Valley profile \cite{V80,HS64}. By integrating atmospheric effects, orbital geometry, and background noise into a unified framework, this work \cite{MRS26} highlights the relative strengths and limitations of these QKD protocols in realistic LEO scenarios, providing practical guidance for protocol selection in future satellite QKD networks.

\section{Introductory Discussion on high dimension QKD protocols for satellite QKD}
\label{ch3int}
QKD ensures information-theoretic security based on the fundamental principles of quantum mechanics rather than computational assumptions. However, traditional QKD protocols such as BB84 \cite{BB84} and B92 \cite{B92} rely on qubits-two-dimensional (2D) quantum systems for encoding information, which imposes certain limitations on key generation rate and noise tolerance. To address these challenges and improve the performance of QKD systems, researchers have been exploring HD extensions of these protocols, wherein information is encoded on qudits-quantum systems with $d>2$ levels \cite{BT00, CBK+02, CDB+19}. HD QKD protocols exploit larger Hilbert spaces to encode more than one bit of information per photon, thereby enhancing channel capacity, improving noise tolerance, and increasing robustness against eavesdropping \cite{SPD+25, HLS+24, LZL+22}. Various photonic degrees of freedom-such as orbital angular momentum, temporal, frequency, and spatial modes-serve as promising candidates for realizing these HD states. With advancements in single-photon sources and WCP technologies, HD-QKD schemes can be effectively implemented in practical systems using decoy-state methods to mitigate PNS attacks \cite{HK05, X05}. In the context of satellite-based QKD, implementing HD protocols becomes particularly advantageous. The satellite-based free-space channel experiences significant attenuation, beam wandering, and atmospheric turbulence, all of which reduce the overall key rate \cite{DW17}. By leveraging HD-QKD schemes, such as the HD-Ext-B92 \cite{IK21} and HD-BB84 \cite{CBK+02} protocols, it becomes possible to generate higher key rates and achieve greater noise resilience even under these challenging conditions. Moreover, the inclusion of decoy states and appropriate modeling of the atmospheric transmittance-such as through the elliptical beam approximation \cite{VSV+17, LKB19} enables a realistic evaluation of the secure key rate for both uplink and downlink configurations. In Chapter 3, our work focuses on a systematic and detailed investigation of the advantages and challenges involved in the satellite-based implementation of two HD QKD protocols, namely the HD-Ext-B92 and HD BB84 schemes. A modified key rate formulation for the HD-Ext-B92 protocol is developed to improve the analytical accuracy and reliability of the performance evaluation. Using this refined model, the variations in secure key rate, PDR, and QBER are analyzed as functions of the system dimension and the depolarizing noise parameter. Furthermore, the average key rate per pulse is studied as a function of the zenith angle and link length for different atmospheric conditions during both day and night scenarios, considering extremely low noise for a HD system with $d = 32$. The analysis employs the elliptical beam approximation to realistically model atmospheric effects such as turbulence, beam divergence, and transmittance fluctuations. In addition, the PDT is utilized to evaluate the average key rate for both uplink and downlink configurations, providing a more realistic understanding of satellite-based quantum communication performance. Our work \cite{DMB+24} demonstrate that in higher dimensions, the HD-BB84 protocol achieves superior performance compared to the HD-Ext-B92 protocol in terms of both secure key rate and noise tolerance. However, the HD-BB84 scheme also exhibits a more pronounced saturation in QBER at large dimensions, highlighting a trade-off between dimensionality and error stability. Overall, our work establishes the effectiveness of HD quantum protocols for satellite-based QKD and provides valuable insights into their optimization under realistic atmospheric conditions.

\section{Discussion on Finite and Asymptotic Key Rate Analysis in CubeSat QKD Systems}
\label{ch4int}
Asymptotic key rate analysis is essential for optimizing the performance of satellite-based QKD systems. For instance, \cite{ABN21} investigates the optimization of signal and decoy intensities to maximize secure key generation. Similarly, \cite{GBL+23} derives asymptotic key rate bounds for both CV and DV QKD under restricted eavesdropping, showing that limiting Eve's access can improve key rates, particularly in high-loss regimes.\\
With the growing feasibility of satellite-based QKD, accurately modeling secure key generation under realistic constraints, particularly in finite-key scenarios, is essential. Advances in finite-key analysis have improved secure key rate estimation by accounting for statistical fluctuations and optimizing security parameters \cite{PKH23}. Early approaches used numerical methods to include statistical uncertainties \cite{BGK+20}, while later SemiDefinite Programming (SDP) frameworks provided generalized finite-key analyses applicable to various QKD protocols, including device-independent schemes \cite{GLL21}. These methods have been extended to satellite QKD, addressing orbital dynamics, channel losses \cite{MB23}, and statistical corrections \cite{SBM+22, ELJ+21, DTR+21}, and more recently incorporating practical constraints such as hardware limitations, link efficiency, and environmental factors \cite{SBM+23}.
Building on recent advances, our work in Chapter 4 analyzes finite-key block sizes using tight statistical techniques for parameter estimation and error correction \cite{LCW+14, SBM+22}, and computes asymptotic key rates for WCP-based efficient and standard BB84 protocols employing the elliptical beam model in a CubeSat QKD downlink scenario \cite{MDB25}. The focus of space-based quantum communication has increasingly shifted toward smaller satellites, particularly CubeSats, which dominate the nanosatellite class. Recent missions, such as Jinan-1 \cite{LCJ+24}, have demonstrated secure quantum communication using compact platforms. Our study in Chapter 4 considers a CubeSat in LEO at 400 km altitude, favored for QKD due to lower channel loss, cost-effectiveness, and practical feasibility \cite{ZSH+24}. Compared to Middle Earth Orbit (MEO) and Geostationary Equatorial Orbit (GEO) satellites, the shorter LEO distances reduce free-space losses and enhance photon detection probabilities, thereby improving key rates \cite{LVM+23, SLM+22, MTP+24}. To model beam propagation realistically, we adopt the elliptical beam approximation \cite{LKB19}, which captures beam dynamics in CubeSat-based QKD links more accurately than circular beam models. Unlike previous studies on CubeSat-based QKD \cite{ZSH+24}, which mainly consider diffraction losses and background noise while neglecting detailed turbulence effects, our approach incorporates beam wandering, elliptical beam deformations, and extinction losses from backscattering and absorption. This enhanced turbulence model allows accurate transmittance evaluation through the receiving aperture by accounting for centroid position, elliptical semi-axes, and orientation angle, enabling realistic assessment of beam fluctuations under various weather conditions \cite{VSV16}. Moreover, while prior works such as \cite{SLM+22} focus on optical ground station implementation and entanglement distribution, they do not consider finite-key rate effects on security. In contrast, our work explicitly addresses finite-key analyses, providing a more comprehensive evaluation of security in CubeSat-based QKD.\\
Chapter 4 focuses on a downlink configuration over an uplink due to its lower transmission losses \cite{LKB19}, making it particularly suitable for CubeSat-based BB84 QKD \cite{GN22}, which is widely implemented owing to its simplicity and proven security \cite{SP00}. Our work \cite{MDB25} considers both efficient and standard BB84 protocols using the elliptical beam approximation for satellite-to-ground links under varying weather conditions. Key rate probability distributions are examined with respect to zenith angles to provide a comprehensive assessment of CubeSat QKD performance. The two decoy-state BB84 protocol is employed for finite and asymptotic key analyses, leveraging its well-established security, ease of implementation, experimental validation \cite{SP00, OKM+24}, and resilience against eavesdropping, as demonstrated in recent CubeSat implementations \cite{ZSH+24, LSL22}. To enhance key generation efficiency while maintaining security, our work implements both efficient and standard BB84 protocols with two decoy states. The efficient BB84 protocol uses a biased basis choice, improving the sifting ratio and raw key generation, which is particularly advantageous for CubeSats facing limited transmission time, high channel loss, and background noise \cite{SBM+22, MTP+24}. Standard BB84 remains relevant due to its symmetric basis selection, rigorous security framework, and broad applicability. Unlike prior work \cite{MTP+24}, which employs a simplified BB84 with a single decoy state and does not quantify the impact of varying atmospheric conditions on key rates, our study incorporates two decoy states, evaluates key rate probability distributions under different weather conditions, and analyzes both finite and asymptotic key rates using the elliptical beam model. This comprehensive approach provides a realistic assessment of CubeSat-based QKD performance, enhancing security, robustness against eavesdropping, and statistical depth.

\section{ Outline Of The Thesis}
\label{summary}
This thesis is structured into five chapters, each contributing to a comprehensive performance analysis framework for satellite-based QKD. \textbf{Chapter 1} presents the theoretical foundation by introducing essential quantum-mechanical concepts-such as qubits, quantum states, density matrices, entanglement, and quantum entropy-and motivates the need for QKD as a secure alternative to classical cryptographic systems threatened by quantum algorithms like Shor’s algorithm. It further highlights the advantages of satellite-based QKD for global-scale secure communication and outlines the motivation for the research present in subsequent chapters. \textbf{Chapter 2} presents a detailed performance analysis of four DV QKD protocols-BB84, B92, BBM92, and E91 within the context of satellite-based quantum communication. Focusing on LEO links due to their lower channel loss and higher detection probability, the study employs the circular beam propagation model to compute transmittance for both uplink and downlink scenarios. Key atmospheric effects such as diffraction, turbulence, attenuation, and pointing errors are included, with uplink turbulence modeled using the Hufnagel-Valley profile whereas downlink modeling included Gaussian-distributed pointing errors. Atmospheric transmittance data from MODTRAN~6 and day/night environmental noise are incorporated to assess their impact on QBER. The chapter derives QBER and asymptotic key-rate expressions for all four protocols and evaluates their performance under realistic mission conditions. The results demonstrate that atmospheric conditions, channel asymmetries, and protocol-specific requirements strongly influence performance and identify the conditions under which each protocol is best suited for satellite-based QKD. The results of this chapter are reported in \cite{MRS26}.
\textbf{Chapter 3} extends the investigation to high-dimensional QKD by analyzing the HD-Ext-B92 and HD-BB84 protocols using the elliptic-beam approximation for realistic turbulence modeling. Motivated by the challenges of LEO satellite links-where turbulence, beam wandering, and diffraction significantly affect transmittance-the study incorporates decoy-state methods and depolarizing noise to evaluate both protocols under practical free-space conditions. A comprehensive numerical analysis of the key rate, QBER, and the key-rate probability distribution across different dimensions, weather conditions, noise levels, and zenith angles shows that HD-BB84 achieves consistently higher and more stable performance than HD-Ext-B92 for both uplink and downlink configurations. The results furthermore reveal that HD-BB84 produces more concentrated and higher-probability regions around favorable key-rate values, indicating greater robustness against channel fluctuations. Overall, this chapter demonstrates that high-dimensional encoding, particularly through the HD-BB84 protocol, offers significant performance advantages for satellite-based QKD systems. The work presented in this chapter is published in Ref. \cite{DMB+24}.
 \textbf{Chapter 4} focuses on finite-key and asymptotic performance analysis of the efficient and standard BB84 protocols for CubeSat-based downlink QKD. Motivated by the rise of CubeSats as practical quantum communication platforms, the study incorporates realistic atmospheric modeling, link geometry, and statistical constraints relevant to short satellite passes. Photon propagation through turbulence is modeled using the elliptical-beam approximation-also employed in Chapter~3-which offers a more accurate description of downlink transmittance than the conventional circular-beam model, particularly due to strong distortions near the ground station. Weather-dependent turbulence and scattering effects (clear, hazy, foggy, and windy conditions) are included, and both protocols are evaluated using a two-decoy-state scheme with weak coherent pulses. The finite-key framework incorporates multiplicative Chernoff bounds and refined error-correction leakage, while the asymptotic analysis provides the fundamental key-rate limits by removing finite-size effects. Key-rate expressions for both regimes are implemented using the PDT to compute zenith-angle-dependent average key rates. Numerical results show that the efficient BB84 protocol consistently outperforms the standard version in both finite and asymptotic settings. Its biased-basis strategy improves sifting efficiency, reduces statistical fluctuations, and enhances robustness under atmospheric loss. Conversely, the standard BB84 protocol suffers from lower rates and higher sensitivity, especially under strong turbulence or fog. Overall, this chapter delivers a complete CubeSat QKD performance study integrating atmospheric modeling, finite-size corrections, PDT-based averaging, and weather-dependent analysis. The results demonstrate that efficient BB84 offers superior key-rate performance, stability, and operational range for CubeSat missions. The work done in chapter 4 is published in Ref.~\cite{MDB25}. \textbf{Chapter 5} summarizes the key findings of the thesis and concludes with a discussion of future research directions motivated by these results.
\label{chap2:Background}



\chapter{Performance analysis of QKD protocols}\label{chap2:Background}
\section{Introduction}
The rapid advancement of quantum communication technologies has established QKD as a cornerstone for achieving unconditional security in data transmission~\cite{SBC09, PAB+20}. Over the years, several QKD protocols have been proposed; however, their practical performance strongly depends on the characteristics of the quantum channel through which photons propagate. Among the various implementation platforms, satellite-assisted QKD provides a promising approach to extend secure communication beyond terrestrial boundaries, enabling intercontinental key exchange~\cite{MB25}. The feasibility and efficiency of such satellite-based systems are primarily influenced by factors such as optical beam propagation, atmospheric attenuation, and background noise, all of which significantly affect the achievable secure key rate~\cite{BMH+13, SB18}. This chapter investigates the performance of four QKD protocols-BB84, B92, BBM92, and E91-representing both prepare-and-measure and entanglement-based schemes discussed in Section \ref{QKD Protocols: Classification and Evolution of Quantum Cryptographic Schemes}, for the computation of asymptotic key rates in satellite-based QKD using the circular beam transmittance model~\cite{BMH+13}. The analysis is carried out for both uplink and downlink configurations, incorporating atmospheric transmittance values obtained from MODTRAN~6 \cite{BCK+14}. Background noise, including stray counts, is considered for both transmission directions under day and night conditions for a low LEO satellite at an altitude of 500~km. Compared to MEO and GEO satellites, the shorter propagation distance in LEO reduces free-space losses, increases photon detection probability, and thereby enhances secure key generation rates. Beam propagation is modeled using the circular beam approximation~\cite{BMH+13}, which provides a realistic representation of optical beam dynamics in LEO-based QKD links. The model accounts for diffraction and Gaussian-distributed pointing errors in downlink scenarios, while in uplink configurations, it additionally incorporates beam broadening caused by atmospheric turbulence, modeled using the Hufnagel-Valley turbulence profile~\cite{V80, HS64}. By incorporating atmospheric modeling, orbital configuration, and background noise effects into a unified framework, this work demonstrates how the resource requirements of different QKD protocols interact with the inherent asymmetries of uplink and downlink satellite channels. The findings \cite{MRS26} provide a clear comparison of the relative strengths and limitations of BB84, B92, BBM92, and E91 in realistic LEO scenarios, offering practical guidelines for protocol selection in future large-scale QKD networks. The remainder of this chapter is organized as follows. Section~\ref{Channel Modeling} describes the channel modeling framework, including the relevant loss mechanisms. Section~\ref{Environmental Photons} discusses environmental noise with a focus on stray photon counts. In Section~\ref{QBER}, the QBER is analyzed for the considered protocols, while Section~\ref{Keyrate} presents the corresponding key rate formulations. Section~\ref{Numerical Results and Discussion} reports the numerical results and offers a detailed discussion. Finally, Section \ref{Conclusion_2} concludes the chapter with a discussion of key insights.
\section{Channel Modeling}\label{Channel Modeling}
 To assess the practicality of satellite QKD, a numerical framework was employed to estimate the overall channel loss. The loss from the diffraction is calculated by the Rayleigh-Sommerfeld diffraction \cite{G96} . After considering the loss, the intensity at the receiver located at position $\vec{l}$, 
\begin{equation}
I_1(\vec{l}) =  \left| \frac{d^2}{\lambda^2}\iint_{S_t} \sqrt{I_0(\vec{l}\,')} \cdot \frac{1}{|\vec{l} - \vec{l}\,'|^2} \cdot \exp\left( \frac{2\pi i}{\lambda} |\vec{l} - \vec{l}\,'| \right) \, dx\,dy \right|^2,
\end{equation}
where $\vec{l'}$ represents a point on the transmitter surface $S_t$, over which the integration is performed. $I_0$ is the intensity distribution at the transmitter, $\lambda$ is the wavelength of the optical beam, and $d$ is the propagation distance between the satellite and ground station. Since the beam exhibits circular symmetry, the intensity can be evaluated along $x=0$ (or $y=0$), from which the radial distribution $I(r)$ is determined, where $r$ denotes the distance from the beam center.
 \\
The loss resulting from pointing errors is analyzed by examining the distribution of the beam center at the receiver. This distribution is modeled as a two-dimensional Gaussian function representing the pointing error, expressed as:
\begin{equation}
g(r) = \frac{1}{2\pi\sigma_p^2} \exp\left(-\frac{r^2}{2\sigma_p^2}\right).
\end{equation}
Here, $\sigma_p$ denotes the standard deviation associated with the pointing error. The resulting beam intensity profile at the receiver, denoted by $I_2(\vec{l})$, is obtained by performing a two-dimensional convolution \cite{B09} of the diffracted beam $I_1$ with the Gaussian distribution $g$, which accounts for the pointing error at the receiver. This can be expressed as:
\begin{equation}
I_2(\vec{l}) = (I_1 * g)(r,\theta) = \int_0^{2\pi} d\theta' \int_0^{\infty} I_1(r')\, g(r - r')\, dr'.
\end{equation}
For uplink transmission, it is important to account for the beam broadening induced by atmospheric turbulence, which evolves over a characteristic timescale of the order of 10-100~ms. When the beam profile is averaged over durations substantially longer than this timescale, the resulting spatial distribution converges to a Gaussian form. Based on the Hufnagel-Valley model of atmospheric turbulence~\cite{AS93} the beam waist $w$ at the receiver is given by:
\begin{equation}
w = \frac{2\sqrt{2}d\lambda}{\pi r_0},
\end{equation}
where $r_0$ is the transverse coherence length, can be expressed as:
\begin{equation}
r_0 = \left[ 1.46 \sec({\phi})\left( \frac{2\pi}{\lambda} \right)^2 \int_0^h C_n^2(z) \left(1 - \frac{z}{h} \right)^{5/3} dz \right]^{-3/5},
\end{equation}
where $\phi$ denotes the zenith angle, $h$ is the altitude of the receiver, and $C_n^2(z)$ represents the refractive-index structure constant as a function of altitude as given by:
\begin{equation}
C_n^2(z) = 0.00594\left(\frac{v}{27}\right)^2(z \times 10^{-5})^{10} e^{-z/1000} + 2.7 \times 10^{-16} e^{-z/1500} + A e^{-z/100}.
\end{equation}
The coefficients $v$ and $A$ characterize the influence of atmospheric turbulence and are defined based on topological atmospheric conditions. For this analysis, we adopt typical nighttime sea-level values of $A = 1.7 \times 10^{-14}~\mathrm{m}^{-2/3}$ and $v = 21~\mathrm{m/s}$~\cite{BMH+13}. After determining the waist of the Gaussian distribution, the overall beam profile $I_3(\vec{v})$ is obtained by applying a two-dimensional convolution of the beam profile $I_2(\vec{v})$ with the turbulence-induced broadening.
Once the beam intensity distribution at the receiver has been determined-whether for an uplink or downlink scenario (where $I_3 = I_2$); the corresponding received optical power $P$ is calculated by integrating the intensity profile $I_3(\vec{v})$ over the receiver aperture:
\begin{equation}
P = \iint_{S_r} I_3(\vec{v})\, dx\, dy,
\end{equation}
where $S_r$ denotes the surface area of the receiving telescope. The resulting optical power is directly proportional to the mean number of detected photons.
The impact of atmospheric transmission and detector efficiency is accounted for by scaling the received optical power with the detector efficiency, $\eta_d$, and the atmospheric transmittance, $\eta_t$. The values of $\eta_t$ for both uplink and downlink scenarios are obtained from MODTRAN 6 \cite{BCK+14} simulations, using the modtran parameters specified in \cite{BMH+13}. Finally, the ratio of the received power to the transmitted power $P_0$ is expressed in decibels to compute the total channel loss. An additional fixed loss of 3~dB is included to account for imperfections in the polarization analyser and intrinsic losses associated with optical components and telescope coupling \cite{BMH+13}. 
\begin{equation}
L = -10 \log_{10} \left( \frac{\eta_t \eta_d P}{P_0} \right) + 3. \label{eq:loss}
\end{equation}

\section{Environmental Photons (stray-photons) \label{Environmental Photons}}
In free-space communication, environmental photons represent a significant noise source. Here, we'll summarize the analysis concerning the impact of these environmental photons on the detector, for both the transmission from a satellite to the ground (downlink) and from the ground to a satellite (uplink), which is crucial in calculating the QBER. We solely focus on the scenario of uplinks during nighttime operation. If the ground station site has a low level of light pollution, the biggest fraction of environmental photons comes from the sunlight reflected first by the moon and then by the earth \cite{LKB19}.
\begin{equation}
    N_{night}^{up} = A_E A_M R_M^2 a^2 \frac{\Omega_{fov}}{d_{EM}^2} B_f \Delta t H_{sun}.
\end{equation}
In this equation, the symbols denote the following: $A_M$ corresponds to the albedo of the Moon, $R_M$ denotes the Moon's radius, $A_E$ represents the Earth's albedo, and $d_{EM}$ signifies the distance between the Earth and the Moon. $H_{sun}$ is an indicator of the solar spectral irradiance measured in photons per second per nanometer per square meter at the specific wavelength under consideration. $\Omega_{fov}$ and $a$ represent the angular field of view (AFOV) and the receiving telescope's radius, respectively. $B_f$ denotes the spectral filtering width, and $\Delta t$ signifies the detection time-window.\\
The assessment of background photons in downlinks varies significantly depending on the location. The formula to express a telescope's receiving power is as follows:
\begin{equation}
    P_b = H_b \Omega_{fov} \pi a^2 B_f.
\end{equation}
As a result of the weather and the hour of the day, the parameter $H_b$ determines the total brightness of the sky background. The number of photons per time window can be calculated from the equation above-
\begin{equation}
    N^{down} = \frac{H_b}{h\nu} \Omega_{fov} \pi a^2 B_f \Delta t, 
\end{equation}
In this equation, the symbol $h$ represents the Planck constant, while the symbol $\nu$ represents the frequency of the background photons, which have been filtered.

\begin{table}
\centering
\begin{tabular}{|c|c|c|}
\hline
 Parameter & Value & Brief description  \\
 \hline
$\lambda$ & $785\,\text{nm}$ & Wavelength of the signal light \\
\hline
$r_T$ & $15\,\text{cm},\,50\,\text{cm}$ & Downlink, Uplink \\
\hline
$a$ & $50\,\text{cm},\,15\,\text{cm}$ & Downlink, Uplink \\
\hline
$\eta_d$ & $0.5$ & Detector efficiency \\
\hline
$p_{dark}$ & $5\times10^{-8}$ & Detector efficiency \\
\hline
$\sigma_P$ & $1.2\times10^{-6}$ & Pointing error \\
\hline
 $H$ & 500 km & Minimum altitude (zenith)  \\
\hline
$H_b$ & $1.5\times10^{-6} W m^{-2} sr^{-1} nm^{-1}$ & Night, clear sky  \\
\hline
 $H_b$ & $1.5\times10^{-3} W m^{-2} sr^{-1} nm^{-1}$ & Day, clear sky  \\
\hline
AFOV, $\Omega_{fov}$ & $(100\times10^{-6})^2 sr$ & Night-time downlink  \\
\hline
AFOV, $\Omega_{fov}$ & $(10\times10^{-6})^2 sr$ &     Day-time downlink  \\
\hline
AFOV, $\Omega_{fov}$ & $(30\times10^{-6})^2 sr$ & Night-time uplink  \\
\hline
 $\Delta t$  & $1 ns$ & Night and day time \\
\hline
$B_f$ & $1 nm$ & Night-time downlink  \\
\hline
 $B_f$ & $0.2 nm$ & Day-time downlink  \\
\hline
 $B_f$ & $1 nm$ & Night-time uplink  \\
\hline
$H_{sun}$ & $4.610\times10^{18}$ $phot$ $s^{-1}$ $nm^{-1}$ $m^{-2}$ & Solar spectral irradiance  \\
\hline
$A_E$ & $0.300$ & Earth's albedo  \\
\hline
$A_M$ & $0.136$ & Moon's albedo  \\
\hline
$R_M$ & $1.737\times10^6 m$ & Moon's radius  \\
\hline
$d_{EM}$ & $3.600\times10^8 m$ & Earth-moon distance  \\

\hline
\end{tabular}
\caption{Parameters related to the atmospheric weather conditions, stray photons and environmental light \cite{LKB19} }
\label{table_1st}
\end{table}
\section{QBER}\label{QBER}
 The QBER is a measure of the ratio of incorrect bit counts to the total number of received bit counts. It is used to quantify the probability of obtaining a false detection in comparison to the total probability of detection per pulse. The QBER is influenced by three main components: the signal component, the dark count component and the stray count component i.e. environmental photons. 
\subsection{ QBER for BB84}
In BB84 protocol, the QBER can be calculated as \cite{L00, LKB19}:
\begin{equation}
    e_{84} = \frac{c \; p_{signal}+\frac{1}{2} \; (p_{dark}+p_{straycounts})}{p_{click}} .
\end{equation}
Here, $c$ corresponds to the error rate associated with depolarization in the encoding degree of freedom or imperfection of the preparation or detection stage leading to incorrect state discrimination. We chose a conservative value of $c=1\%$ \cite{L00}, $p_{click}$ represents the overall anticipated probability that Bob will observe the detection of a photon during a specific pulse. Typically, $p_{click}$ is determined by considering three distinct sources that can independently trigger a detection event. These sources encompass photons transmitted by Alice, background dark counts and straycounts.
\begin{equation}
    p_{click} = p_{signal}+p_{dark}+p_{straycounts}.\label{eq:pclick}
\end{equation}
The probability of Bob's detector firing due to a photon emitted by Alice's source is denoted as $p_{signal}$ and is given by-
\begin{equation}
     p_{signal}=1-\exp(-\eta_d \eta \mu), 
\end{equation}
where $\eta_d$ is the detector efficiency, $\eta$ is the total transmittance efficiency and is calculated using Eq.\eqref{eq:loss} and $\mu$ is the average number of photons per pulse.
On the other hand, $p_{dark}$ represents the probability of a dark count occurring in Bob's detector \cite{L00}. $p_{straycounts}$ is the probability of occuring straycounts which can be calculated using the expressions from Sec. \ref{Environmental Photons}. It is important to note that the Eq.\eqref{eq:pclick} assumes the neglect of simultaneous occurrences of signal, dark count and straycounts events when all three $p_{signal}$, $p_{dark}$ and $p_{straycounts}$  are small.\par
\subsection{ QBER for B92}
In the B92 protocol, the logical bits are encoded using two non-orthogonal basis states. As a result, in $50\%$ of the cases the coding and decoding are performed in the same basis, while in the remaining $50\%$ the bases differ. The number of usable bits equals $25\%$ of the total received bits \cite{RFH+11}. The QBER for the B92 protocol can be calculated as: 

\begin{equation}
    e_{92} = \frac{c \; p_{signal}+\frac{1}{4} \; (p_{dark}+p_{straycounts})}{p_{click}} .\\  
\end{equation}

Here, the parameters have  the same definitions for $p_{signal}$,  $p_{dark}$, $p_{straycounts}$ and  $p_{click}$  as in the BB84 protocol.
\subsection{ QBER for BBM92}\label{QBER for BBM92}
The QBER for BBM92 depends on various factors such as the properties of the quantum channel, the quality of the detectors used, and the presence of any eavesdroppers \cite{WZY02}. In one beam splitter with transmission, we combine all losses to each receiver from the channel, detectors, and optics.
\begin{equation}
    \alpha_L = \eta_d*\eta.
\end{equation}
The parameter $\eta_d$ represents the detector efficiency of the system. The $p_{coin}$ that represents the coincidence probability is divided into three components: $p_{true}$, which denotes the probability of a genuine coincidence between a pair of entangled photons, $p_{false}$  which represents the probability of a false coincidence and  $p_{straycounts}$ which represents the probability of straycounts.
\begin{equation}
    p_{coin}= p_{true}+p_{false}+p_{straycounts}.
\end{equation}
We must choose a location for the source. Setting the source at a distance of $(L-x)$ from Bob and $x$ from Alice, we get
\begin{equation}
    p_{true}= \alpha_x\alpha_{L-x} =\eta_d\alpha_L.
\end{equation}
\begin{equation}
    p_{false}= 4\alpha_x p_{dark}+4\alpha_{L-x}p_{dark}+ 16p_{dark}^2.
\end{equation}
Keeping only terms which are second order in $\alpha_x$ and $p_{dark}$, it can be observed that the probability of a true coincidence remains constant with respect to $x$, while the false coincidence rate changes. By performing a straightforward optimization, it can be determined that the false coincidence rate reaches its minimum value at a distance halfway between Alice and Bob. The value of this minimum false coincidence rate can be calculated using the given formula:
\begin{equation}
    p_{false}=8\alpha_{L/2}\; p_{dark} + 16\; p_{dark}^2 .   
\end{equation}
The QBER is given by \cite{WZY02}
\begin{equation}
    e_{M92} = \frac{c \; p_{true}+\frac{1}{2} \; (p_{false}+p_{straycounts})}{p_{coin}}.
\end{equation}
\subsection {QBER for E91}
QBER for $E91$ is given by \cite{EK91, WZY02}
\begin{equation}
     e_{91} = \frac{c \; p_{true}+\frac{2}{9} \; (p_{false}+p_{straycounts})}{p_{coin}}.
\end{equation}
Here, the parameters have  the same definitions for $p_{true}$,  $p_{false}$, $p_{straycounts}$ and  $p_{coin}$  as in the BBM92 protocol. In the E91 protocol, Alice and Bob randomly choose among three measurement bases. Among the nine possible basis combinations, only two yield correlated outcomes suitable for key generation, while the others are discarded or used for Bell inequality verification. Therefore, the sifting factor is $\tfrac{2}{9}$~\cite{EK91} and  QBER will be impacted by a sifting factor of $\frac{2}{9}$.

\section{Keyrate}\label{Keyrate}
Keyrate, in the context of QKD, refers to the rate at which a secure cryptographic key can be generated and shared between two communicating parties, typically referred to as Alice and Bob.  The key rate is measured in bits per pulse and is influenced by factors such as the quality of transmitted quantum states, detection efficiency, channel losses, and potential eavesdropping attempt. The keyrate serves as a benchmark for evaluating the effectiveness and practicality of protocols using QKD. In the key rate calculation, the QBER, denoted by $e$, is taken into account. The QBER reflects the influence of the total transmittance efficiency and incorporates the effects of various loss mechanisms, such as diffraction loss, pointing error and turbulence loss, as well as other factors affecting the security and performance of the QKD system, as discussed above. Accordingly, the key rate provides a comprehensive measure of the effectiveness and efficiency of QKD protocols.

\subsection{Keyrate for BB84}\label{Keyrate for BB84}
The secure key generation rate against individual attack for the BB84 protocol is given by \cite{L00}:
\begin{equation}
    R_{BB84} = \frac{1}{2} p_{click} \{(1-\tau(e_{84})+f(e_{84})(e_{84}\log_2(e_{84})+(1-e_{84})\log_2(1-e_{84}))\}. \label{k_BB84}
\end{equation}
\
Here, $f(e_{84})$ is error correction factor \cite{L00},
$\tau$ is fraction of the key to be discarded during privacy amplification,
$\tau(e_{84})$= $\log_2(1+4e_{84}-4e_{84}^2)$ if $e_{84}< 1/2$ and $\tau(e_{84})=1$ if $e_{84}>1/2$.

\subsection{Keyrate for B92}
The secure key generation rate of the B92 protocol against individual attack can be formulated as \cite{RFH+11}:
\begin{equation}
    R_{B92} = \frac{1}{4} p_{click} \{(1-\tau(e_{84})+f(e_{92})(e_{92}\log_2(e_{92})+(1-e_{92})\log_2(1-e_{92}))\}. \label{K_B92}
\end{equation}
In B92, only $25\%$ of the bits transmitted will be detected by Bob, i.e., only $25\%$ of the raw key bits should be kept.
Hence, $\frac{1}{4}$ is the sifting factor. All the other parameters are defined in Sec. \ref{Keyrate for BB84}.
\subsection{Keyrate for BBM92}
 The keyrate for BBM92 protocol against double blinding attack is given by \cite{WZY02}:
 \begin{equation}
     R_{BBM92} = \frac{p_{coin}}{2} \{\tau(e_{M92})+f(e_{M92})(e_{M92}\log_2(e_{M92})+(1-e_{M92})\log_2(1-e_{M92}))\}.\label{K_BBM92}
 \end{equation}
 Due to the double-blinding attack discussed in Appendix A , Alice and Bob are unable to detect the presence of Eve, resulting in a complete elimination of information leakage. In other words, the measure of information leakage, denoted as  $\tau$ becomes zero in this scenario,
$f(e_{M92})$ is error correction factor,
$p_{coin}$ is the coincidence probability which has already been explained in Sec. \ref{QBER for BBM92}.\\

\subsection{Keyrate for E91}
The keyrate for E91 protocol against double blinding attack is given by

 \begin{equation}
     R_{E91} = \left( \frac{2}{9} \, p_{\text{coin}} \right)   \{\tau(e_{91})+f(e_{91})(e_{91}\log_2(e_{91})+(1-e_{91})\log_2(1-e_{91}))\}\label{K_E91}.
 \end{equation}
In the E91 protocol, since Alice and Bob each choose among three measurement bases, only two out of the nine possible basis combinations contribute to the raw key, leading to a sifting factor of $\tfrac{2}{9}$
 \cite{EK91},  The parameters used in Eq. \eqref{K_E91} have been described  above.

\section{Numerical Results and Discussion }\label{Numerical Results and Discussion}

This section examines the QBER and secret key rate as functions of the zenith angle for the BB84, B92, E91, and BBM92 protocols, considering uplink (night) and downlink (day and night) scenarios. To accurately model atmospheric losses, we employ the circular beam model, which effectively accounts for diffraction losses and pointing errors. In the case of uplink transmission, atmospheric turbulence is also significant, as it primarily affects the initial stage of propagation. Since turbulence is concentrated within the lower 20~km of the atmosphere and is strongest near the Earth's surface, its impact can be neglected for downlink transmission, where it occurs only near the end of the optical path \cite{BMH+13}. The simulations incorporate the experimental parameters specified in Table \ref{table_1st}.


\begin{figure}[htbp]
\centering

\setcounter{subfig}{0}
\refstepcounter{subfig}
\label{2a}

\begin{tikzpicture}
\node[anchor=south west,inner sep=0] (img1)
{\includegraphics[width=0.8\textwidth]
{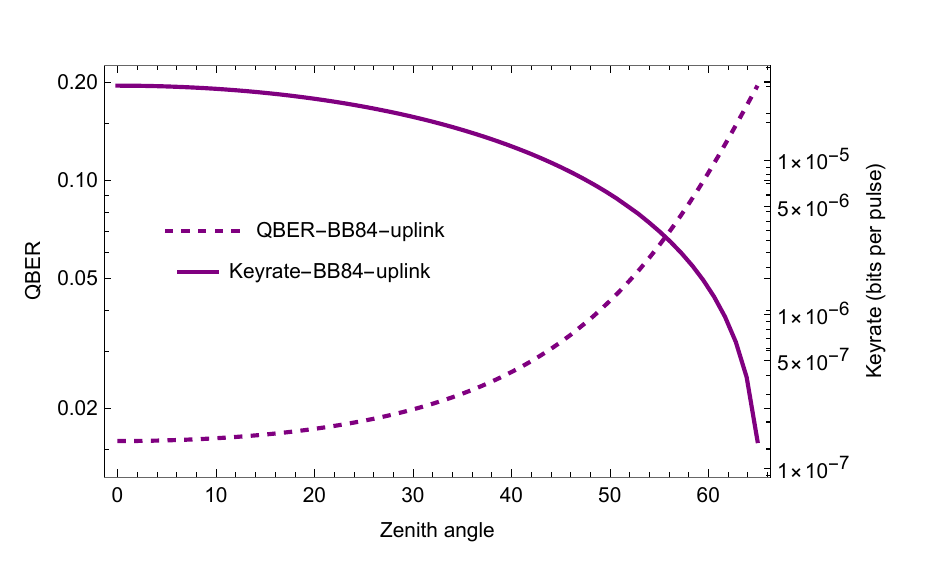}};

\node[
anchor=north east,
font=\bfseries\small
] at ([xshift=-70pt,yshift=-25pt]img1.north east)
{(\alph{subfig})};
\end{tikzpicture}

\vspace{-1cm}

\refstepcounter{subfig}
\label{2b}

\begin{tikzpicture}
\node[anchor=south west,inner sep=0] (img2)
{\includegraphics[width=0.8\textwidth]
{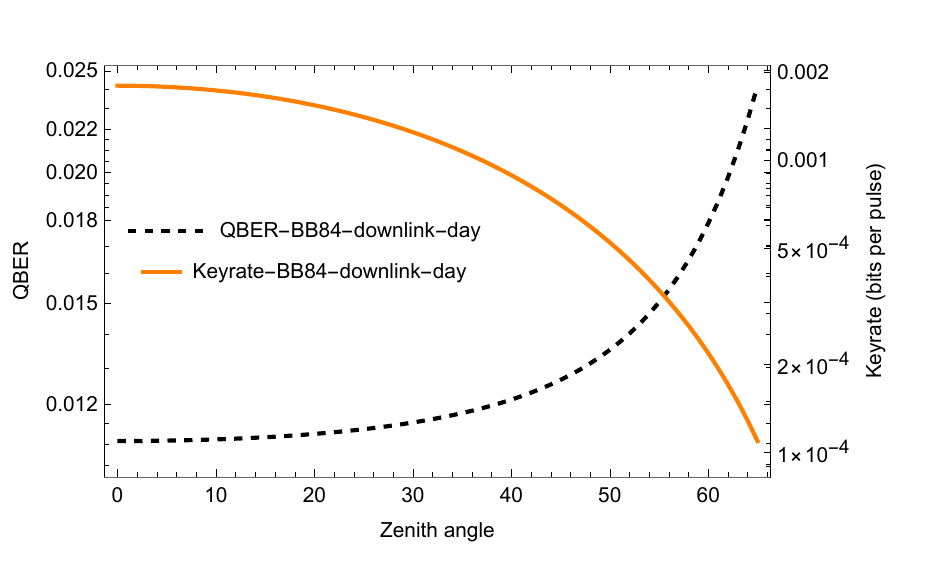}};

\node[
anchor=north east,
font=\bfseries\small
] at ([xshift=-70pt,yshift=-25pt]img2.north east)
{(\alph{subfig})};
\end{tikzpicture}

\vspace{-1cm}

\refstepcounter{subfig}
\label{2c}

\begin{tikzpicture}
\node[anchor=south west,inner sep=0] (img3)
{\includegraphics[width=0.8\textwidth]
{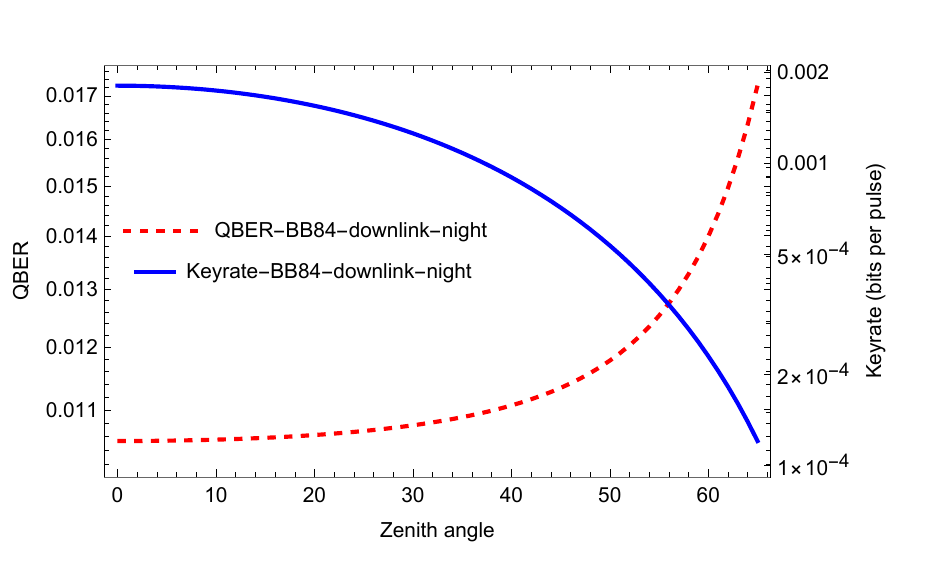}};

\node[
anchor=north east,
font=\bfseries\small
] at ([xshift=-70pt,yshift=-25pt]img3.north east)
{(\alph{subfig})};
\end{tikzpicture}

\caption{(a) QBER and key rate versus zenith angle for the BB84 protocol in the uplink (night-time). (b) and (c) QBER and key rate versus zenith angle for the BB84 protocol in the downlink (day-time) and downlink (night-time), respectively.}
 
\label{QKB84}

\end{figure}
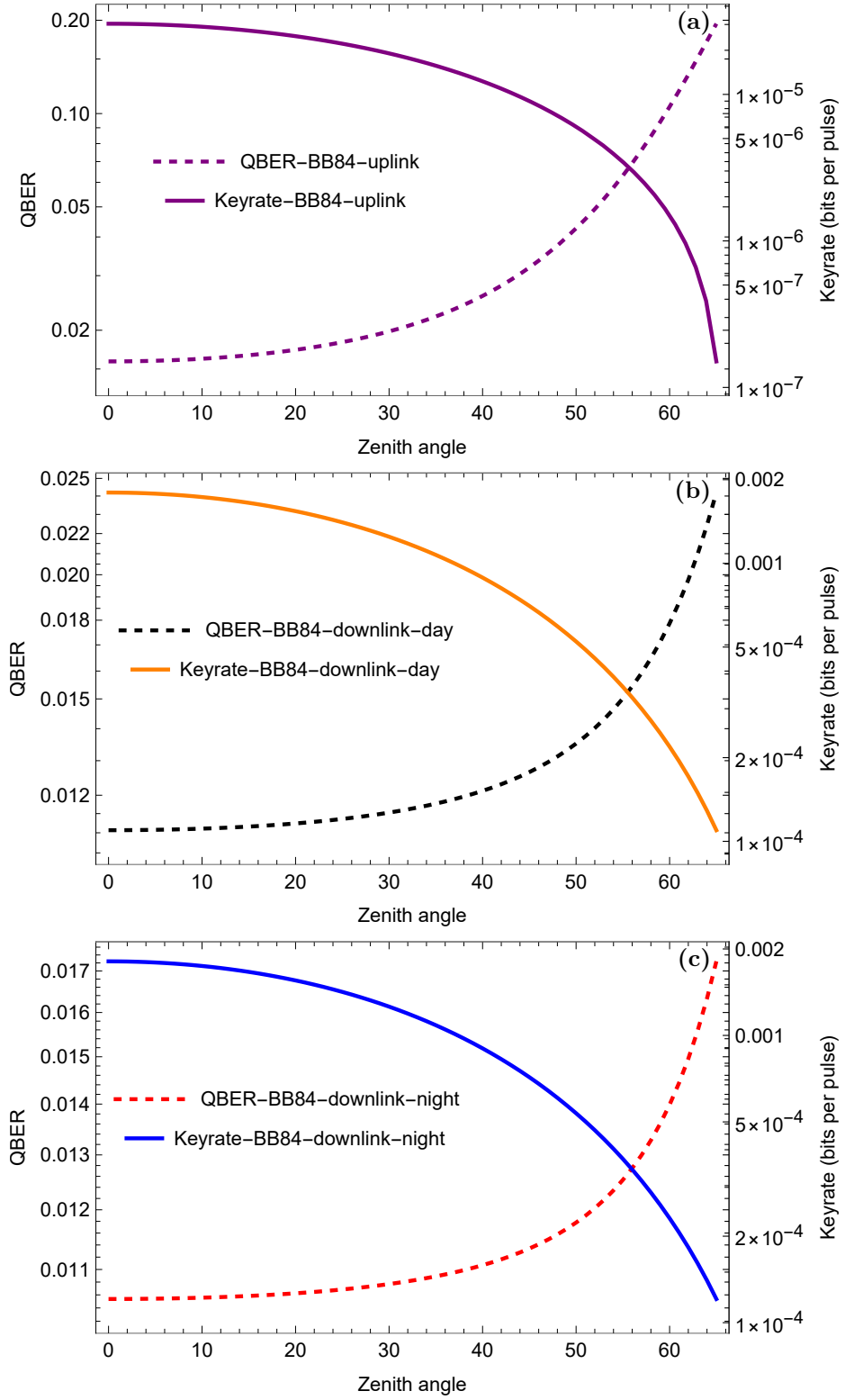
\begin{figure}[htbp]
\centering
\setcounter{subfig}{0}

\refstepcounter{subfig}
\label{3a}

\begin{tikzpicture}
\node[anchor=south west,inner sep=0] (img1)
{\includegraphics[width=0.8\textwidth]
{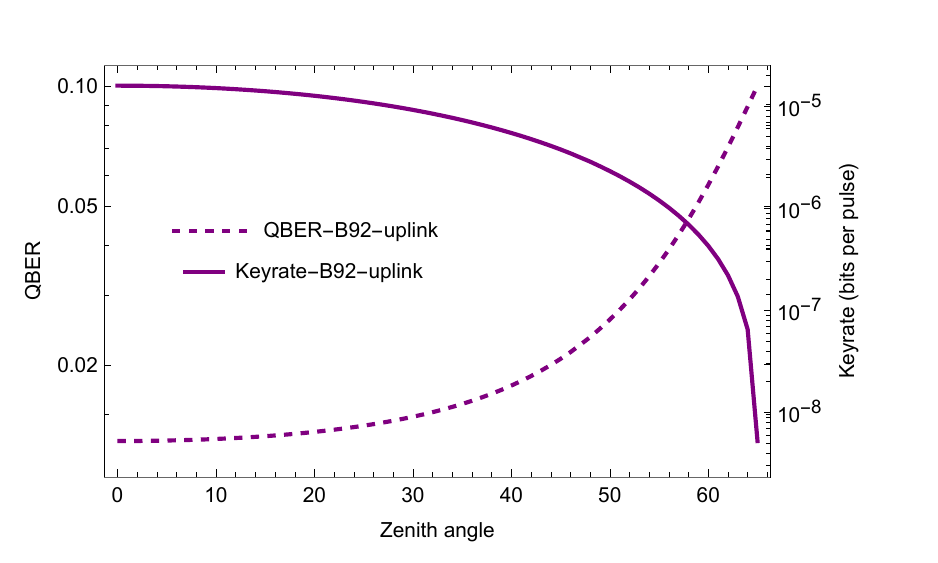}};

\node[
anchor=north east,
font=\bfseries\small
] at ([xshift=-70pt,yshift=-25pt]img1.north east)
{(\alph{subfig})};
\end{tikzpicture}

\vspace{-1cm}

\refstepcounter{subfig}
\label{3b}

\begin{tikzpicture}
\node[anchor=south west,inner sep=0] (img2)
{\includegraphics[width=0.8\textwidth]
{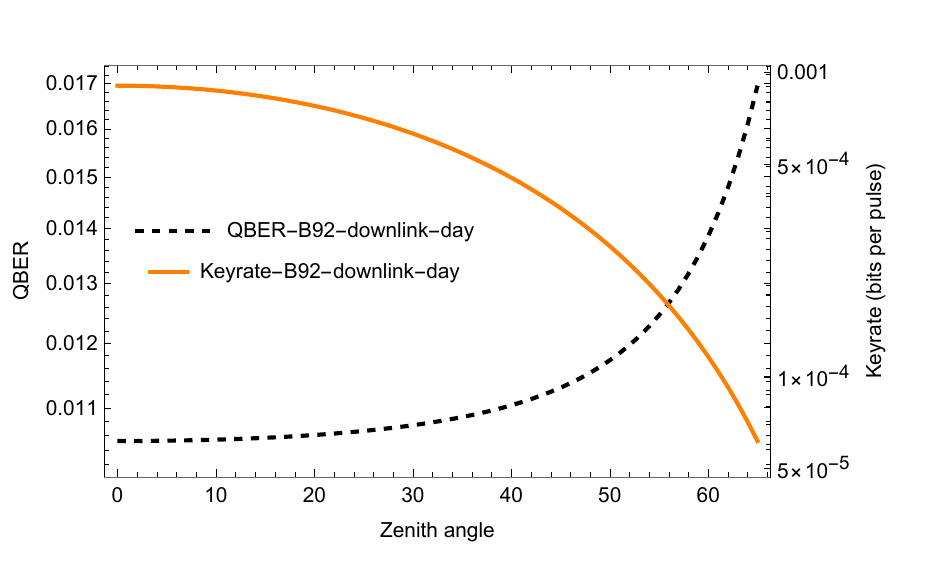}};

\node[
anchor=north east,
font=\bfseries\small
] at ([xshift=-70pt,yshift=-25pt]img2.north east)
{(\alph{subfig})};
\end{tikzpicture}

\vspace{-1cm}

\refstepcounter{subfig}
\label{3c}

\begin{tikzpicture}
\node[anchor=south west,inner sep=0] (img3)
{\includegraphics[width=0.8\textwidth]
{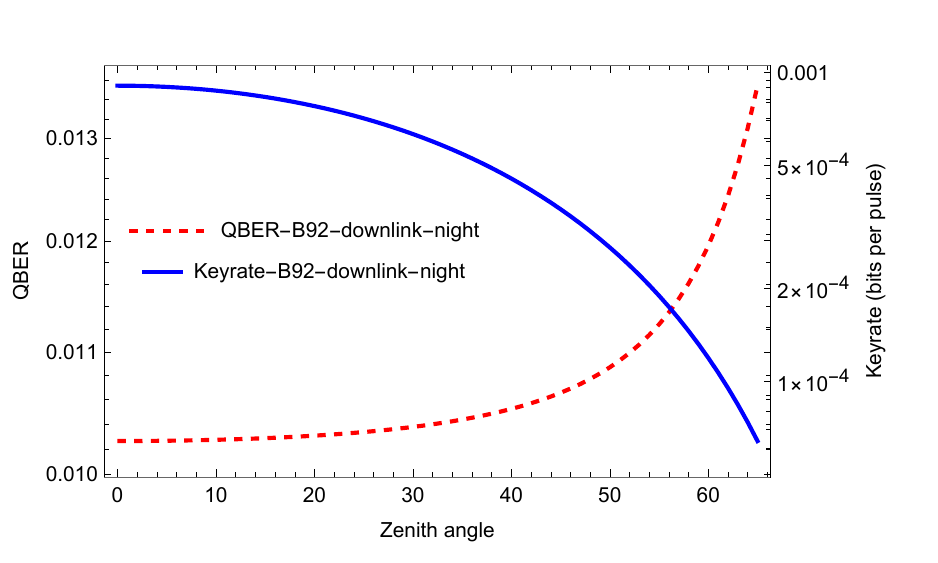}};

\node[
anchor=north east,
font=\bfseries\small
] at ([xshift=-70pt,yshift=-25pt]img3.north east)
{(\alph{subfig})};
\end{tikzpicture}

\caption{(a) QBER and key rate versus zenith angle for the B92 protocol in the uplink (night-time). (b) and (c) QBER and key rate versus zenith angle for the B92 protocol in the downlink (day-time) and downlink (night-time), respectively.}
\label{QKB92}
\end{figure}
\begin{figure}[htbp]
\centering

\setcounter{subfig}{0}

\refstepcounter{subfig}
\label{5a}

\begin{tikzpicture}
\node[anchor=south west,inner sep=0] (img1)
{\includegraphics[width=0.8\textwidth]
{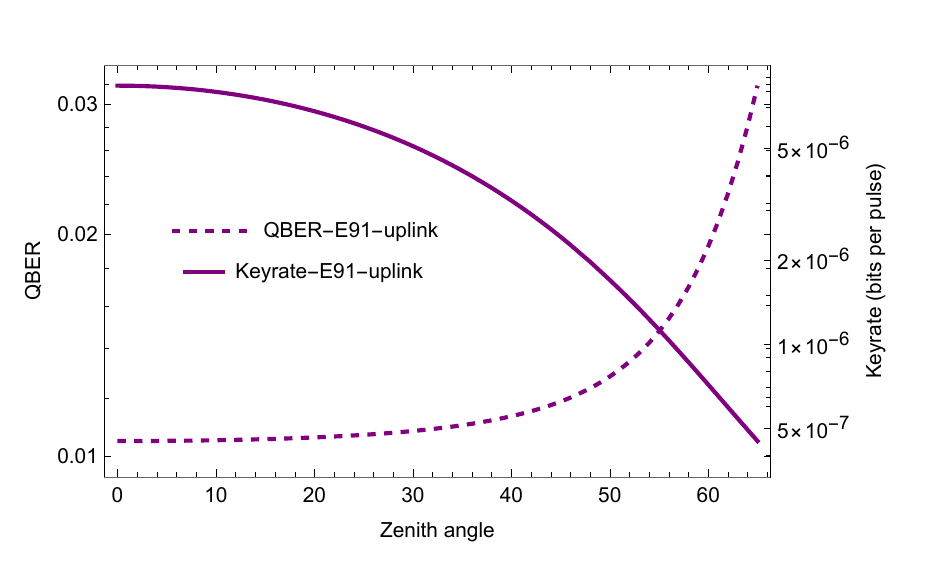}};

\node[
anchor=north east,
font=\bfseries\small
] at ([xshift=-70pt,yshift=-25pt]img1.north east)
{(\alph{subfig})};
\end{tikzpicture}

\vspace{-1cm}

\refstepcounter{subfig}
\label{5b}

\begin{tikzpicture}
\node[anchor=south west,inner sep=0] (img2)
{\includegraphics[width=0.8\textwidth]
{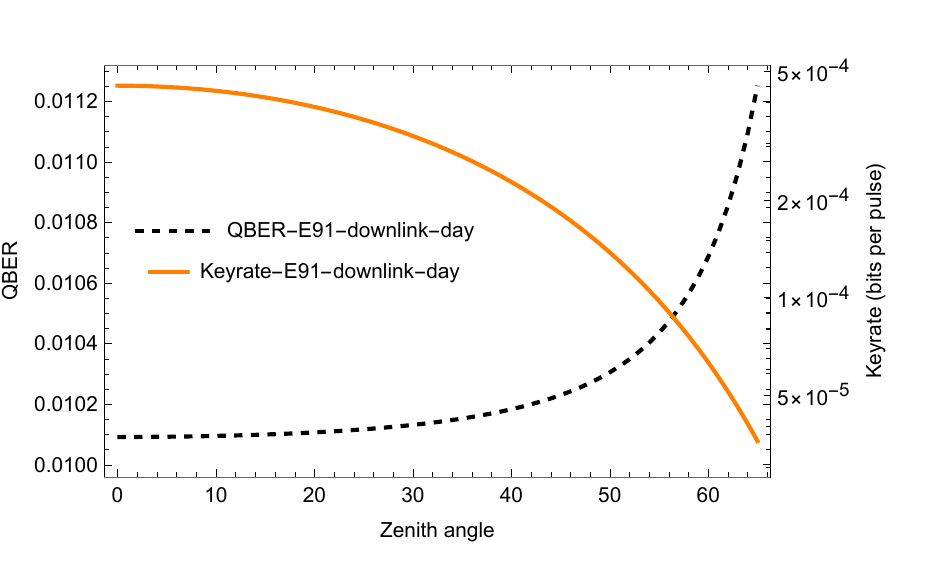}};

\node[
anchor=north east,
font=\bfseries\small
] at ([xshift=-70pt,yshift=-25pt]img2.north east)
{(\alph{subfig})};
\end{tikzpicture}

\vspace{-1cm}

\refstepcounter{subfig}
\label{5c}

\begin{tikzpicture}
\node[anchor=south west,inner sep=0] (img3)
{\includegraphics[width=0.8\textwidth]
{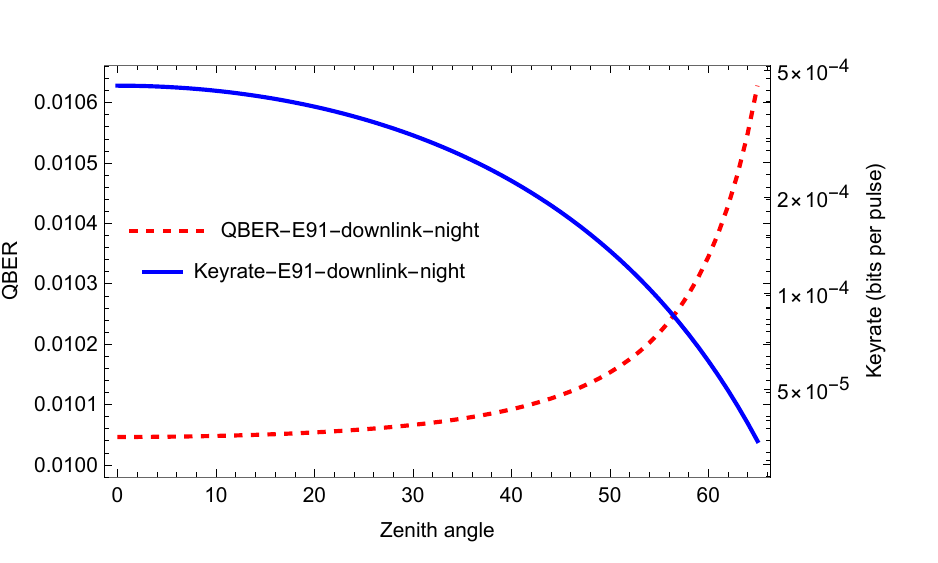}};

\node[
anchor=north east,
font=\bfseries\small
] at ([xshift=-70pt,yshift=-25pt]img3.north east)
{(\alph{subfig})};
\end{tikzpicture}

\caption{(a) QBER and key rate versus zenith angle for the E91 protocol in the uplink (night-time). (b) and (c) QBER and key rate versus zenith angle for the E91 protocol in the downlink (day-time) and downlink (night-time), respectively.}
\label{QKE91}
\end{figure}

\begin{figure}[htbp]
\centering

\setcounter{subfig}{0}

\refstepcounter{subfig}
\label{4a}

\begin{tikzpicture}
\node[anchor=south west,inner sep=0] (img1)
{\includegraphics[width=0.8\textwidth]
{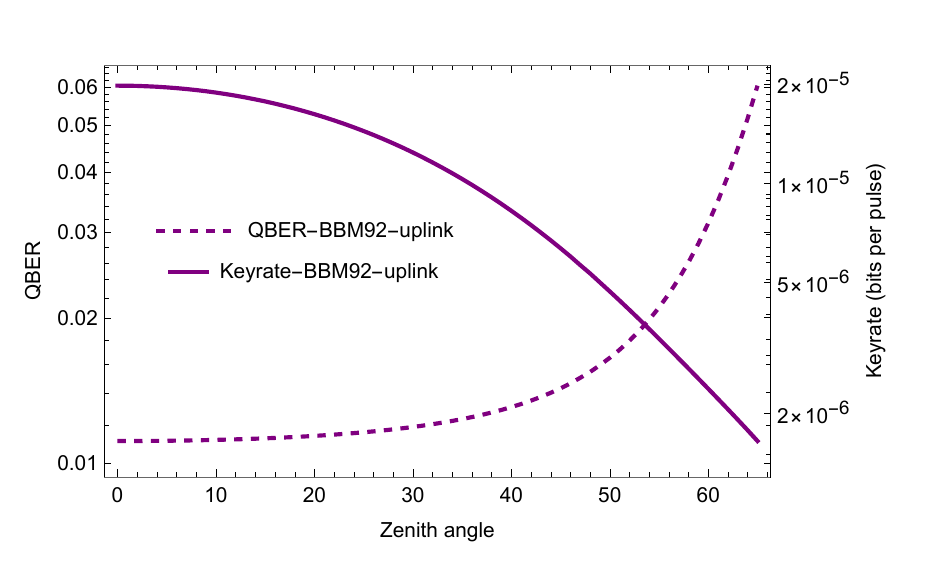}};

\node[
anchor=north east,
font=\bfseries\small
] at ([xshift=-70pt,yshift=-25pt]img1.north east)
{(\alph{subfig})};
\end{tikzpicture}

\vspace{-1cm}

\refstepcounter{subfig}
\label{4b}

\begin{tikzpicture}
\node[anchor=south west,inner sep=0] (img2)
{\includegraphics[width=0.8\textwidth]
{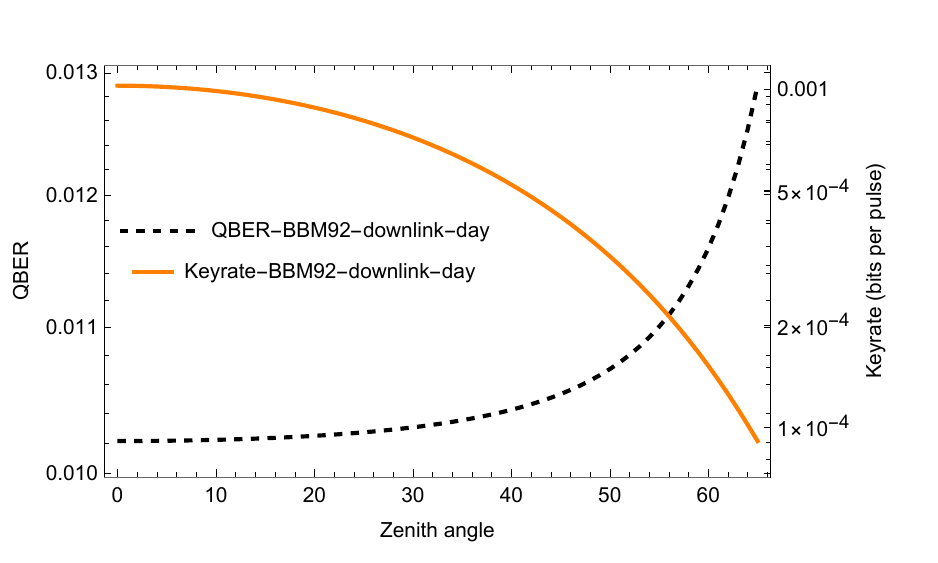}};

\node[
anchor=north east,
font=\bfseries\small
] at ([xshift=-70pt,yshift=-25pt]img2.north east)
{(\alph{subfig})};
\end{tikzpicture}

\vspace{-1cm}

\refstepcounter{subfig}
\label{4c}

\begin{tikzpicture}
\node[anchor=south west,inner sep=0] (img3)
{\includegraphics[width=0.8\textwidth]
{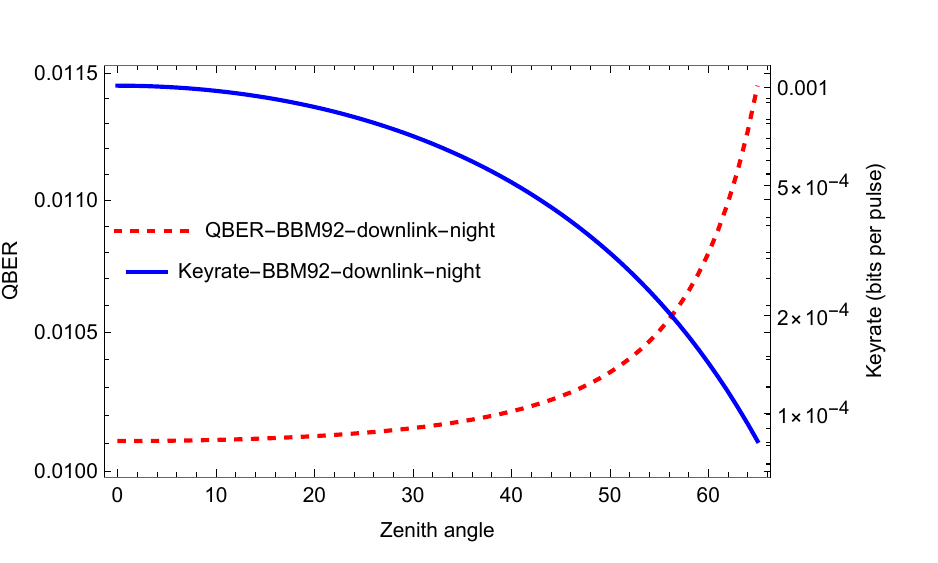}};

\node[
anchor=north east,
font=\bfseries\small
] at ([xshift=-70pt,yshift=-25pt]img3.north east)
{(\alph{subfig})};
\end{tikzpicture}

\caption{(a) QBER and key rate versus zenith angle for the BBM92 protocol in the uplink (night-time). (b) and (c) QBER and key rate versus zenith angle for the BBM92 protocol in the downlink (day-time) and downlink (night-time), respectively.}
\label{QKM92}
\end{figure}
\begin{figure*}[htbp]
\centering


\begin{tikzpicture}

\node (a) at (0,0)
{
\begin{tikzpicture}
\node[anchor=south west,inner sep=0] (img1)
{\includegraphics[width=0.7\textwidth]
{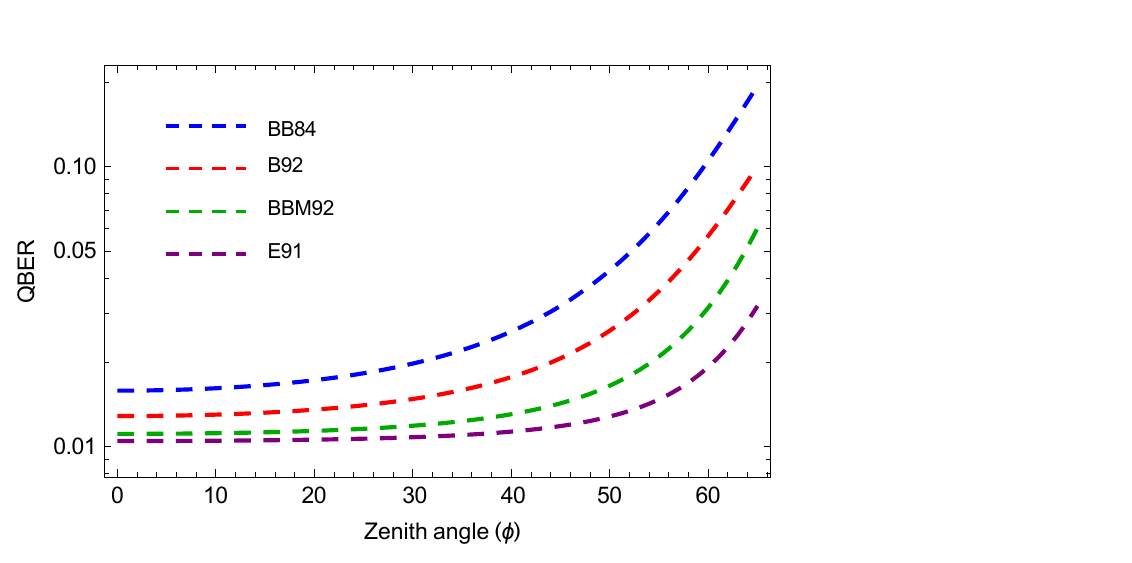}};

\node[
anchor=north east,
font=\bfseries\small
] at ([xshift=-260pt,yshift=-20pt]img1.north east)
{(a)};
\end{tikzpicture}
};

\node[right=-3.15cm of a] (d)
{
\begin{tikzpicture}
\node[anchor=south west,inner sep=0] (img2)
{\includegraphics[width=0.7\textwidth]
{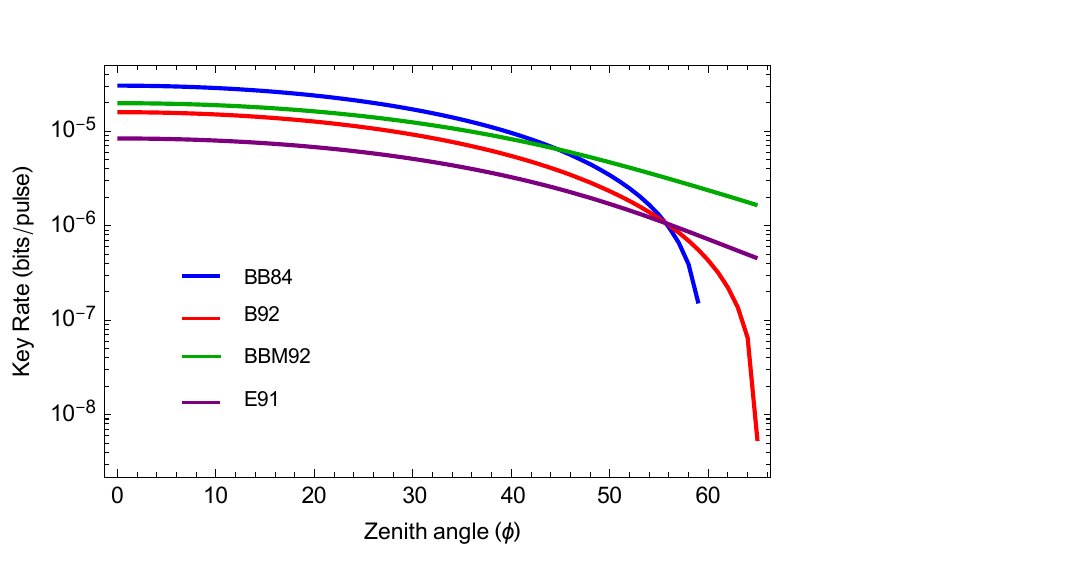}};

\node[
anchor=north east,
font=\bfseries\small
] at ([xshift=-100pt,yshift=-20pt]img2.north east)
{(d)};
\end{tikzpicture}
};

\end{tikzpicture}

\vspace{-0.2cm}

\begin{tikzpicture}

\node (b) at (0,0)
{
\begin{tikzpicture}
\node[anchor=south west,inner sep=0] (img3)
{\includegraphics[width=0.7\textwidth]
{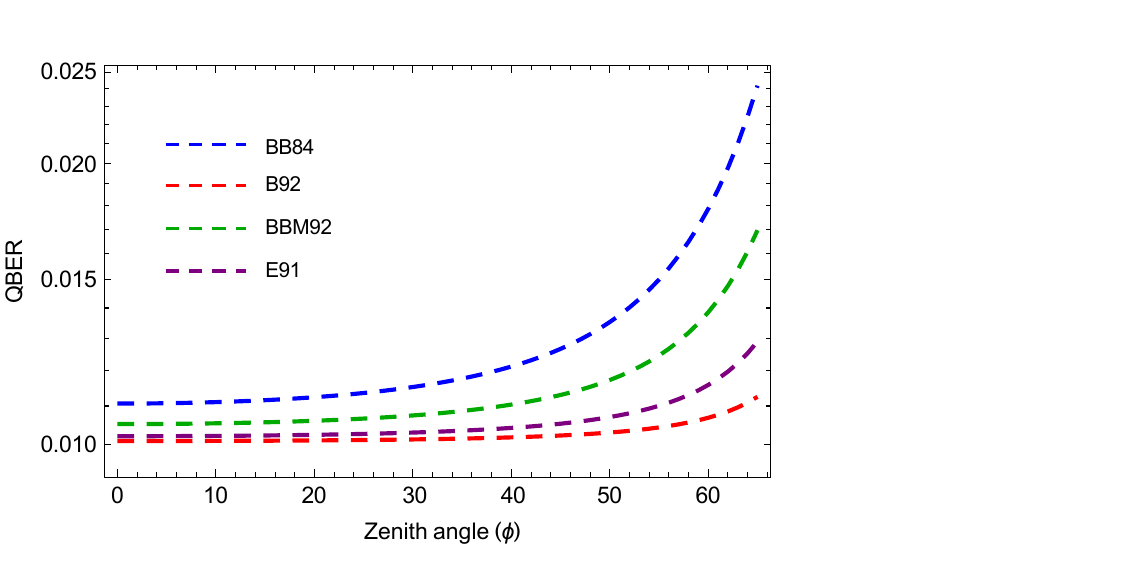}};

\node[
anchor=north east,
font=\bfseries\small
] at ([xshift=-260pt,yshift=-20pt]img3.north east)
{(b)};
\end{tikzpicture}
};

\node[right=-3.15cm of b] (e)
{
\begin{tikzpicture}
\node[anchor=south west,inner sep=0] (img4)
{\includegraphics[width=0.7\textwidth]
{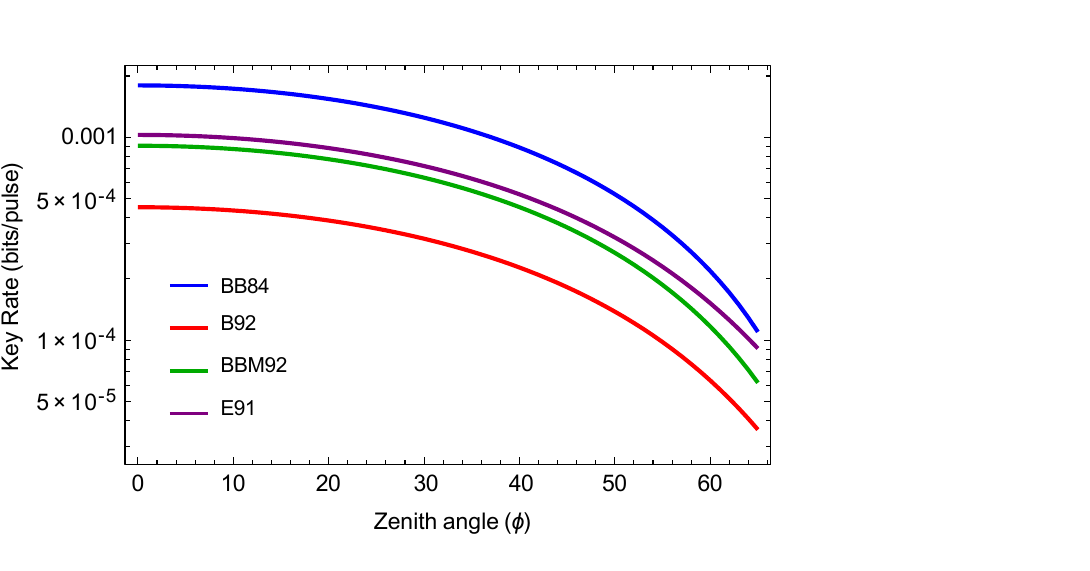}};

\node[
anchor=north east,
font=\bfseries\small
] at ([xshift=-100pt,yshift=-20pt]img4.north east)
{(e)};
\end{tikzpicture}
};

\end{tikzpicture}

\vspace{-0.2cm}

\begin{tikzpicture}

\node (c) at (0,0)
{
\begin{tikzpicture}
\node[anchor=south west,inner sep=0] (img5)
{\includegraphics[width=0.7\textwidth]
{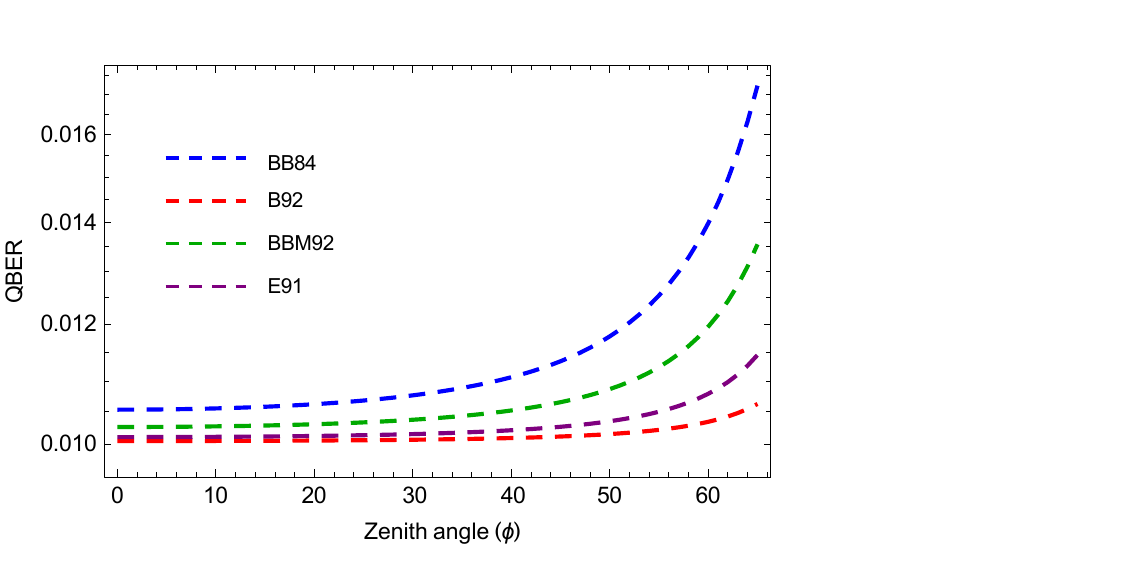}};

\node[
anchor=north east,
font=\bfseries\small
] at ([xshift=-260pt,yshift=-20pt]img5.north east)
{(c)};
\end{tikzpicture}
};

\node[right=-3.15cm of c] (f)
{
\begin{tikzpicture}
\node[anchor=south west,inner sep=0] (img6)
{\includegraphics[width=0.7\textwidth]
{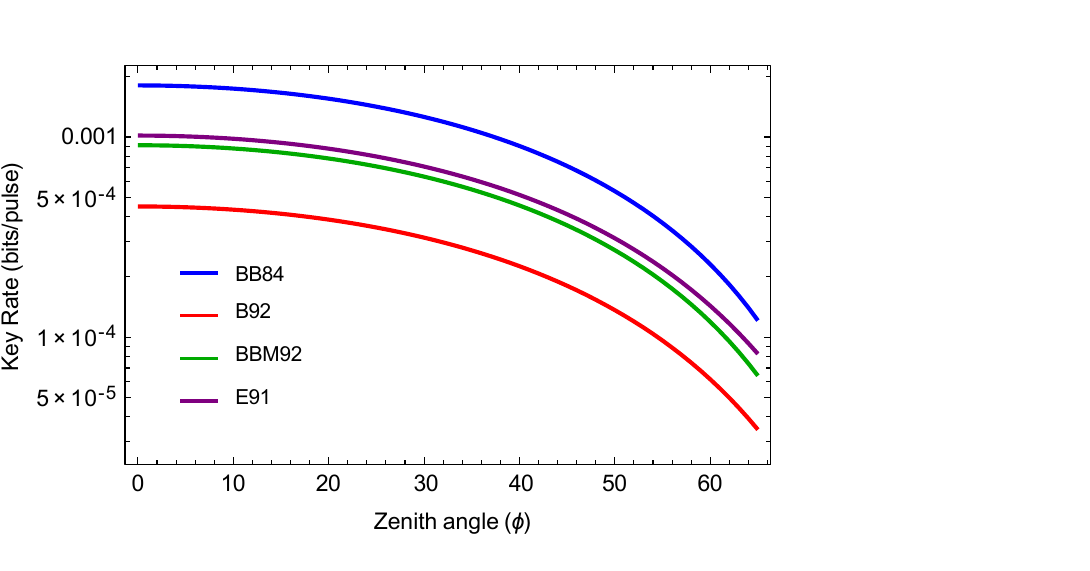}};

\node[
anchor=north east,
font=\bfseries\small
] at ([xshift=-100pt,yshift=-20pt]img6.north east)
{(f)};
\end{tikzpicture}
};

\end{tikzpicture}

\caption{
QBER and key rate versus zenith angle for all considered QKD protocols:
(a,d) uplink (night),
(b,e) downlink (day), and
(c,f) downlink (night).
}
\label{fig:combined_all}

\end{figure*}














Equations. \eqref{k_BB84} and \eqref{K_B92} give the asymptotic key rate formulas for the BB84 and B92 protocols, while Eqs. \eqref{K_BBM92} and \eqref{K_E91} give the formulas for the entanglement-based protocols BBM92 and E91. These formulas are combined with the circular beam model, which includes the effects of atmospheric transmittance and stray counts for both day and night. This allows us to calculate the key rates for LEO-based quantum communication systems. Figure. \ref{QKB84} shows how the QBER and key rate changes with the zenith angle, taking into account both the transmittance model and stray counts. The analysis is done for uplink (night) and downlink (day and night) cases. In Fig. \ref{QKB84}a, at the zenith position i.e at $0^\circ$ zenith angle, the BB84 uplink exhibits a QBER of $0.0159$ and a key rate of $3.04\times10^{-5}$ bits per pulse. At a higher zenith angle of $55^\circ$, the QBER increases to $0.0631$, while the key rate decreases to $1.30\times10^{-6}$ bits per pulse.  As the zenith angle increases, the signal traverses a longer atmospheric path, leading to greater attenuation due to absorption and scattering. Furthermore, enhanced beam divergence reduces the overlap with the receiving telescope aperture, further lowering the detected photon count. Therefore QBER increases and key rate decreases wih increasing zenith angle.  Figures. \ref{QKB84}b and \ref{QKB84}c depict the QBER and secret-key rate for the BB84 protocol in the downlink day and downlink night configurations, respectively. In Fig. \ref{QKB84}b downlink day, at the zenith position i.e., at $0^\circ$ zenith angle, the QBER is $0.0111$ with a secret-key rate of $1.80\times10^{-3}$ bits per pulse. At a zenith angle of $55^\circ$, the QBER increases to $0.0150$ and the secret-key rate decreases to $3.59\times10^{-4}$ bits per pulse. In Fig. \ref{QKB84}c downlink night, at $0^\circ$ zenith angle, the QBER is $0.0105$ with a secret-key rate of $1.81\times10^{-3}$ bits per pulse. At $55^\circ$, the QBER is $0.0125$, while the secret-key rate is $3.72\times10^{-4}$ bits per pulse.
 In both cases Fig. \ref{QKB84}b and \ref{QKB84}c, the downlink configuration yields a lower QBER and a higher secret-key rate than the uplink scenario shown in Fig. \ref{QKB84}a. When comparing the uplink and downlink scenarios, the uplink experiences beam propagation through the turbulent atmosphere, leading to significant broadening, while the downlink only encounters turbulence during the final stage, resulting in less spreading. As a result, the attenuation in the uplink is more pronounced compared to the downlink due to these differing propagation conditions. Day-time uplink was not considered due to excessive background light \cite{LKB19}. Furthermore, in the night-time downlink scenario Fig. \ref{QKB84}c, the atmospheric transmittance is reduced due to the higher moisture content compared to daytime conditions. The contrast between day and night operations is significant: during the day, elevated temperatures generate stronger winds and greater mixing between atmospheric layers, which enhances turbulence effects, thereby increasing the QBER and reducing the key rate. In contrast, on clear nights, the lower atmosphere generally exhibits reduced moisture levels relative to daytime, resulting in less beam spreading from scattering particles. Consequently, the night-time scenario achieves a marginally higher key rate and a slightly lower QBER compared to the daytime case.\par

Now, comparing the BB84 and B92 performance for both uplink and downlink configurations, it is observed that BB84 consistently achieves higher secret-key rates while maintaining QBER values that are slightly higher than those of B92. This is because the BB84 protocol employs four quantum states, enabling a larger fraction of detection events to contribute to the sifted key. In contrast, the B92 protocol relies on only two nonorthogonal states, which often lead to inconclusive results. Consequently, its sifting factor is 0.25, lower than that of BB84, which is 0.5. Hence, even with a marginally higher QBER, BB84 retains more sifted bits and tolerates higher error thresholds, resulting in a greater final key rate than B92 under the same channel conditions. In the uplink scenario at the zenith position 0$^\circ$ (Fig.~ \ref{QKB84}a), BB84 records a QBER of about 0.016 with a key rate of approximately $3.04\times10^{-5}$~bits per pulse, whereas B92 (Fig.~\ref{QKB92}a) exhibits slightly lower QBER of around 0.012 and lower key rate of $1.59\times10^{-5}$~bits per pulse, i.e., lower than BB84. For downlink day and night operations, both protocols exhibit lower QBER than in the uplink case; however, BB84 has a clear key rate advantage-for instance, in night-time downlink at 0$^\circ$ (Fig.~\ref{QKB84}c), BB84’s key rate is about $1.81\times10^{-3}$~bits per pulse,  higher than that of B92 (Fig.~\ref{QKB92}c). This performance gap persists at higher zenith angles. For instance, in night-time downlink at 55$^\circ$, BB84 achieves $3.72\times10^{-4}$~bits per pulse versus $1.90\times10^{-4}$~bits per pulse for B92, while in day-time downlink at the same angle, BB84 records $3.59\times10^{-4}$~bits per pulse compared to $1.87\times10^{-4}$~bits per pulse for B92. The sightly lower QBER and higher key rates observed during night-time are primarily due to the absence of background sunlight, which reduces detector noise and photon count errors, as well as the lower atmospheric moisture levels compared to daytime, leading to reduced beam spreading from scattering particles and thereby enhancing both the accuracy of received bits and the overall key generation efficiency. These results \cite{MRS26} highlight BB84’s superior resilience to atmospheric losses and noise, with its four-state encoding enabling more robust and reliable secure key distribution in LEO-based quantum communication across diverse channel conditions.

For the uplink configuration, the BBM92 protocol exhibits QBER values ranging from 0.01 to 0.06 with corresponding secret key rates between $1.64\times10^{-6}$ and $2\times10^{-5}$~bits per pulse (Fig.~\ref{QKM92}a), whereas the E91 protocol records slightly lower QBER values in the range of 0.01-0.03 with key rates between $4.52\times10^{-7}$ and $8.39\times10^{-6}$~bits per pulse (Fig.~\ref{QKE91}a). The relatively higher key rates observed in BBM92 compared to E91 can be attributed to its simpler sifting process and better tolerance to background noise, enabling it to extract more secure bits per detected entangled pair, while the basis correlation requirements in E91 lead to a smaller fraction of usable bits after sifting. In the downlink night-time scenario, BBM92 achieves QBER values in the range of 0.0100-0.0115 with key rates from about $9.09\times10^{-5}$ to $1.03\times10^{-3}$~bits per pulse (Fig.~\ref{QKM92}c), while E91 shows slightly lower QBER values between 0.0100 and 0.0106 with key rates ranging from $3.44\times10^{-5}$  to $4.50\times10^{-4}$~bits per pulse (Fig.~\ref{QKE91}c). In the downlink day-time case, both protocols maintain low QBER values, with BBM92 ranging between 0.010-0.013 and key rates from $9.09\times10^{-5}$ up to $1.03\times10^{-3}$~bits per pulse (Fig.~\ref{QKM92}b), while E91 shows QBER values of 0.0100-0.0112 and key rates in the range of $3.62\times10^{-5}$ to $4.51\times10^{-4}$~bits per pulse (Fig.~\ref{QKE91}b). Overall, E91 generally exhibits marginally lower QBER and lower key rates than BBM92. For both protocols, the key rate decreases with increasing zenith angle because the longer atmospheric slant path increases optical losses and turbulence, reducing the number of entangled photon pairs detected at the receiver.
The QBER and key rate for uplink (night), downlink (day), and downlink (night) are illustrated for all QKD protocols in combined plots for improved comparison (Fig.~\ref{fig:combined_all}).

\section{Conclusion}
\label{Conclusion_2}
This chapter presented a comparative performance analysis of four QKD protocols-BB84, B92, BBM92, and E91 over LEO satellite-based optical links, considering uplink (night) and downlink (day and night) configurations. By incorporating the circular beam model with diffraction, pointing errors, atmospheric transmittance, and stray counts, we evaluated the QBER and secret-key rate as functions of the zenith angle. The results \cite{MRS26} show that downlink configurations consistently outperform uplink in both QBER and key rate due to reduced turbulence impact. Among the prepare-and-measure protocols, BB84 achieves higher key rates than B92 owing to its four-state encoding, which yields a larger sifted key fraction despite slightly higher QBER. Similarly, in entanglement-based schemes, BBM92 attains higher key rates than E91 due to its simpler sifting process and better tolerance to background noise, though E91 exhibits marginally lower QBER. Across all protocols, key rates decrease with increasing zenith angle as a result of longer atmospheric paths and greater optical losses. Night-time operations yield better performance than daytime due to reduced background light and detector noise. These findings highlight BB84 and BBM92 as more suitable choices for high-rate, long-distance satellite QKD, while also providing insights into the trade-offs between protocol design, link configuration, and atmospheric conditions in space-based quantum communication.



\chapter{High-Dimensional QKD Protocols: BB84 and Extended B92 via Satellite}\label{chap2:Background}

\section{Introduction}
The pursuit of secure quantum communication has evolved considerably with the introduction of HD-QKD, which extends the encoding space beyond traditional two-dimensional qubit systems. While conventional QKD protocols such as BB84~\cite{BB84} and B92~\cite{B92} have demonstrated unconditional security grounded in the fundamental principles of quantum mechanics, their reliance on qubit-level encoding imposes intrinsic limitations on achievable key rates and robustness against noise. To overcome these constraints, HD-quantum systems (qudits), characterized by a Hilbert space of dimension $d>2$, have emerged as a promising alternative~\cite{BT00, C00, CLB+19}. By enabling the encoding of multiple bits of information per photon, HD-QKD enhances the information capacity, noise resilience, and security margins of quantum communication systems~\cite{SPD+25, HLS+24, LZL+22}. These improvements stem from the exploitation of additional photonic degrees of freedom, such as orbital angular momentum (OAM), temporal, spatial, or frequency modes, which enable a more efficient utilization of the quantum channel. Combined with the decoy-state method~\cite{HK05, X05}, these protocols provide enhanced protection against PNS attacks, allowing for their practical realization using WCP sources. In the context of satellite-based QKD, adopting a HD-encoding framework offers significant advantages. Free-space quantum channels, unlike optical fibers, experience substantial losses due to beam diffraction, pointing errors, and atmospheric turbulence, particularly in uplink configurations where the transmitted signal first passes through the dense layers of the atmosphere~\cite{DW17}. HD-QKD protocols are inherently more resilient to such degradations, allowing higher key generation rates even in the presence of depolarization and background noise. The elliptical beam approximation~\cite{VSV+17, LKB19} provides a realistic model for describing the effect of turbulence on beam propagation, enabling the estimation of channel transmittance and corresponding key rates under varying atmospheric conditions. This chapter presents a comprehensive analysis of two prominent HD-QKD protocols, namely the HD-Ext-B92~\cite{IK21} and HD-BB84~\cite{C00} schemes, tailored for satellite-based free-space quantum communication. A refined analytical model for the key rate of the HD-Ext-B92 protocol is developed to enhance the accuracy of performance assessment compared to previously established formulations. Utilizing this improved expression, the secure key rate and QBER are systematically evaluated as functions of system dimension ($d$) and channel noise. Furthermore, the average key rate per pulse is investigated with respect to zenith angle, link distance, and atmospheric conditions during both day and night scenarios, employing the elliptical beam approximation to account for beam broadening, turbulence-induced fluctuations, and transmittance variability. The PDT is incorporated into the analysis to provide a statistically robust evaluation of performance in both uplink and downlink configurations \cite{DMB+24}. The results demonstrate that the HD-BB84 protocol generally outperforms HD-Ext-B92 in terms of secure key generation and noise tolerance across various dimensions, albeit with a higher saturation in QBER at large $d$, indicating a trade-off between dimensionality and error sensitivity. Conversely, HD-Ext-B92 exhibits relatively stable performance under moderate noise conditions with reduced implementation complexity. Overall, this chapter establishes the superiority and practicality of HD quantum communication schemes in satellite-based environments. The findings provide quantitative insights into optimizing system parameters, including beam characteristics, dimensionality, and environmental conditions, for achieving enhanced security and efficiency in global-scale quantum networks.\\
The remainder of this chapter is structured as follows.
Section \ref{sec:II} provides a detailed exploration of the HD-Ext-B92 and HD-BB84 protocols, alongside an extensive analysis of how atmospheric conditions affect satellite communication links and the elliptical approximation of beam deformation at the receiver. We also investigate the key rate and QBER under varying noise parameters to determine
the noise tolerance of these higher-dimensional protocols. Section \ref{sec:III} presents a thorough evaluation of the performance of these high-dimensional protocols, supported by illustrative results from simulations. Finally, we summarize our paper with the findings being consolidated and deliberated upon in Section \ref{sec:IV}.

\section{Description of the Implemented Protocols and the Elliptic Beam Approximation \label{sec:II}}

Numerous researchers have extensively investigated the unconditional
security of QKD-based protocols, and their research, (see for examples,
\cite{TKI03,M13}) has consistently revealed increasingly robust results.
For instance, in \cite{M13}, a noise tolerance of 6.5\% was reported
for the B92 protocol. Depending on the user's selected key encoding
states, the noise tolerance for this B92 protocol can extend up to
11\% in the asymptotic scenario, as demonstrated in \cite{LGT09}.
This level of noise tolerance is comparable to that of BB84. In scenarios
with a finite key length, as indicated in \cite{AK20}, the protocol
still maintains a minimum noise tolerance of 7\%. In this context,
we summarize the key-rate analysis for HD-Ext-B92 and HD-BB84 protocols.
We modify the calculation for HD-Ext-B92 using a theorem to eliminate
any additional free parameters as detailed in Appendix B. Additionally,
we briefly delve into the methodology of elliptical beam approximation,
designed to encompass satellite-based connections while accounting
for signal losses in various real-world scenarios, including diverse
weather conditions. This methodology is particularly tailored for
application in LEO satellite contexts. 

\subsection{High-dimensional extended B92 protocol and high-dimensional BB84
protocol \label{subsec:HD-Ext-B92=000026HD-BB84}}

Before going into the intricacies of higher-dimensional
protocols, let's briefly discuss the higher-dimensional quantum states
utilized in performing HD-QKD schemes \cite{HZL+20, CLB+19}. Traditional
two-level quantum systems, represented by discrete variable states
\cite{BPM+97, FDI+04, OMM+09}, and continuous variable states \cite{FSB+98, YBF07, LBT+11}
within a single degree of freedom (such as polarization), have historically
been employed for communicating information as qubits. However, there
is a growing interest in exploring quantum information within larger
Hilbert spaces, achieved either by increasing the number of qubits
or by utilizing d-level quantum systems, known as qudits. The decision
to expand into higher dimensions depends on the specific objectives
of the task at hand. The overarching aim is to enhance the available
dimensions to transmit more than one bit per photon from one party
(Alice) to another (Bob). Various photonic degrees of freedom, such
as OAM \cite{DLB+11}, temporal mode \cite{MGT+17},
frequency mode \cite{KRR+17}, and spatial mode \cite{HXL+20, VSP+20},
inherent to single photons, are natural candidates for realizing HD systems. However, implementing HD-QKD protocols ideally necessitates
a reliable single-photon on demand source. While significant experimental
efforts have been directed towards constructing such sources (see
\cite{LP21,TS21} and references therein), WCPs
generated by attenuating laser outputs are commonly used as an approximate
single-photon source in many commercial products. The quantum state
of a WCP resulting from laser attenuation can be characterized as
follows:

\[
|\alpha\rangle 
= \left|\sqrt{\mu}\, e^{i\theta}\right\rangle
= \sum_{k=0}^{\infty}
\left( \frac{e^{-\mu}\mu^{k}}{k!} \right)^{1/2}
e^{ik\theta} |k\rangle,
\]
here, the symbol $|k\rangle$ denotes a Fock state ($k$ photon state)
and the mean photon number is denoted as $\mu=|\alpha|^{2}\ll1$.
Essentially, Alice generates a quantum state that can be conceptualized
as a superposition of Fock states, characterized by a Poissonian photon
number distribution expressed as $p(k,\mu)=\frac{e^{-\mu}\mu^{k}}{k!}$.
When utilizing the WCP source for signal state generation, users encounter
a probability of multi-photon pulses within the signal state. In this
scenario, Eve may execute a PNS attack.
Eve initiates the attack by employing a QND to determine the photon number, subsequently obstructing the
single-photon pulses and retaining one photon from the multi-photon
pulses. To counter the threat of PNS attacks, the decoy state method
is implemented \cite{H03,LMC05}. Notably, different intensities are
employed for generating the signal particles and the decoy state,
resulting in distinct photon number distributions. The security procedure
involves intentional and random replacement of signal pulses with
multi-photon pulses (decoy pulses) by legitimate users. Subsequently,
they assess the loss of the decoy pulses. If the loss of decoy pulses
is anomalously lower than that of signal pulses, the entire protocol
is aborted. Conversely, if the decoy pulse loss aligns with certain
expectations, the protocol continues. The estimation of signal multi-photon
pulse loss is then conducted based on the decoy pulse loss, assuming
similar values for the two losses. Within HD-QKD protocols, the decoy
state serves as a crucial tool for scrutinizing potential eavesdropping
activities and ensuring channel security.

\subsubsection{HD-Ext-B92}

Here, we summarize the HD-Ext-B92 protocol and recap some important
steps involved in the parameter estimation process proposed
in Ref. \cite{IK21}. In fact, in this section, after briefly discussing
the HD-Ext-B92 protocol we modify the derivation of the asymptotic
key rate given in \cite{IK21} (see
Appendix B). It is apt to note that negotiation
efficiency stands as a crucial parameter in determining
the security and accuracy of the final cryptographic key \cite{ZHG18, ZWF+23, CDG+21, FLL+21, BZJ+21, ZM20}.
This parameter encompasses several procedural steps, including quantum
state preparation-measurement, data reconciliation, QBER estimation,
parameter estimation, error correction, and privacy amplification\footnote{For simplicity, while evaluating the performance
of the satellite-based HD-Ext-B92 and HD-BB84 protocols, we are intentionally
excluding the incorporation of error correction and privacy amplification
measures.}. Before explaining the protocol, we would like
to introduce the notations used and the methodology for achieving
key rate. $|m\rangle$ and $|n\rangle$ are the fixed d-dimensional
states and defined from d-dimensional computational basis states $\{Z\in|1\rangle,\cdots,|m\rangle,|n\rangle,\cdots,|{\rm d}\rangle\}$,
and $|\psi\rangle=\frac{1}{\sqrt{2}}\left(|m\rangle+|n\rangle\right)$
is a fixed state which is chosen from d-dimensional diagonal basis
($X$-basis) states. As previously elucidated, a
photon can manifest as a HD system through the utilization
of different photonic degrees of freedom. It is pertinent to mention
that $|m\rangle$ and $|n\rangle$ denote specific higher-dimensional
systems that signify distinct states within the photonic degrees of
freedom. These states can be precisely measured by Bob using d-dimensional
computational basis states. In the context of the HD-Ext-B92 protocol,
Alice is only required to transmit three HD states.
Simultaneously, Bob's task entails conducting either a computational
basis measurement or a partial basis measurement in an alternative
basis (POVM $X$). It is noteworthy that this partial measurement
need only discriminate a specific superposition state as defined in
the protocol and is not obliged to discern all the possible states
(d states).

\emph{State preparation and transmission: }Alice randomly chooses
key-round and test-round. The key-round is employed for generating
raw key bit and test-round is employed to estimate error for this
protocol that will help to improve the negotiation
efficiency. Alice generates a sequence using states $|m\rangle$ and
$|\psi\rangle$ to represent classical bit values 0 and 1 during the
key-round, respectively. To assess channel noise and security, she
randomly includes decoy states in the sequence\footnote{The decoy state is not required to be a higher-dimensional
state; it can be a two-dimensional quantum state, qubit.}. Subsequently, Alice transmits the enlarged sequence
to Bob while maintaining confidentiality of the basis information.
During test-round, she uniformly prepares states $|m\rangle$, $|n\rangle$,
or $|\psi\rangle$ with a random selection, inserts decoy states,
and transmits the sequence to Bob. The basis information remains confidential
until Bob performs measurements on the sequence.

\emph{Estimation of channel noise and loss:}
Alice communicates the position and basis information of the decoy
state to Bob through a public classical announcement. Bob performs
measurements on the decoy state and publicly discloses the obtained
results. The comparison of these results allows for the calculation
of channel noise. If the noise falls within the predetermined threshold,
the protocol advances to the next step; otherwise, the protocol is
aborted. The measurement of decoy states also serves to assess channel
loss. In both the key round and the test round, the loss of signal
particles corresponds to the loss occurring for the decoy state. This
ensures the security of the channel against PNS attack.

\emph{Measurement and classical announcement:} Following
the security check conducted using the decoy state, Bob will proceed
to measure each state within the received sequence. This measurement
involves the elimination of decoy states, accomplished either using
the $Z$ basis or by a POVM bases defined by $\left\{ |\psi\rangle\langle\psi|,I-|\psi\rangle\langle\psi|\right\} $,
and referred to as POVM $X$. Here, the symbol $I$ represents the
d-dimensional identity operator. Bob sets the bit value as $1$ when
he observes $I-|m\rangle\langle m|$ by using measurement basis $Z$,
i.e., any measurement outcome in $Z$ basis other than $|m\rangle\langle m|$\footnote{This process resembles the B92 protocol \cite{B92}.
Alice utilizes the $|0\rangle$ and $|+\rangle$ states to encode
0 and 1, respectively. Bob deciphers 0 and 1 based on his measurement
outcomes, which correspond to the $|-\rangle$ and $|1\rangle$ states,
respectively.}; and he sets bit value $0$ when his measurement outcome using POVM
$X$ is other than $|\psi\rangle\langle\psi|$. All other results
are not taken into account as conclusive measurements. Alice and Bob
discard the iteration for inconclusive outcomes in key-round, and
determine the channel error rate in test-round by announcing their
basis choices and measurement results using an authenticated classical
channel. To enhance negotiation efficiency, we conduct
parameter estimation and QBER analysis, taking into account a depolarizing
channel with the consideration of the noise parameter $q$. In instances
where a round is inconclusive or does not qualify as a key round,
the obtained results are employed for parameter estimation. Finally,
they run the error correction and privacy amplification protocols
to get the final secure key. Here are some crucial
formulations pertaining to parameter estimation, which contribute
to deriving the key rate equation of the HD-Ext-B92 protocol. Depolarizing
noise is a very general noise of the Pauli class of noise channels
and can be obtained by twirling them in two dimensional quantum state.
Any quantum channel can be twirled by the depolarizing channel. Our
aspiration to expand randomized benchmarking to d-dimensions led
to the selection of the depolarizing channel in the protocol.

\[
{\mathcal{D}_{{\rm q}}(\rho)=\left(1-\frac{{\rm d}}{{\rm d}-1}{\rm q}\right)\rho+\frac{{\rm q}}{{\rm d}-1}I.}
\]
The observable statistics can be expressed within
the context of a depolarizing channel scenario.

\[
{\begin{array}{lccclclcl}
 &  & p_{mm} & = & p_{nn} & = & p_{\psi\psi} & = & 1-{\rm q},\\
 &  & p_{mc} & = & p_{nc} & = & p_{\psi c} & = & \frac{{\rm q}}{{\rm d}-1},\\
p_{m\psi} & = & p_{n\psi} & = & p_{\psi m} & = & p_{\psi n} & = & \frac{1}{2}\left(1-\frac{{\rm q}\,{\rm d}}{{\rm d}-1}\right)+\frac{{\rm q}}{{\rm d}-1}.
\end{array}}
\]
Supposing $p_{ij}$ represents the joint probability
associated with Alice's and Bob's raw bits being $i$ and $j$, considering
the scenario without eliminating that specific iteration. The values
of observable probabilities under the simulated channel for parameter
estimation are,

\[
{\begin{array}{lcl}
p_{00} & = & \frac{1}{2M}\left(1-p_{m\psi}\right),\\
p_{01} & = & \frac{1}{2M}\left(1-p_{mm}\right),\\
p_{10} & = & \frac{1}{2M}\left(1-p_{\psi\psi}\right),\\
p_{11} & = & \frac{1}{2M}\left(1-p_{\psi i}\right).
\end{array}}
\]

In \cite{IK21}, authors proposed a collective attack by Eve in which
she can independently and identically attack each round of the protocol.
Eve also can delay measurement on her register (quantum memory) after
completion of the protocol. The Devetak Winter key rate equation \cite{DW05, RGK05}
is used to compute the key rate in the asymptotic limit\footnote{For instance, we are interested in seeing the performance of satellite-based
communication in the infinitely generated raw key scenarios.}:

\begin{equation}
R\left(a,b,E\right)=\underset{N\longrightarrow\infty}{{\rm lim}}\frac{l}{N}={\rm inf}\left[S\left(a|E\right)-H\left(a|b\right)\right],\label{eq:Key-rate equantion}
\end{equation}
this analysis helps us to obtain the minimum value of the key rate
by subtracting conditional Shannon entropy $H\left(a|b\right)$ from
conditional von Neumann $S\left(a|E\right)$. Here, $S\left(a|E\right)$
is defined as the entropy or the uncertainty present in Alice's classical
register $a$ given Eve's quantum memory $E$ and $H\left(a|b\right)$
denotes the entropy present in Alice's register $a$ given Bob's classical
register $b$. Here, $l$ is the number of secret key
bits over the transmission of $N$ number of raw key. In Eq. (\ref{eq:Key-rate equantion}),
$R$ elucidates the infimum value of key rate under all collective
attacks performed by Eve. We apply Theorem
\cite{K16, IK21} introduced by Krawec and analyze
the parameter estimation to derive $S(a|E)$ and $H(a|b)$, consequently
enhancing negotiation efficiency, thereby contributing to the overall
improvement of the security and correctness of the final key. This
formulation also facilitates the determination of QBER for the HD-Ext-B92
protocol. Using these findings, we can establish the minimum value
for the key rate by employing Equations (\ref{eq:Value of S(A|E)})
and (\ref{eq:Value of H(a|b)}) from Appendix B in Equation (\ref{eq:Key-rate equantion}).

\subsubsection{HD-BB84}

In a two-level system, BB84 \cite{BB84} protocol is well studied
both in theoretical and experimental domains. Essentially  qubits are used to realize this scheme for QKD which
uses two mutually unbiased bases randomly. In a more general scenario,
higher dimensional quantum systems (qudits) can
be used to realize the same task (i.e., QKD), and such a modified
version of BB84 protocol is referred to as \emph{qudit}- (i.e., a
quantum state in ${\rm d}$-dimensional Hilbert space) based BB84
protocol or HD-BB84 protocol. Here, we briefly discuss the HD-BB84
protocol \cite{CBK+02} and the necessary formulae to compute the
secret key rate.

In this protocol, Bob generates a sequence of qudits,
where each qudit represents a higher-dimensional quantum state. These
states are prepared based on a randomly selected basis, chosen from
two mutually complementary bases: $Z=\left\{ |0\rangle,|1\rangle,\cdots,|D-1\rangle\right\} $
and $X=\left\{ |x_{0}\rangle,|x_{1}\rangle,\cdots,|x_{D-1}\rangle\right\} $.
The qudit sequence is generated using a WCP source, with the possibility
of a PNS attack. To counter the threat of a PNS attack, Bob strategically
introduces decoy states at random positions within the qudit sequence,
expanding it. Upon receiving the enlarged sequence, Bob communicates
the positions and basis information of the decoy states to Alice.
Subsequently, Alice performs measurements on the decoy states and
publicly announces the measurement outcomes. Both parties then compute
the losses and errors incurred during the communication channel. If
the error rate falls within the acceptable threshold, the protocol
advances to the next stage. After discarding the decoy particles from
the enlarged sequence, Alice proceeds with a measurement operation
on the qudits. The measurement is performed by randomly selecting
one of the two d-dimensional bases, namely $Z$ and $X$. This process
ensures the security and reliability of the quantum communication
protocol, especially in the presence of potential PNS attacks. Subsequently,
they announce their bases choice in a public authenticated classical
channel \cite{CBK+02} and obtain correlated ${\rm d}$-ary random
variables when they use the same bases. With $\frac{1}{2}$ probability,
Alice and Bob use different bases and yield uncorrelated results which
are considered as discarded data after key-sifting sub-protocol. It
is crucial to emphasize that the loss should be consistent for both
decoy states and signal states, thereby ensuring the security of the
channel. This method ensures that any effort made by an,
Eve (who is unaware of the chosen basis), to obtain information about
Bob's state will result in an error in transmission, which can subsequently
be detected by the legitimate parties.

To ensure a smooth comprehension of readers we would like to provide
a concise overview of key points discussed in Ref. \cite{BCC+10}, authors have modified the Maassen and Uffink
bound \cite{K87,MU88} to establish a new bound on the uncertainties
associated with the measurement results, contingent on the amount
of entanglement between the measured particle $(A)$, and the quantum
memory ($B)$. This relationship can be expressed mathematically as,

\begin{equation}
S\left(Z|B\right)+S\left(X|B\right)\ge\log_{2}\frac{1}{C}+S\left(A|B\right),\label{eq:First_Bound}
\end{equation}
where, $Z$ and $X$ are two possible observable like measurement
bases and $A$ refers to the qudit measured by Alice which is sent
by Bob and $B$ refers to the qudit which represents the quantum memory
of Bob. $S$ represents von Neumann entropy and $S\left(A|B\right)$
quantifies the amount of entanglement between $A$ and $B$. $C:={\rm max_{i,j}|\langle\phi_{i}|\psi_{j}\rangle|^{2}}$,
where $|\phi_{{\rm i}}\rangle$ and $|\psi_{{\rm j}}\rangle$ are
the eigenvectors of $Z$ and $X$, respectively. Using a result established
by Devetak and Winter \cite{DW05}, the minimum limit on the quantity
of key that Alice and Bob can extract from each state can be expressed
as $S\left(Z|E\right)-S\left(Z|B\right)$\footnote{Here, Z and X can be employed in a similar manner or with a similar
effect.}. This limit is applied when the eavesdropper is trying to obtain
the key from the composite quantum system\footnote{Eve performs an entanglement operation using her ancillary state $E$
with Alice's state ($A$) and Bob's quantum memory ($B$).} $\rho_{ABE}$, where $A$ is Alice's particle, $B$ is Bob's quantum
memory, and $E$ is Eve's ancillary state. Equation (\ref{eq:First_Bound})
may be reformulated as $S\left(Z|E\right)+S\left(X|B\right)\ge{\rm \log_{2}}\frac{1}{C}$
(see Supporting Information of ref. \cite{BCC+10}), and
the key rate equation may be written as,

\begin{equation}
\begin{array}{lcl}
r\left(A,B,E\right) & \ge & S\left(Z|E\right)-S\left(Z|B\right)\\
 & \ge & \left[{\rm \log_{2}}\frac{1}{C}-S\left(X|B\right)\right]-S\left(Z|B\right).
\end{array}\label{eq:KeyRate_HD_BB84_Before_Fano}
\end{equation}
Both parties involved in the HD-BB84 protocol utilize complete bases
elements ($Z$ and $X$ bases) within a d-dimensional Hilbert space.
In this context, the lower limit for parameter estimation can be directly
achieved from Fano's inequality. It is possible to consider an arbitrary
quantum channel with a noise probability denoted as $q$. Fano's inequality
states that $S\left(Z|B\right)\le h\left(q\right)+q\log_{2}\left({\rm d}-1\right)$.
By applying this relation in Eq. (\ref{eq:KeyRate_HD_BB84_Before_Fano})
and considering the condition that ${\rm \log_{2}}\frac{1}{C}$ cannot
exceed ${\rm \log_{2}}{\rm d}$, we can derive

\begin{equation}
\begin{array}{lcl}
r & \ge & \log_{2}{\rm d}-2\left(h\left(q\right)+q\log_{2}\left({\rm d}-1\right)\right)\end{array}.\label{eq:Key_rate_HD_BB84}
\end{equation}
For the binary encoding and decoding scheme the conditional entropy
of Alice's measurement outcome given Bob's measurement result is equal
to $h\left(\varepsilon\right)$, where $\varepsilon$ is QBER \cite{CRE_04,BMA+09}. Here, $q$ is the depolarizing
channel parameter i.e., the probability that outcome of the $Z$ by
Alice and Bob is not equal and $h$ is binary entropy. Through
an examination of the aforementioned formulas, we ascertain the lower
bound of the key rate for the HD-BB84 protocol. This analysis involves
the application of Fano's inequality, which establishes bound on parameter
estimation. Additionally, the determination of the QBER is crucial
for enhancing negotiation efficiency, thereby playing a pivotal role
in augmenting the overall security and correctness of the final key
in the HD-BB84 protocol.

Before delving
into the formal analysis of the aforementioned formalism, it is crucial
to elucidate the relationship among key rate, QBER, noise and negotiation
efficiency. The noise introduced in the quantum channel ($q$), results
in QBER ($\varepsilon$) in the raw key sequence after executing the
quantum protocol. The conditional entropy of Alice's measurement outcome
given Bob's measurement outcome, as well as the conditional von Neumann
entropy of Alice's quantum state given Bob's quantum state, can be
expressed as functions of $\varepsilon$. Additionally, it is evident
that the conditional entropy is directly influenced by the noise in
the quantum channel. We may now explicitly analyze the previously
mentioned formulae, considering negotiation efficiency and QBER within
the key rate equation. First, we need to elaborate on all the elements
in Eq. (\ref{eq:Key-rate equantion}). From Appendix B, $S(a|E)$
represents the conditional von Neumann entropy, defined as follows:

\[
S(a|E)\ge\underset{c\ne m,c\ne n}{\sum}\left(\frac{K_{c}^{0}+K_{c}^{1}}{M}\right)S_{c}+\left(\frac{K_{m}^{0}+K_{m}^{1}}{M}\right)S_{m}+\left(\frac{K_{n}^{0}+K_{n}^{1}}{M}\right)S_{n},
\]
where,

\[
\begin{array}{lclcclcl}
K_{c}^{0} & := & \langle E_{c}^{m}|E_{c}^{m}\rangle, &  &  & K_{c}^{1} & := & \frac{1}{2}\,\langle F_{c}|F_{c}\rangle,\forall\,c\ne m,n\\
\\
K_{m}^{0} & := & \frac{1}{4}\,\langle E_{m}^{m}|E_{m}^{m}\rangle, &  &  & K_{m}^{0} & := & \frac{1}{8}\,\langle F_{m}|F_{m}\rangle,\\
\\
K_{n}^{0} & := & \frac{3}{4}\,\langle E_{n}^{m}|E_{n}^{m}\rangle, &  &  & K_{n}^{1} & := & \frac{3}{8}\,\langle F_{n}|F_{n}\rangle.
\end{array}
\]
and
\[
\begin{array}{lcl}
S_{c} & = & h\left(\frac{K_{c}^{0}}{K_{c}^{0}\,+\,K_{c}^{1}}\right)-h\left(\frac{1}{2}+\frac{\sqrt{\left(K_{c}^{0}\,-\,K_{c}^{1}\right)^{2}+4\,{\rm Re^{2}}\langle E_{c}^{m}|\frac{1}{\sqrt{2}}F_{c}\rangle}}{2\,\left(K_{c}^{0}\,+\,K_{c}^{1}\right)}\right),\\
S_{m} & = & h\left(\frac{K_{m}^{0}}{K_{m}^{0}\,+\,K_{m}^{1}}\right)-h\left(\frac{1}{2}+\frac{\sqrt{\left(K_{m}^{0}\,-\,K_{m}^{1}\right)^{2}+4\,{\rm Re^{2}}\langle\frac{1}{2}E_{m}^{m}|\frac{1}{2\sqrt{2}}F_{m}\rangle}}{2\,\left(K_{m}^{0}\,+\,K_{m}^{1}\right)}\right),\\
S_{n} & = & h\left(\frac{K_{n}^{0}}{K_{n}^{0}\,+\,K_{n}^{1}}\right)-h\left(\frac{1}{2}+\frac{\sqrt{\left(K_{n}^{0}\,-\,K_{n}^{1}\right)^{2}+4\,{\rm Re^{2}}\frac{3}{4\sqrt{2}}\langle E_{n}^{m}|F_{n}\rangle}}{2\,\left(K_{n}^{0}\,+\,K_{n}^{1}\right)}\right).
\end{array}
\]
Further, $H\left(a|b\right)$ represents the conditional entropy of
Alice's measurement outcome given Bob's measurement outcome, defined
as follows:

\[
\begin{array}{lcl}
H\left(a|b\right) & = & H\left(p_{00},\,p_{01},\,p_{10},\,p_{11}\right)-h\left(p_{00}+p_{10}\right).\end{array}
\]
We have previously defined the values $p_{00},p_{01},p_{10}$ and
$p_{11}$ (also detailed in Appendix B) through the analysis of parameter
estimation for the HD-Ext-B92 protocol. These terms depend on $q$.
Moreover, for binary encoding and decoding, $H(a|b)\equiv h(\varepsilon)$.
Therefore, it can be concluded that the conditional entropy and key
rate depend on the value of $\varepsilon$ and, consequently, on $q$
as $\varepsilon$ depends on $q$. Now, if we incorporate negotiation
efficiency\footnote{Negotiation efficiency is defined as the effectiveness
with which the steps involved in establishing a secure key are executed.
These steps generally encompass sifting, error correction and privacy
amplification. High negotiation efficiency indicates that these processes
are conducted in a way that optimizes the conversion of raw key bits
into secure key bits \cite{ZHG18,ZWF+23,BZJ+21}.} ($\xi$) into the key rate for the HD-Ext-B92 protocol
(as shown in Eq. (\ref{eq:Key-rate equantion})), the secure key rate
equation can be written as,

\[
R_{\xi}\equiv\xi\,R\left(a,b,E\right)=\xi\left({\rm inf}\left[S\left(a|E\right)-H\left(a|b\right)\right]\right).
\]
Here, $H(a|b)$ depends on noise and the QBER value. To reduce QBER,
legitimate users perform error correction and privacy amplification,
which increase the mutual information between Alice and Bob. This
leads to a a decrease in $H(a|b)$ and maximizes the proportion of
raw key bits successfully converted into secure key bits. Therefore,
analyzing QBER of a quantum protocol enhances the \emph{correctness}
and \emph{security} of the raw
key and optimizes the conversion of raw key into secure key more efficiently,
thereby improving the negotiation efficiency. The QBER analysis during
the test rounds is a crucial aspect of quantum communication protocols,
as it helps to improve negotiation efficiency. This logic applies
similarly to the HD-BB84 protocol. From Eqs. (\ref{eq:KeyRate_HD_BB84_Before_Fano})
and (\ref{eq:Key_rate_HD_BB84}), it is clear that the conditional
von Neumann entropy $S\left(Z|B\right)$ depends on $q$ (as Fano's
inequality states that $S\left(Z|B\right)\le h\left(q\right)+q\log_{2}\left({\rm d}-1\right)$),
which affects the QBER value as well. Considering negotiation efficiency
in the key rate for the HD-BB84 protocol (Eq. (\ref{eq:Key_rate_HD_BB84})),
the secure key rate equation can be expressed as follows:

\[
r_{\xi}\equiv\xi\,r\ge\xi\left(\log_{2}{\rm d}-2\left(h\left(q\right)+q\log_{2}\left({\rm d}-1\right)\right)\right).
\]
Here, the legitimate parties perform parameter estimation and error
correction on the raw key sequence, which leads to a reduction in
QBER and a decrease in $S\left(Z|B\right)$. As a result, the key
rate increases. Consequently, this process increases the proportion
of raw key bits that are successfully converted into secure key bits.
More specifically, analyzing the QBER during the test rounds of a
quantum communication protocol improves the correctness and security
of the raw key, making the conversion process to secure key more efficient.
This optimization of the secure key conversion improves the negotiation
efficiency, $\xi$. From this analysis, we can conclude that QBER
analysis is crucial for improving the negotiation efficiency of a
quantum protocol.

The preceding discussion, along with an analysis
of the formulations, elucidates that Eq. (\ref{eq:Key-rate equantion}),
which incorporates Eqs. (\ref{eq:Value of S(A|E)}) and (\ref{eq:Value of H(a|b)})
from Appendix B, and Eq. (\ref{eq:Key_rate_HD_BB84}), representing
the key rate formula for HD-Ext-B92 and HD-BB84 protocols, inherently
encompasses the essential steps of parameter negotiation
efficiency. By employing these key rate equations,
coupled with the transmittance in satellite-based communication (cf.
Eqs. (\ref{eq:Transmittance}) and (\ref{eq:PDT Equation})), we can
derive the average key rate for LEO satellite-based quantum communication.
The computation of the average key rate using Eq. (\ref{eq:Average key-rate})
facilitates the determination of the probability distribution of key
rates and the variation of key rates concerning various parameters
in satellite quantum communication, as detailed in Sec.\ref{sec:III}.
Before delving further this section, we conduct an in-depth analysis
of these key rate equations for both HD-QKD protocols in the subsequent
paragraph. Before proceeding,
it is crucial to emphasize the definition of noise tolerance and its
relationship with QBER. Noise tolerance in a quantum communication
protocol refers to the ability of the protocol to function correctly
and securely despite the presence of noise. Specifically, the value
of noise tolerance for a protocol is determined by the point at which
the secure key rate approaches zero. Noise can originate from various
sources, including environmental disturbances, imperfections in quantum
devices, and potential eavesdropping activities. Noise tolerance and
QBER are interdependent factors in quantum communication protocols.
Effective noise management ensures that QBER remains below the critical
threshold, thereby enabling secure and efficient quantum communication.
Error correction techniques are employed to correct errors in the
raw key, and their efficiency depends on the QBER. Higher QBER necessitates
more robust error correction, which can diminish the efficiency of
the key generation process. Privacy amplification is used to reduce
the information an Eve might have obtained; the amount
of privacy amplification required increases with higher QBER, further reducing the final key length. Maintaining
noise within the noise tolerance limit helps to keep QBER below a
secure threshold value, which is specific to the protocol. In such
cases, extensive privacy amplification and error correction may not
be strictly necessary.

\begin{figure}[h]
\begin{centering}
\includegraphics[width=\linewidth]{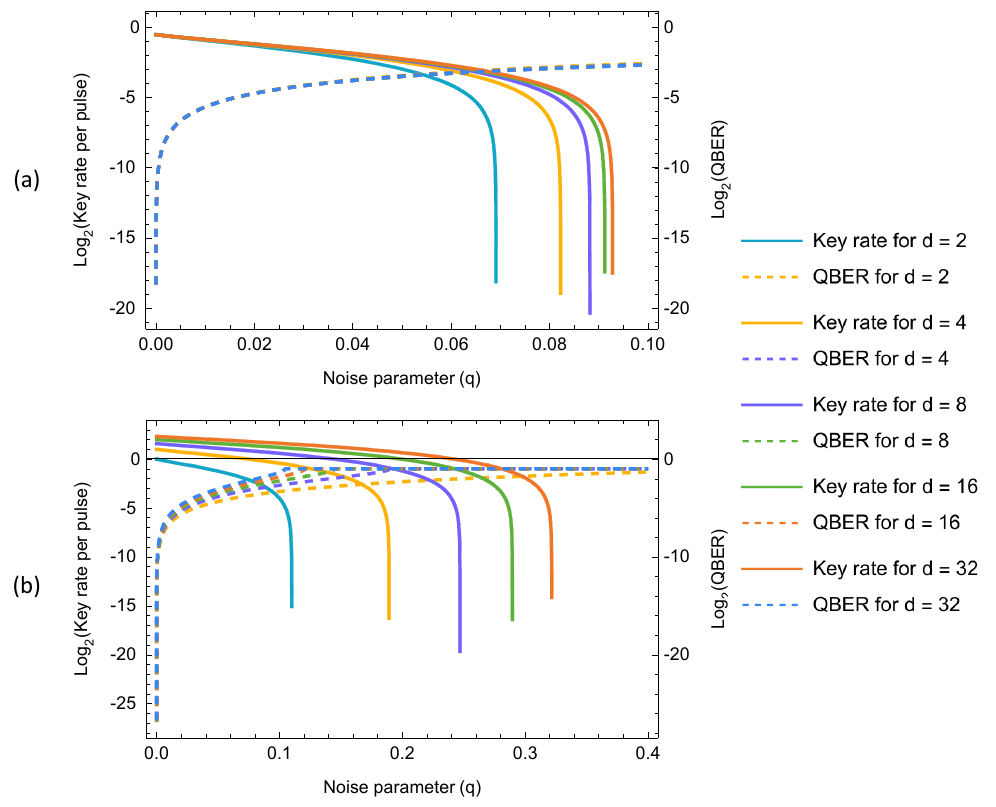}
\par\end{centering}
\caption{\label{fig:Keyrate_Noise_QBER} Plot of variation of
key-rate and QBER with channel noise parameter (both plots share a
common legend): (a) key-rate and QBER analysis of HD-Ext-B92 protocol
with noise parameter (q) for different dimensions in Hilbert space
(the plot lines representing QBER for all dimensions are superimposed),
(b) key-rate and QBER analysis of HD-BB84 protocol with noise parameter
(q) for different dimensions in Hilbert space.}
\end{figure}

To analyze the behavior of the key rate per pulse and the QBER concerning
the noise parameter in both the above-discussed HD protocols, we utilize
the key rate equations (refer to Eqs. (\ref{eq:Key-rate equantion}),
(\ref{eq:Key_rate_HD_BB84}), and (Appendix B) and the binary QBER
function $\left(h\left(\varepsilon\right)\right)$. Now, we analyze
the result illustrated in Fig. \ref{fig:Keyrate_Noise_QBER} for
HD-Ext-B92 and HD-BB84 schemes. We can observe in the HD-Ext-B92 protocol
that events with mismatched bases are not disregarded, which occurs
when Alice and Bob employ different measurement bases. These events
can significantly enhance key generation rates \cite{IK21, BHP93, WMU08, MW08, TCK+14},
and therefore noise tolerance is also increased for this scheme which
is evident from the graph. We plot the variation of the key rate of
HD-Ext-B92 protocol with noise parameter (${\rm q}$) in a depolarizing
channel in the ${\rm d}$-dimensional Hilbert space; we also depict
the variation of QBER with the same noise parameter $({\rm q})$.
It may be observed that as the value of ${\rm d}$ increases, the
tolerance for noise also increases, showing a rise from $7\%$ to
$10\%$. It is apt to note that, the maximum tolerable noise is dependent
on the choice of both the depolarizing channel and of Eve's ancilla
state, since, these two factors significantly impact parameter estimation
and consequently affect the key rate. Nevertheless, our analysis is
confined to a specific choice of these two factors, which have been
outlined in Appendix B. In Fig. \ref{fig:Keyrate_Noise_QBER} (a),
it becomes evident that the QBER remains constant across different
${\rm d}$ values. The plots representing the distinct ${\rm d}$
values (i.e., ${\rm d}=2,4,8,16,32$) overlap in QBER analysis, indicating
consistent outcomes for higher-dimensional cases of the HD-Ext-B92
protocol. Additionally, the graph (represented by a dotted line) demonstrates
that the QBER reaches a saturation point for a particular ${\rm q}$
value (for HD-Ext-B92). We have computed the initial point where the
QBER begins to rise for various ${\rm q}$ values and observed a physically
reasonable variation. For instance, when ${\rm q}$ is $\sim0.005$,
the QBER is approximately $0.015$. As the ${\rm q}$ value increases
to $\sim0.1$, the QBER saturates at approximately at $0.088$. Now,
we analyze the plots for HD-BB84 in Fig. \ref{fig:Keyrate_Noise_QBER}
(b) and undertake a comprehensive comparison with HD-Ext-B92. A numerical
assessment reveals that the key rate increases as the value of ${\rm d}$
rises for HD-BB84. Conversely, in HD-Ext-B92, the minimum key rate
remains fairly consistent for all ${\rm d}$ values which is around
$0.7$. Further, the noise tolerance is increased significantly with
a greater value of ${\rm d}$ in HD-BB84. For instance, the tolerable
noise is $\sim11\%$ for ${\rm d}=2$ (for qubit) and with the increased
value of ${\rm d}=32$, this limit increases to $\sim32\%$. More simply,
the HD-Ext-B92 protocol operates securely and correctly with a maximum
noise tolerance ranging from $7\%$ to $10\%$, whereas the HD-BB84
protocol can tolerate up to $11\%$ noise for ${\rm d}=2$ and up
to 32\% noise for ${\rm d}=32$, as can be inferred from Eq. (\ref{eq:Key_rate_HD_BB84}). This
outcome demonstrates the advantage of opting for the HD-BB84 protocol
over HD-Ext-B92 when considering aspects like key rate and noise tolerance.
HD-BB84 surpasses the HD-Ext-B92 protocol. It is worth mentioning
that in the original scheme of HD-Ext-B92 \cite{IK21}, authors do
not employ two complete bases as HD-BB84 does. In their approach,
they utilize a simplified version in which Alice's requirement is
reduced to transmitting just three states, and Bob only needs to carry
out partial measurements within the second basis \cite{IK21}. Additionally,
it is important to highlight that they did not select an optimal basis
configuration. Alternate choices for the encoding state might yield
greater key rates for the HD-Ext-B92 protocol, as shown in cases involving
qubits \cite{LGT09,K16}. If we examine the QBER aspect within the
context of HD-BB84, it becomes apparent that the variation of QBER
with the noise parameter ($q$) rapidly converges to a saturation
value ($\sim0.25$) as the dimension of qudit increases. In contrast
to the HD-Ext-B92 protocol, the susceptibility of QBER to noise is
notably more vulnerable in the HD-BB84 protocol. Moreover, as depicted
in Fig. \ref{fig:Keyrate_Noise_QBER}, when considering ${\rm d}=32$,
the saturation point of noise tolerance is attained in the HD-Ext-B92
protocol. In contrast, in the HD-BB84 protocol, the rate at which
noise tolerance increases becomes progressively lower as ${\rm d}$
increases. It is noteworthy that at ${\rm d}=32$, the QBER has not
yet reached its saturation point (for HD-BB84); this point will be
reached at higher values of ${\rm d}$. From Fig. \ref{fig:Keyrate_Noise_QBER} (b), we observe that when
the key rate reaches zero, it indicates the noise tolerance values,
${\rm q}$, with QBER lines reaching their maximum value for the respective
dimensions. The conditional entropy of Alice's measurement outcome
given Bob's measurement outcome (and the conditional von Neumann entropy
of Alice's quantum state given Bob's quantum state) is directly influenced
by the noise in the quantum channel. These conditional entropies can
be expressed as function of QBER. Mathematically, the relationship
between noise tolerance and QBER in a quantum communication protocol
is illustrated by the effect of QBER on the secure key rate. A protocol
has a capability to handle noise up to a specific QBER threshold,
allowing error correction and privacy amplification to still generate
a secure key. The binary entropy function $h({\rm QBER})$ and the
specific security function $f({\rm QBER})$ measure the information
loss due to errors and potential eavesdropping, respectively.

\subsection{Satellite-based optical links: model used for the elliptic beam approximation\label{subsec:Elliptic_Beam_Model}}

In this chapter, we aim to analyze the performance of key rates in
various situations of HD-Ext-B92 and HD-BB84 protocols. The channel
transmission $\eta$ for the light propagation through atmospheric
links using elliptic-beam approximation as introduced by Vasylyev
et al. \cite{VSV16,VSV+17} will be employed to perform the analysis.
Further, in what follows, we impose the generalized approach\footnote{Using non-uniform link between a satellite and the ground station,
referred to in Eq. (\ref{eq:Down-Link and Up-Link condition}).} and different weather conditions as introduced in \cite{LKB19}.
This method yields an impact on the value of transmittance as the
transmittance is determined by beam parameters along with the diameter
of the receiving aperture. To provide readers with a clearer understanding
of both the elliptic beam approximation and its modified version in
a more comprehensive manner, in this section, we offer a succinct
explanation of the underlying theory.

\begin{figure}[h]
\centering{}\includegraphics[scale=0.6]{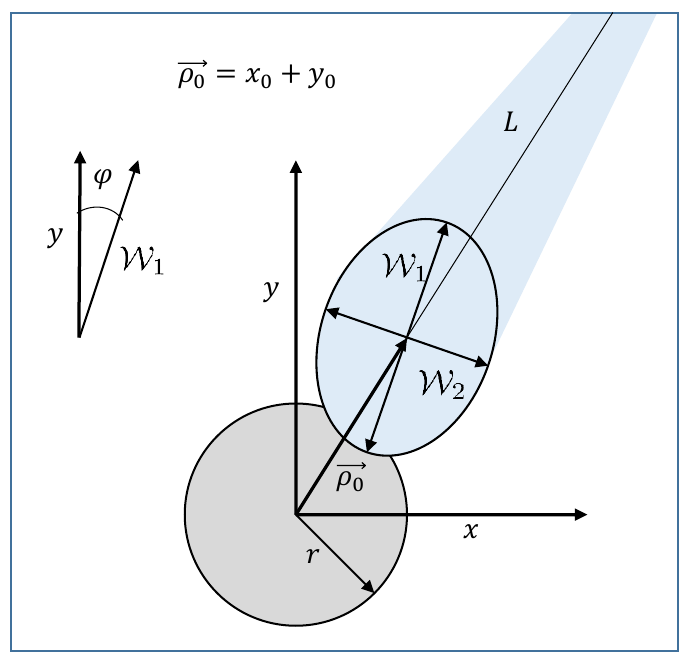}\caption{\label{fig:Elliptic_beam_impinge_circular_aperture}
Diagram illustrating the received beam and the receiving aperture.
${\rm L}$ is the total link-length in the propagation direction,
$r$ represents the radius of the receiving aperture, $\rho_{0}=(x_{0},y_{0})$
signifies the position of the beam centroid, $\mathcal{W}_{1}$ and
$\mathcal{W}_{2}$ are the principal semi axes of elliptic beam profile,
and $\varphi$ is the orientation angle of the elliptic beam.}
\end{figure}

Temporal and spatial fluctuations in temperature and pressure within
turbulent atmospheric flows result in random variations of the air's
refractive index. Consequently, the atmosphere introduces losses to
transmitted photons, which are detected at the receiver through a
detection module featuring a limited aperture. The transmitted signal
undergoes degradation due to phenomena like beam wandering, broadening,
deformation, and similar effects. We can examine this scenario by
focusing on a Gaussian beam propagating along the z axis, reaching
the aperture plane positioned at a distance $z={\rm L}$. In
this analysis, we observe that assuming perfect Gaussian beams emitted
by the transmitter is not entirely realistic. Standard telescopes
typically produce beams with intensity distributions that closely
resemble a circular Gaussian profile with some deviations, often caused
by truncation effects at the edges of optical elements. One notable
consequence of these imperfections is the inherent broadening of the
beam due to diffraction. In our model, we can address this phenomenon
by adjusting the parameter representing the initial beam width $\left(\mathcal{W}_{0}\right)$,
thereby accounting for the increased divergence in the far-field resulting
from the imperfect quasi-Gaussian beam. To capture this effect, we
incorporate the transmission of the elliptical beam through a circular
aperture and consider the statistical characteristics of the elliptical
beam as it propagates through turbulence using a Gaussian approximation.
However, it's important to mention certain restriction for simplifications
in our approach, particularly the assumption of isotropic atmospheric
turbulence. For a more detailed formulation, readers are referred
to the Supporting Information of Ref. \cite{VSV16}. That quasi-Gaussian
beam is directed through a link that spans both the atmosphere and
vacuum, originating from either a transmitter situated in orbit or
a ground station. The link is characterized by non-uniform conditions.
Generally, the varying intensity transmittance of such a signal (received
beam) via a circular aperture of radius $r$ of the receiving telescope
is expressed as follows \cite{VSV12, VSV16}:

\begin{equation}
\begin{array}{lcl}
\eta & = & \int_{\left|\rho\right|^{2}=r^{2}}{\rm d^{2}\boldsymbol{\rho}\left|u\left(\mathbf{\boldsymbol{\rho}},L\right)\right|^{2},}\end{array}\label{eq:Transmittance}
\end{equation}
where $u\left(\mathbf{\boldsymbol{\rho}},{\rm L}\right)$ represents
the beam envelope at the receiver plane, located at a distance ${\rm L}$
from the transmitter, and $\left|u\left(\mathbf{\boldsymbol{\rho}},{\rm L}\right)\right|^{2}$
is the normalized intensity with respect to full $\boldsymbol{\rho}$
plane, where $\boldsymbol{\rho}$ denotes the position vector within
the transverse plane. The vector parameter ${\rm \boldsymbol{v}}$
fully characterizes the state of the beam at the receiver plane (see
Fig. \ref{fig:Elliptic_beam_impinge_circular_aperture}),

\begin{equation}
{\rm \boldsymbol{v}}=\left(x_{0},y_{0},\mathcal{W}_{1},\mathcal{W}_{2},\varphi\right),\label{eq:Vector of beam-parameters}
\end{equation}
$x_{0}$, $y_{0}$, $\mathcal{W}_{1/2}$, and $\varphi$ imply the
beam centroid coordinates, the principal semi-axes of the elliptic
beam profile, and the orientation angle of the elliptic beam, respectively.
The transmittance is determined by these beam parameters along with
the radius of the receiving aperture ($r$).

\begin{figure}[!htbp]
    \centering
    \includegraphics[scale=0.5]{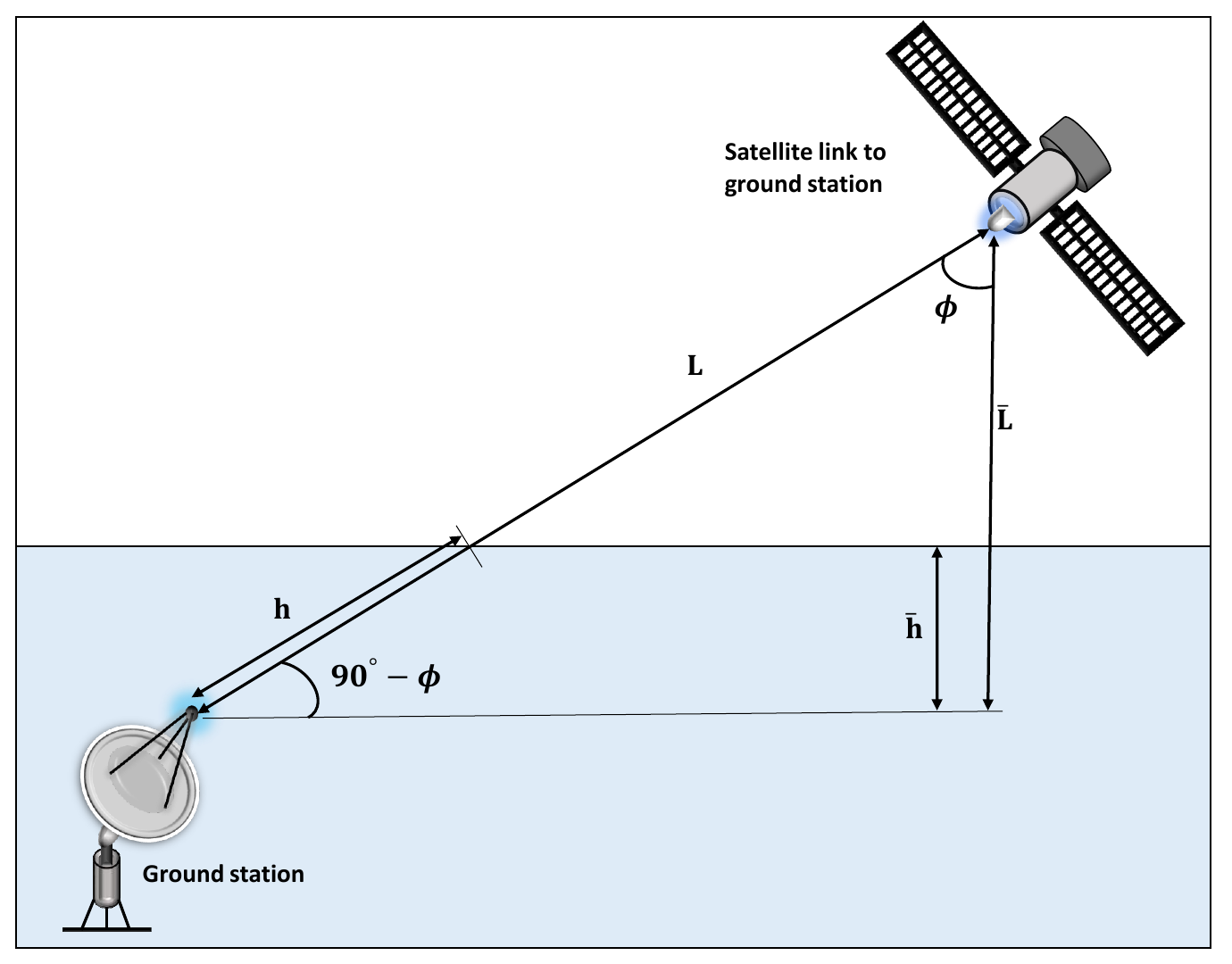}
    \caption{This figure depicted that the non-uniform free-space link between the satellite
and the ground station. The diagram highlights key parameters, $\overline{{\rm h}}$
is the thickness of the atmosphere, $\overline{{\rm L}}$ is the altitude
of the satellite, ${\rm h}$ represents length of the propagation
of light inside atmosphere, ${\rm L}$ is the link length between
satellite and ground station, $\phi$ denotes the zenith angle. Uplink
(downlink) configuration represents the transmission of light from
ground station to satellite (satellite to ground station).}
    \label{fig:Non-uniform_free_space_link}
\end{figure}

In general, the atmosphere can be categorized into distinct layers,
each characterized by various physical parameters such as air density,
pressure, temperature, the presence of ionized particles, and more.
The arrangement of these layers varies according to location, particularly
concerning the extent of each layer's thickness. Without loss of generality,
we adopt a simplified model of a satellite-based optical link \cite{LKB19}.
This model entails a uniform atmosphere up to a specific altitude
denoted as $\overline{{\rm h}}$, beyond which a vacuum extends all
the way to the satellite situated at an altitude marked as $\overline{{\rm L}}$,
as illustrated in Fig. \ref{fig:Non-uniform_free_space_link}. Rather
than dealing with a continuous range of values characterizing physical
quantities as a function of altitude, this approach involves just
two key parameters. These parameters encompass the value of the physical
quantity within the uniform atmosphere and the effective altitude
range, $\overline{{\rm h}}$. This simplification is likely to be
quite accurate because atmospheric influences are predominantly significant
only within the initial $10$ to $20$ kilometers above the Earth's
surface. This is particularly relevant considering that the standard
orbital height for LEO satellites is above 400 kilometers . In our
analysis, we set the value of $\overline{{\rm L}}$ to $500$ km,
and assume that the zenith angle falls within the range of $\left[0^\circ,80^\circ\right]$.
Under these conditions, the range of the satellite's orbit suitable
for key distribution is approximately ${\rm L}\in\left[500,2000\right]$
km\footnote{The correlation between total link length and zenith angle is, ${\rm L}=\overline{{\rm L}}\sec\phi$.}.
The given context mandates that the effective atmospheric thickness
$\overline{{\rm h}}$ remains constant at 20 km, by the aforementioned
factors. We extend the discussion by maintaining the premise that
the parameters quantifying the influence of atmospheric effects remain
constant (with values greater than $0$) within the atmosphere and
are set to $0$ outside it. In this context, we can make use of the
assumption that,

\begin{equation}
\begin{array}{cl}
{\rm Down-link} & \begin{cases}
C_{n}^{2}\left(z\right) & =C_{n}^{2}\,\text{\textohm}\left(z-\left({\rm L}-{\rm h}\right)\right),\\
n_{0}\left(z\right) & =n_{0}\,\text{\textohm}\left(z-\left({\rm L}-{\rm h}\right)\right),
\end{cases}\\
\\
{\rm Up-link} & \begin{cases}
C_{n}^{2}\left(z\right) & =C_{n}^{2}\,\text{\textohm\ensuremath{\left({\rm h}-z\right)},}\\
n_{0}\left(z\right) & =n_{0}\,\text{\textohm}\left({\rm h}-z\right).
\end{cases}
\end{array}\label{eq:Down-Link and Up-Link condition}
\end{equation}
Here, $C_{n}^{2}$ represents the refractive index structure constant\footnote{Several altitude-dependent models describing the refractive index
structure constant $C_{n}^{2}$ have been documented \cite{V80, HS64, LC06, FSV+10}.
Among these, the parametric fit proposed by Hufnagel and Valley is
widely adopted and faithfully captures the characteristics of $C_{n}^{2}$
in climates characteristic of mid-latitudes \cite{HS64, V80}.}, and $n_{0}$ denotes the density of scattering particles \cite{TP88, T84}.
The function $\text{\textohm}\left(z\right)$ corresponds to the Heaviside
step-function\footnote{The value of this function is zero for negative arguments and one
for positive arguments. This function falls within the broader category
of step functions.}. As stated above, the parameter $z$ signifies the longitudinal coordinate,
while ${\rm L}$ stands for the overall length of the link. Additionally,
${\rm h}$ represents the distance covered within the atmosphere,
as illustrated in the accompanying Fig. \ref{fig:Non-uniform_free_space_link}.

Now, let's consider the transmittance, as defined in Eq. (\ref{eq:Transmittance}),
for an elliptic beam that strikes a circular aperture with a radius
of $r$. This transmittance can be expressed as follows \cite{VSV16}:

\begin{equation}
\begin{array}{lcl}
\eta\left(x_{0},y_{0},\mathcal{W}_{1},\mathcal{W}_{2},\varphi\right) & = & \frac{2\,\chi_{{\rm ex}t}}{\pi\mathcal{W}_{1}\mathcal{W}_{2}}\int_{0}^{r}\rho\,{\rm d}\rho\int_{0}^{2\pi}{\rm d}\theta{\rm e^{-2A_{1}\left(\rho cos\theta-\rho_{0}\right)^{2}}}{\rm e^{-2A_{2}\rho^{2}sin^{2}\theta}}e^{-2{\rm A}_{3}\left(\rho{\rm cos}\theta-\rho_{0}\right)\rho{\rm sin}\theta}.\end{array}\label{eq:PDT Equation}
\end{equation}
In this context, $r$ represents the radius of the aperture, while
$\rho$ and $\theta$ denote the polar coordinates of the vector $\boldsymbol{\rho}$,

\[
\begin{array}{lcl}
x & = & \rho{\rm \,cos}\theta,\\
y & = & \rho{\rm \,sin}\theta,
\end{array}
\]
here, $\rho_{0}$ and $\theta_{0}$ represent polar coordinates corresponding
to the vector $\boldsymbol{\rho}_{0}$,

\[
\begin{array}{lcl}
x_{0} & = & \rho_{0}\,{\rm cos}\theta_{0},\\
y_{0} & = & \rho_{0}{\rm \,sin}\theta_{0},
\end{array}
\]
and

\[
\begin{array}{lcl}
{\rm A}_{1} & = & \left(\frac{{\rm cos}^{2}\left(\varphi-\theta_{0}\right)}{\mathcal{W}_{1}^{2}}+\frac{{\rm sin}^{2}\left(\varphi-\theta_{0}\right)}{\mathcal{W}_{2}^{2}}\right),\\
{\rm A}_{2} & = & \left(\frac{{\rm sin}^{2}\left(\varphi-\theta_{0}\right)}{\mathcal{W}_{1}^{2}}+\frac{{\rm cos}^{2}\left(\varphi-\theta_{0}\right)}{\mathcal{W}_{2}^{2}}\right),\\
{\rm A}_{3} & = & \left(\frac{1}{\mathcal{W}_{1}^{2}}-\frac{1}{\mathcal{W}_{2}^{2}}\right){\rm sin\,2\left(\varphi-\theta_{0}\right).}
\end{array}
\]
These expressions can be employed for numerical integration, as described
in Eq. (\ref{eq:PDT Equation}), through the Monte Carlo method or
another effective technique for the same purpose. To simplify the
process of integration using the Monte Carlo method, it requires the
generation of $N$ sets of values for the vector ${\rm \boldsymbol{v}}$
(see Eq. (\ref{eq:Vector of beam-parameters})). It is assumed that
the angle $\left(\varphi-\theta_{0}\right)$ follows a uniform distribution
over the interval $[0,\frac{\pi}{2}]$ and other parameters\footnote{To compute transmittance, first one has to evaluate $\mathcal{W}_{{\rm i}}$
from $\Theta_{{\rm i}}$ using relation $\begin{array}{lcl}
\Theta_{{\rm i}} & = & \ln\left(\frac{\mathcal{W}_{{\rm i}}^{2}}{\mathcal{W}_{{\rm 0}}^{2}}\right),\end{array}$ ${\rm i}=1,2.$ Here, $\mathcal{W}_{0}$ is the beam spot radius
at the transmitter.} ($x_{0},y_{0},\Theta_{{\rm 1}},\Theta_{{\rm 2}}$) follow the normal
distribution \cite{WHW+18}. Substitution of the simulated values
of ${\rm \boldsymbol{v}}$ into Eq. (\ref{eq:PDT Equation}) makes
it feasible to perform the numerical integration. The outcome of this
process also involves the \emph{extinction factor}\footnote{The parameter $\chi_{{\rm ext}}(\phi)$ denotes the extinction losses
caused by atmospheric back-scattering and absorption. It varies depending
on the elevation angle $\left(90^{\circ}-\phi\right)$ or zenith
angle $(\phi)$ \cite{BSH+13,VBB2000}.}\emph{,} $\chi_{{\rm ext}}$, thereby producing $N$ atmospheric transmittance
values, denoted as $\eta\left({\rm \boldsymbol{v}_{i}}\right)$, where
$i$ ranges from $1$ to $N$. The necessary parameters for simulation
are described in Appendix C which are calculated according to our
model. These expressions are different for uplink and downlink configuration
as different expressions mentioned in Eq. (\ref{eq:Down-Link and Up-Link condition})
are used for uplink and downlink configuration.

In the next section, we will evaluate the effectiveness of the HD
protocols selected by us in the satellite-based links. To conduct
this assessment, we need average key rates over the PDT\footnote{Some authors followed the relation $\eta_{\delta}=10^{-\frac{\delta}{10}}$
with $\delta=\alpha_{1}{\rm L}$ $[{\rm dB}]$ to represent the channel
transmittance with the form of attenuation, here, ${\rm L}$ total
link length and $\alpha_{1}$ is loss in the channel transmission
${\rm dB/km}$.} computed for different link lengths and configurations. The same
can be expressed as \cite{LKB19},

\begin{equation}
\begin{array}{lclcl}
\bar{R} & = & \intop_{0}^{1}R(\eta)\,P(\eta)\,{\rm d}\eta & = & \sum_{i=1}^{N_{\text{bins}}}{\sum}R(\eta_{{\rm i}})\,P(\eta_{{\rm i}}),\end{array}\label{eq:Average key-rate}
\end{equation}
where, $\bar{R}$ represents the average key rate, while $R(\eta)$
signifies the key rate corresponding to a specific transmittance value.
The PDT is denoted as $P(\eta)$. To compute the integral average,
the interval $[0,1]$ is divided into $N_{bins}$ bins, each centered
at $\eta_{{\rm i}}$ for $i$ ranging from $1$ to $N_{bins}$, and
is evaluated by combining the weighted sum of the rates. The estimation
of $P(\eta_{{\rm i}})$ relies on random sampling, as explained in
the earlier paragraph.\textcolor{red}{{} }The formulations for the distinct
implementations key rates $R(\eta)$ can be found in Sec. \ref{subsec:HD-Ext-B92=000026HD-BB84}.

\section{Performance analysis of protocols after simulation \label{sec:III}}

In this section, we elaborately analyze the impact of PDT\footnote{See PDT in Figures 3 and 4 in Ref. \cite{LKB19} after random sampling
of beam parameters ${\rm \boldsymbol{v}}$ for a down-link and an
up-link, respectively.} on key rate after the weighted sum, as well as the PDR concerning the HD-Ext-B92 and HD-BB84 protocols.
The minimum separation between Alice and Bob (i.e., altitude of the
satellite) remains constant at a distance of $\overline{{\rm L}}=500$
km, as the primary focus is on scenarios involving LEO satellites
like the Chinese satellite Micius \cite{LCL+17, YCL+17, RXY+17, YCLL+17}.
We present outcomes of numerical simulation for satellite-based HD-Ext-B92
and HD-BB84 schemes under asymptotic conditions \cite{DMB+24}. The simulation incorporates
the experimental parameters outlined in Table \ref{tab:Parameters-associated-with-link-length}
\cite{MFR12,
XXL14,
LKB19}. The parameters $C_{n}^{2}$,
$n_{0}$, and $h$ are typically determined by fitting experimental
data. However, for the sake of establishing a predictive model, we
parameterize these values in a rational manner. We conduct simulations
under varying atmospheric conditions, encompassing clear, slightly
foggy, and moderately foggy nights, as well as non-windy, moderately
windy, and windy days \cite{LKB19}. A particularly noteworthy aspect
is the comparison between nighttime and daytime operations. In daytime
conditions, elevated temperatures result in stronger winds and heightened
mixing across atmospheric layers, leading to more pronounced turbulence
effects and consequently higher values of $C_{n}^{2}$ compared to
nighttime conditions. Nevertheless, on average, during clear days,
the lower atmosphere exhibits reduced moisture content compared to
nighttime, resulting in diminished beam spreading due to scattering
particles. Conversely, nighttime conditions, characterized by lower
temperatures, yield a less turbulent atmosphere. Additionally, the
formation of haze and mist contributes to higher values of $n_{0}$
compared to daytime conditions. In such scenarios, the impact of scattering
over particulate matter can surpass the effects induced by turbulence. The crucial factors in this scenario include
not only those associated with atmospheric influences but also the
radii of the transmitting and receiving telescopes, along with the
wavelength of the signal. For the satellite in orbit, we opted for
a radius of $r_{{\rm sat}}=15$ cm ($\mathcal{W}_{0}$), while the
ground station telescope has a radius of $r_{{\rm grnd}}=0.5$ m,
and the signal wavelength is $\lambda=785$ nm. Based on Eq. (\ref{eq:Down-Link and Up-Link condition}),
it is evident that a downlink pertains to satellite-to-ground communication,
where atmospheric effects become significant only in the latter part
of the propagation process, i.e., when $z$ exceeds $({\rm L-h})$.
On the other hand, for uplinks, these effects are relevant only when
$z$ is below ${\rm h}$.

\begin{table}[h]
\begin{centering}
\begin{tabular}{>{\centering}p{2.5cm}>{\centering}p{4cm}>{\centering}p{5cm}}
\toprule 
Parameter & Value & Short description\tabularnewline
\midrule
$\mathcal{W}_{0}$ & 15 cm, 50 cm & Down-link, up-link\tabularnewline
$r$ & 50 cm, 15 cm & Down-link, up-link\tabularnewline
$\lambda$ & 785 nm & Wavelength of the signal light\tabularnewline
$\beta$ & 0.7 & Parameter in $\chi_{{\rm ext}}(\phi)$\tabularnewline
$\alpha$ & $2\times10^{-6}$ rad & Pointing error\tabularnewline
$\overline{{\rm h}}$ & 20 km & Atmosphere thickness\tabularnewline
$\overline{{\rm L}}$ & 500 km & Minimum altitude (at zenith)\tabularnewline
$n_{0}$ & 0.61 ${\rm m^{-3}}$ & Night-time condition 1\tabularnewline
$n_{0}$ & 0.01 ${\rm m^{-3}}$ & Day-time condition 1\tabularnewline
$n_{0}$ & 3.00 ${\rm m^{-3}}$ & Night-time condition 2\tabularnewline
$n_{0}$ & 0.05 ${\rm m^{-3}}$ & Day-time condition 2\tabularnewline
$n_{0}$ & 6.10 ${\rm m^{-3}}$ & Night-time condition 3\tabularnewline
$n_{0}$ & 0.10 ${\rm m^{-3}}$ & Day-time condition 3\tabularnewline
$C_{n}^{2}$ & $1.12\times10^{-16}$ ${\rm m^{-\frac{2}{3}}}$ & Night-time condition 1\tabularnewline
$C_{n}^{2}$ & $1.64\times10^{-16}$ ${\rm m^{-\frac{2}{3}}}$ & Day-time condition 1\tabularnewline
$C_{n}^{2}$ & $5.50\times10^{-16}$ ${\rm m^{-\frac{2}{3}}}$ & Night-time condition 2\tabularnewline
$C_{n}^{2}$ & $8.00\times10^{-16}$ ${\rm m^{-\frac{2}{3}}}$ & Day-time condition 2\tabularnewline
$C_{n}^{2}$ & $1.10\times10^{-15}$ ${\rm m^{-\frac{2}{3}}}$ & Night-time condition 3\tabularnewline
$C_{n}^{2}$ & $1.60\times10^{-15}$ ${\rm m^{-\frac{2}{3}}}$ & Day-time condition 3\tabularnewline
\bottomrule
\end{tabular}
\par\end{centering}
\caption{\label{tab:Parameters-associated-with-link-length}Parameters associated
with the optical and technical characteristics of the link and different
atmospheric weather conditions.}

\end{table}

From Appendix C, it becomes evident, as expected that the impact of
atmospheric effects is considerably more pronounced in the case of
uplinks compared to downlinks. The underlying phenomena at play
here, namely beam deflection and broadening, encompass angular effects.
These effects play a role in determining the ultimate size of the
beam, thus influencing the channel losses. Their magnitude is directly
proportional to the distance covered after the initiation of the effect
known as \emph{kick in effect}. For uplinks, these effects manifest
near the transmitter, resulting in beam broadening spanning hundreds
of kilometers before detection at the satellite. Conversely, in the
downlink scenario, the majority of the beam's trajectory occurs within
a vacuum, with atmospheric effects coming into play only during the
final fifteen to twenty kilometers before reaching the receiver. A
secondary distinction lies in the origin of fluctuations in the position
of the beam centroid, denoted as $(x_{0},y_{0})$. In uplinks, the
atmosphere-induced deflections tend to be significantly more influential
than pointing errors ($\varphi$), which is disregarded. On the other
hand, in downlinks, the beam dimensions are already substantially
larger than any turbulent irregularities at the top of the atmosphere.
As a consequence, the resulting beam wandering due to atmospheric
effects can be neglected, rendering pointing errors the dominant contributing
factor.

\begin{figure}[htbp]
\begin{centering}
\hspace*{-1.3cm}
\includegraphics[scale=0.42]{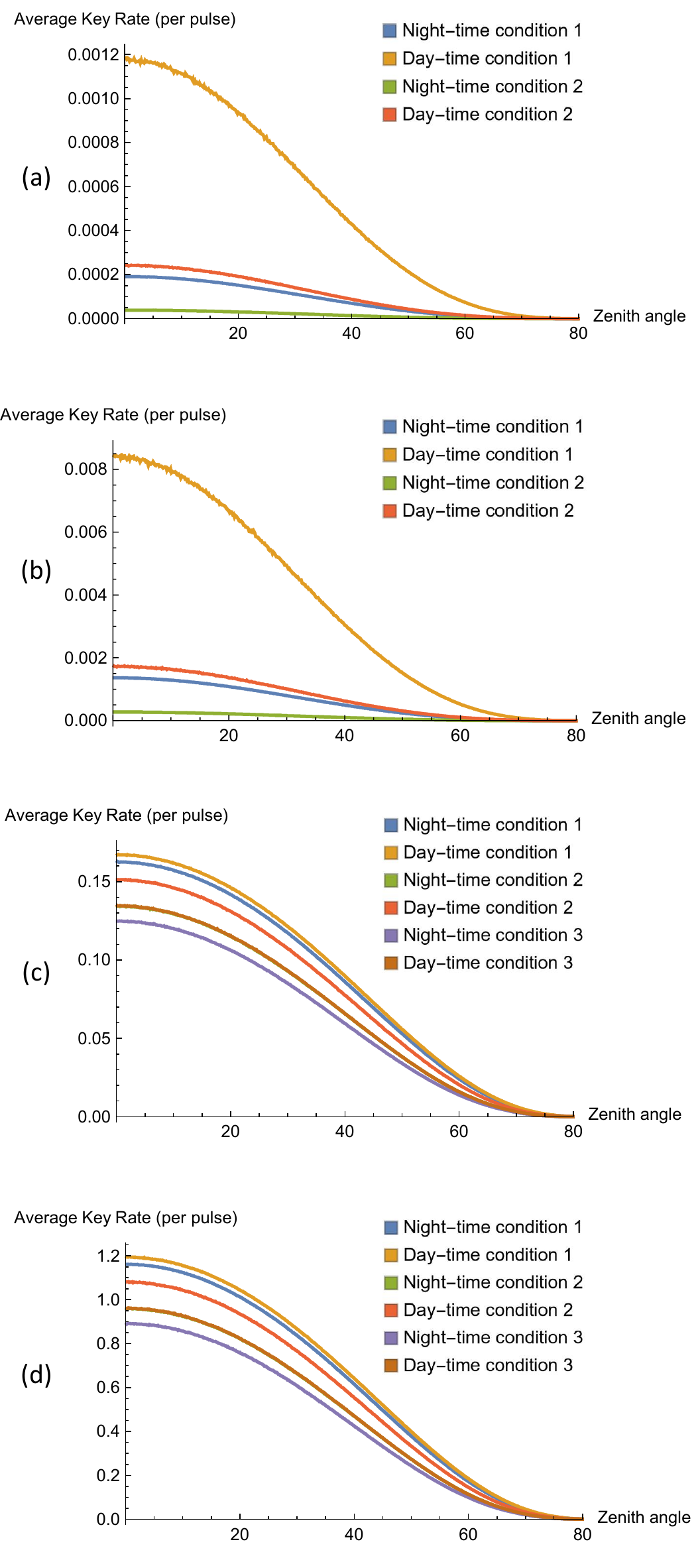}
\par\end{centering}

\captionof{figure}{\label{fig:AKR_Zenith_Different_Weather} Plot of variation
of average key rate (per pulse) with zenith angle in different weather
conditions considering minimal noise (${\rm d}=32$), i.e., day-time
conditions 1, 2 and 3 correspond to not windy, moderate windy and
windy, respectively (described as Day-time condition 1/2/3) and night-time
conditions 1, 2, 3 correspond to clear, slightly foggy and moderate
foggy, respectively (described as Night-time condition 1/2/3). The
upper row corresponds to the uplink scenario and the lower row corresponds
to the downlink scenario: (a) Average key rate generated by HD-Ext-B92
protocol as a function of zenith angle for uplink configuration under
four different weather conditions (Day 1-2 and Night 1-2), (b) Average
key rate generated by HD-BB84 protocol as a function of zenith angle
for uplink configuration under four different weather condition (Day
1-2 and Night 1-2), (c) Average key rate generated by HD-Ext-B92 protocol
as a function of zenith angle for downlink configuration under six
different weather condition (Day 1-2-3 and Night 1-2-3), (d) Average
key rate generated by HD-BB84 protocol as a function of zenith angle
for downlink configuration under six different weather condition
(Day 1-2-3 and Night 1-2-3).}
\end{figure}

Utilizing Equation (\ref{eq:Key-rate equantion}),
which integrates Equations (\ref{eq:Value of S(A|E)}) and (\ref{eq:Value of H(a|b)})
from Appendix B, and Eq. (\ref{eq:Key_rate_HD_BB84}), which represent
as the key rate formula for HD-Ext-B92 and HD-BB84 protocols, along
with PDT in satellite-based communication (refer to Eqs. (\ref{eq:Transmittance})
and (\ref{eq:PDT Equation})), enables the computation of the average
key rate for LEO satellite quantum communication. Employing Eq. (\ref{eq:Average key-rate})
in this calculation facilitates the determination of the probability
distribution of key rates and the assessment of key rate variations
with respect to different parameters in the context of satellite quantum
communication. Now, we aim to investigate the average key rate as
a function of zenith angle, considering minimal noise. Figures \ref{fig:AKR_Zenith_Different_Weather}
illustrate the average key rate using the PDT concerning the angle
relative to the zenith. This analysis is carried out for both uplinks
and downlinks across various weather conditions for dimension\footnote{The weather data information is used from Ref. \cite{LKB19}. We also
mention the required information in Table \ref{tab:Parameters-associated-with-link-length}.}, ${\rm d}=32$\textcolor{red}{{} }(see Table \ref{tab:Parameters-associated-with-link-length}).
Each data point on the graph is derived from $10,000$ parameter samples
in Eq. (\ref{eq:Vector of beam-parameters}) and computed using Eq.
(\ref{eq:PDT Equation}). In Figures \ref{fig:AKR_Zenith_Different_Weather}
(a) and \ref{fig:AKR_Zenith_Different_Weather} (b), the graphs reveal
that during daytime condition 1, the highest average key rate is yielded
in the zenith position ($\sim0.0012$ and $\sim0.008$) for HD-Ext-B92
and HD-BB84 protocols, respectively, in the uplink configuration.
Notably, the key rate\footnote{For ease of reference, we will refer to the average key rate as the
\textquotedbl key rate\textquotedbl .} is slightly greater for HD-BB84 which corresponds to the expected
result. For the same configuration, the key rate sharply diminishes
under other conditions (Day 2 and Night 1-2). Comparatively, for HD-Ext-B92,
the maximum value of the key rate ($\sim0.0002$) is nearly ten times
lower than that of the HD-BB84 protocol ($\sim0.002$) corresponding
to the day condition 2. A similar comparison holds for night 1/2 conditions.
For these conditions, the key rate becomes approximately zero at zenith
angle $50^{\circ}$. It may be noted that in night-time condition
1, the key rate is lower than in day-time condition 2 for both schemes
within the same configuration. Based on these observations, we can
infer that daytime transmission in the uplink configuration performs
more favorably than nighttime transmission. Due to the very low key
rate during night-time condition 2, we have chosen to negate condition
3, both in night-time and day-time, from the graphical representation. Additionally, in the uplink configuration, the simulation
results reveal a tenfold disparity in key rates between HD-Ext-B92
and HD-BB84 during day-time condition 2. In contrast, during day-time
condition 1, the difference is less pronounced, approximately five
fold. This discrepancy is attributed to the non-windy nature of day-time
condition 1, while day-time condition 2 experiences moderate wind,
resulting in a lower value of $C_{n}^{2}$ for the former condition
compared to the latter. Moreover, the absence of windy conditions
indicates a lower moisture content in the lower atmosphere. Consequently,
the scattering particle density, denoted as $n_{0}$, is lower in
day-time condition 1 compared to day-time condition 2 (see Table \ref{tab:Parameters-associated-with-link-length}).
The down-link configuration is depicted in Fig. \ref{fig:AKR_Zenith_Different_Weather}
(c) and \ref{fig:AKR_Zenith_Different_Weather} (d). As previously
discussed, the influence of atmospheric effects is comparatively reduced
in the downlink configuration compared to the uplink configuration.
Consequently, the performance of the link transmittance is superior
for downlink as compared to uplink. This is supported by Fig.
\ref{fig:AKR_Zenith_Different_Weather} (c) and \ref{fig:AKR_Zenith_Different_Weather}
(d), which further highlight the enhanced key rate. From these two
figures, the overall plot patterns can be seen to be (sequential arrangement
of plots representing different weather conditions) consistent for
both protocols. The sequence of different weather conditions that
yield higher key rate values follows this order: day-time condition
1, night-time condition 1, day-time condition 2, day-time condition
3, night-time condition 2, and night-time condition 3. Additionally,
it can be seen that similar to the uplink scenario, the daytime conditions
favor channel transmission over the nighttime conditions. This pattern
remains consistent across both scenarios. Of particular interest is
the comparison between operations during night-time and day-time.
During daylight hours, higher temperatures facilitate stronger winds
and heightened mixing across distinct atmospheric layers. This generates
more prominent turbulence effects. However, on average, clear days
witness a reduced moisture content in the lower atmosphere compared
to night-time conditions. Consequently, the scattering of particles
causes less pronounced beam spreading. Conversely, during night-time,
the cooler temperatures result in an atmosphere with lower turbulence
levels, coupled with the formation of mist and haze. In such circumstances,
scattering tends to have a more substantial impact at night-time than
the effects induced by turbulence at day-time. In the downlink scenario,
during day-time condition 1, the highest achievable key rates are
$0.165$ and $1.2$ for HD-Ext-B92 and HD-BB84 protocols, respectively.
Conversely, in night-time condition 3, the highest attainable key
rates are $0.125$ and $0.9$. The key rate ratio, in the downlink
scenario, between the HD-BB84 and HD-Ext-B92 protocols is $7.27$
for the maximum scenario and $7.2$ for the minimum scenario. This
observation substantiates the anticipated outcome that HD-BB84 consistently
outperforms HD-Ext-B92. Furthermore, the key rate decreases significantly
within the zenith angle range of $70^{\circ}$ to $80^{\circ}$
for the downlink scenario, whereas for the uplink scenario, this
reduction begins at a zenith angle of $50^{\circ}$. Intuitively,
downlink transmission exhibits a higher tolerance for larger zenith
angles compared to uplink transmission.

\begin{figure}[htbp]
\begin{centering}
\hspace*{-1.5cm}

\includegraphics[scale=0.65]{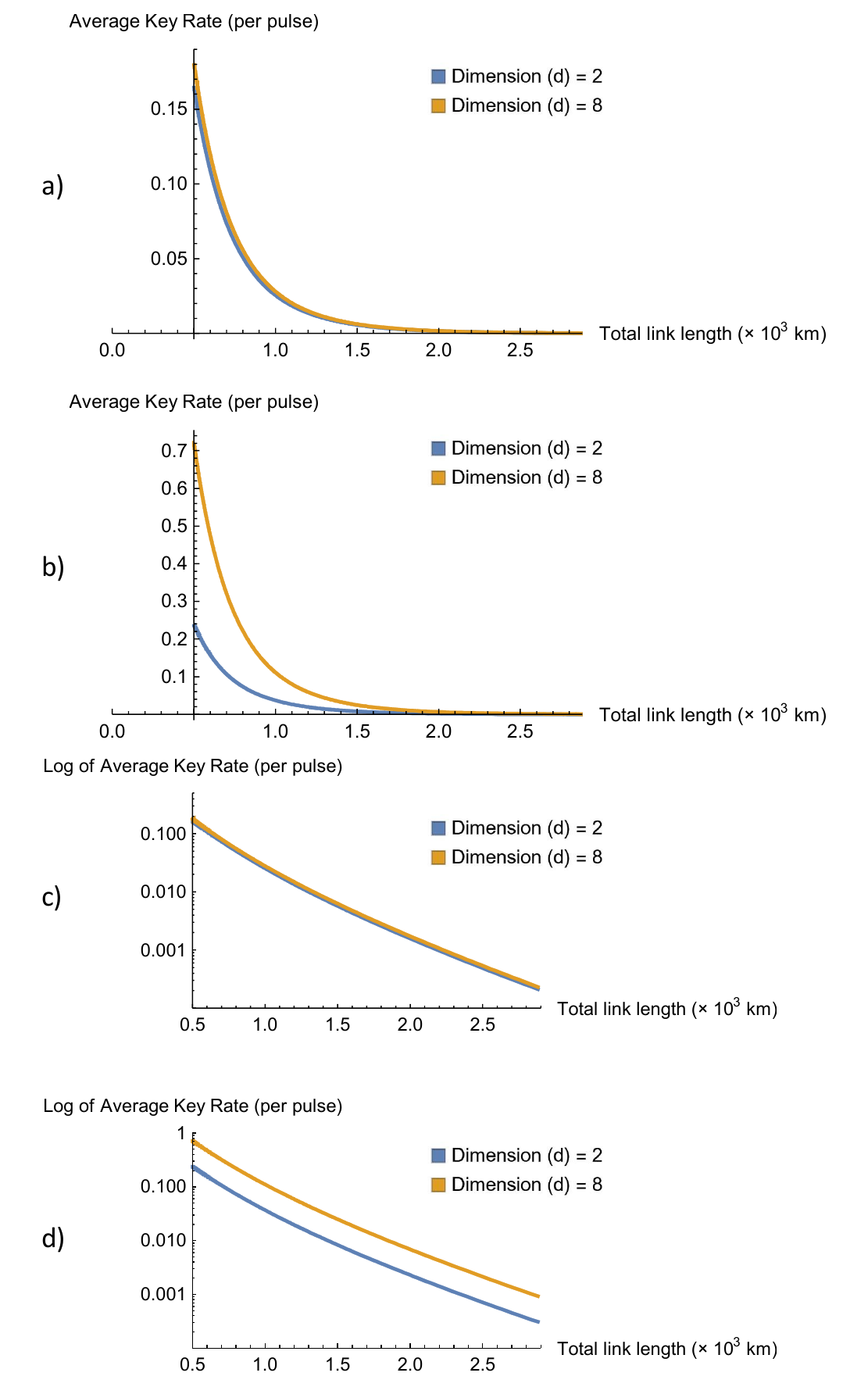}
\par\end{centering}
\begin{minipage}{\textwidth}
\caption{\label{fig:AKR_LinkLength_DownLink_Day1} Plot of variation
of average key rate (per pulse) with total link length under condition
Day-1 utilizing different dimensions of qudit (${\rm d}=2$ and ${\rm d}=8$):
(a) Average key rate generated by HD-Ext-B92 protocol as a function
of total link length (${\rm L}$) for downlink configuration, (b)
Average key rate generated by HD-BB84 protocol as a function of total
link length (${\rm L}$) for downlink configuration,
(c) and (d) illustrate the same results as depicted in (a) and (b),
respectively, but for a better visualization of the impact of link
length on the average key rate, here a logarithmic scale is used along
the $y$-axis.}
\end{minipage}
\end{figure}

To obtain the best possible results, hereafter we focus on the downlink
configuration under optimal weather conditions where the average key
rate is highest (cf. Fig. \ref{fig:AKR_Zenith_Different_Weather}).
Specifically, we analyze and illustrate the variation of key rate
with total link length (${\rm L}$) in day-time condition 1 within
downlink configuration, assuming an extremely low noise. In this
scenario, the HD-Ext-B92 protocol yields maximum key rates of 0.17
and 0.155 for qudit dimensions 8 and 2, respectively, as illustrated
in Fig. \ref{fig:AKR_LinkLength_DownLink_Day1} (a). Notably, the
key rate of the HD-BB84 protocol exhibits notable fluctuations across
different dimensions. As can be seen from Fig. \ref{fig:AKR_LinkLength_DownLink_Day1}
(b), for qudit dimensions 8 and 2, the maximum key rates are 0.7 and
0.24, respectively. Furthermore, the key rate decreases
almost linearly for both the HD-QKD protocols and across both dimensions
when plotted on a logarithmic scale. Consequently, it can be inferred
that the decrease in key rate follows an exponential pattern. Specifically,
at a higher zenith angle of $80^\circ$, with a total link distance
of $2900$ km, the key rate of the HD-Ext-B92 protocol is approximately
$10^{-4}$ for both dimensions. In contrast, at the same link distance,
the key rates for HD-BB84 are $10^{-3}$ and $10^{-4}$ for dimensions
$8$ and $2$, respectively. The HD-BB84 protocol outperforms at higher
dimensions, consistent with the findings depicted in the accompanying
Fig. \ref{fig:AKR_Zenith_Different_Weather}.

\begin{figure}[htbp]
\begin{centering}
\hspace*{-2.2cm}
\includegraphics[scale=0.8]{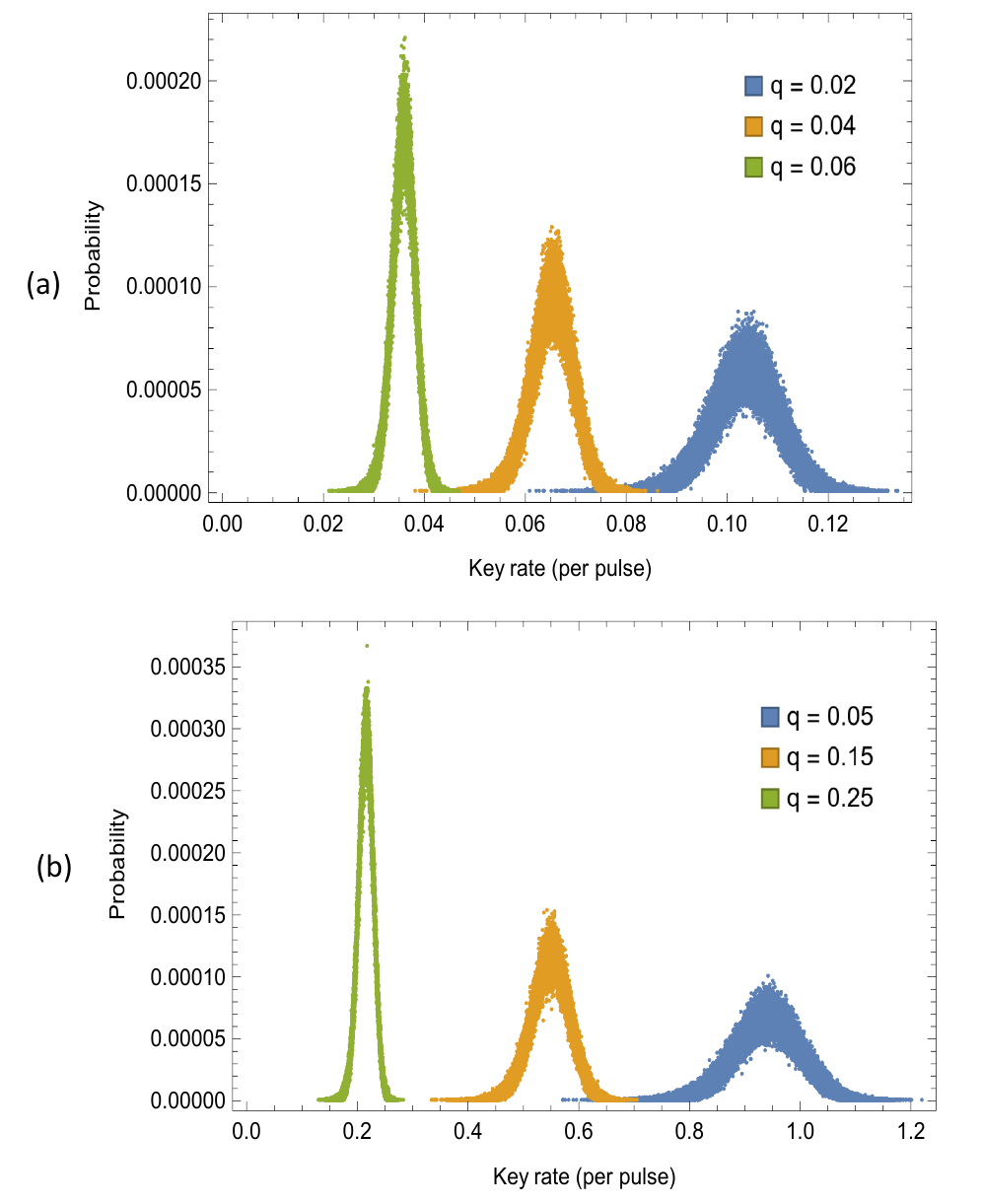}
\par\end{centering}
\caption{\label{fig:Probability_Keyrate_Different_Noise} Plot
of the distribution of key-rate variation for different channel noise
parameters (q) at the zenith position under condition Day-1 utilizing
qudit of dimension 32: (a) Probability distribution of key-rate for
HD-Ext-B92 protocol, (b) Probability distribution of key-rate for
HD-BB84 protocol under downlink configuration.}
\end{figure}

\begin{figure}[htbp]
\begin{centering}
\hspace*{-2.2cm}
\includegraphics[scale=0.8]{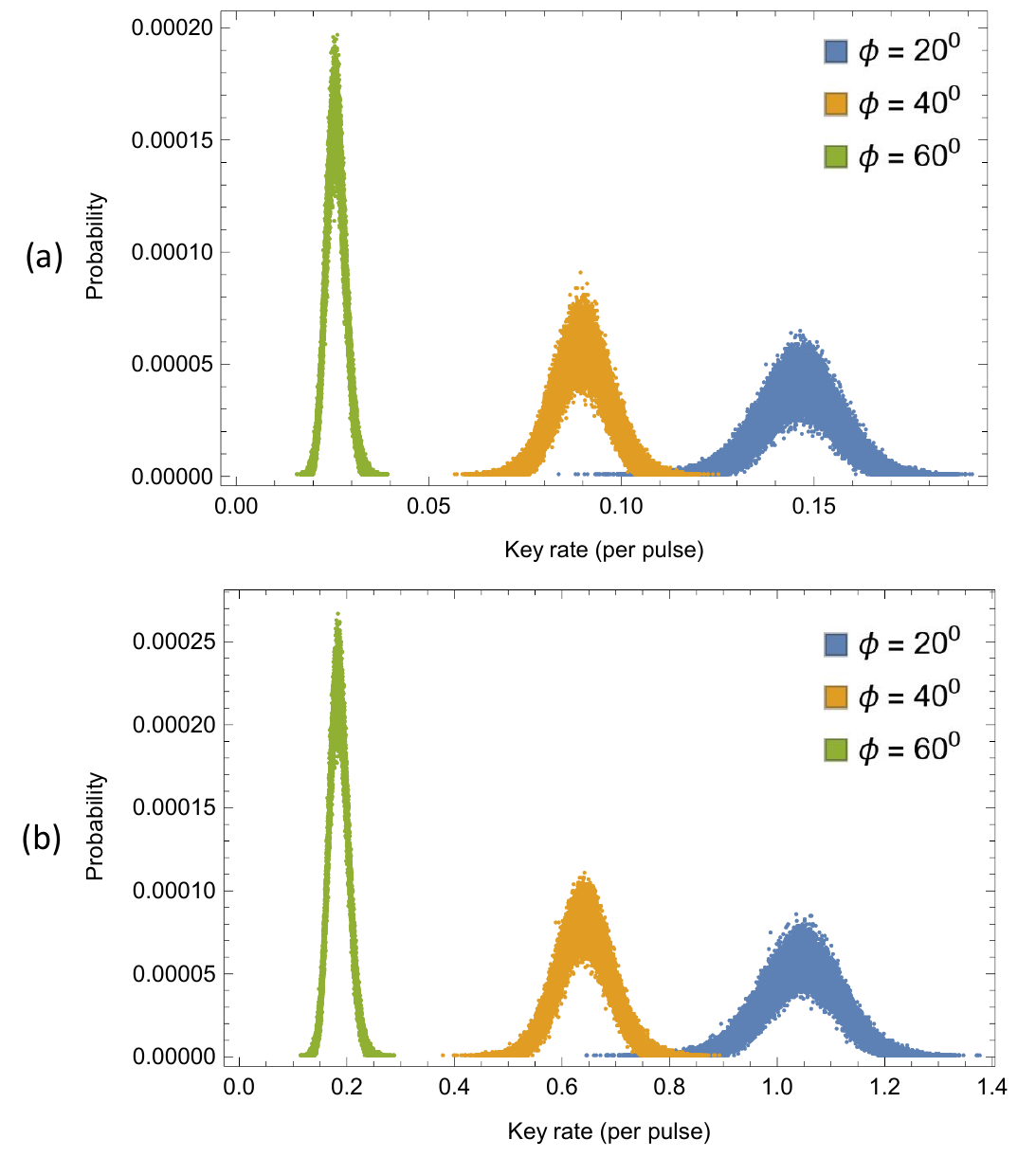}
\par\end{centering}
\caption{\label{fig:Probability_Keyrate_Different_Zenith_Angle}
Plot of distribution of key-rate variation for different zenith angles
($\phi$) considering minimal noise, characterized by the parameter
${\rm q}\ll1$, under condition Day-1 utilizing qudit of dimension
32: (a) Probability distribution of key-rate for HD-Ext-B92 protocol,
(b) Probability distribution of key-rate for HD-BB84 protocol under
downlink configuration.}
\end{figure}

In Fig. \ref{fig:Probability_Keyrate_Different_Noise}, we present
the PDR with different values of noise parameter (${\rm q}$) at the
zenith position ($\phi=0^{\circ}$) under downlink configuration.
In this context, we employ the optimal performance scenario during
day-time condition 1 utilizing qudit dimension of 32. We have used
a data set of $10^{6}$ beam parameters to simulate the values of
the average key rate and approximate the results to six (five) decimal
places\footnote{This is a good choice of approximation to represent, well-suited for
PDR representation.} to get PDR plots for HD-Ext-B92 (HD-BB84). Within the HD-Ext-B92
protocol, comparing the cases of ${\rm q}=0.02$ and ${\rm q}=0.06$
(in Fig. \ref{fig:Probability_Keyrate_Different_Noise} (a)), we
observe a higher key rate for ${\rm q}=0.02$, while the maximum value
of probability of key rate is greater for ${\rm q}=0.06$. The maximum
values of probability are consistently greater with greater values
of noise parameter. Notably, a higher key rate corresponds to a lower
value of probability of occurrence. A specific shape of PDT (as is
the case here) implies that the shape of the PDR would remain the
same with different noise parameters and different zenith angles (or
equivalently with different distances). For example, see that the
shape of the PDR remains same for HD-Ext-B92 protocol and HD-BB84
protocol, although the density of data points are more in the case
of HD-BB84 (see Fig. \ref{fig:Probability_Keyrate_Different_Noise}
(a) and (b)). However, this protocol (HD-BB84) exhibits significantly
elevated key rate values as well as higher probabilities compared
to HD-Ext-B92. Subsequently, we also plot the PDR with different zenith
angles in Fig. \ref{fig:Probability_Keyrate_Different_Zenith_Angle},
considering extremely low noise characterized by the parameter ${\rm q}\ll1$
at the zenith position under condition Day-1 with the same configuration
(downlink). Notably, the shapes of the PDR curves remain consistent
across both the protocols; however, the data points on the plot appear
more densely concentrated in the HD-BB84 protocol. In this case, we
have utilized a dataset of $10^{6}$ beam parameters to simulate the
values of the average key rate and approximate the results to six
(five) decimal places to get PDR plots for HD-Ext-B92 (HD-BB84). The
peak values of the probability of key rates in the PDR graph for distinct
zenith angles are different for both the protocols. Moreover, for
different zenith angles, the peak values of probability in the PDRs
are consistently greater in HD-BB84 compared to HD-Ext-B92. The higher uncertainity of keyrates at lower zenith angles (e.g., $\phi = 0^\circ$) in PDR is primarily due to the combined effect of high transmittance and residual atmospheric fluctuation. At small zenith angles, the shorter propagation path results in strong signal transmission and higher average keyrates. However, in this regime, the key rate becomes more sensitive to fluctuations caused by turbulence, beam wandering and pointing errors. Because the signal strength is relatively high in this regime, even small variations in atmospheric conditions produce noticeable variations in the received photon statistics, which directly impacts the estimated keyrate. Consequently, this results in more uncertainity at low zenith angles. 
In contrast, at higher zenith angles (e.g., $\phi = 20^\circ$ and $\phi = 40^\circ$ ), the optical path length increases significantly, leading to stronger attenuation that results reduced transmittance which decreases the effect of turbulence . In this regime, the received signal is consistently weak, and the keyrate is predominantly limited by channel loss rather than fluctuations. As a result, the key rate values tend to cluster around lower values, producing a narrower and more concentrated distribution with reduced apparent uncertainity. Therefore, the increased uncertainity at lower zenith angles is not due to poorer channel conditions, but rather due to the greater sensitivity of higher keyrates to stochastic channel variations. In conclusion,
we deduce that the PDR curves maintain a uniform shape across varying
zenith angles as PDT considered here has a fixed shape.

\section{Conclusion\label{sec:IV}}

In this chapter, we study two protocols for QKD in higher dimensions.
We analyze the key rates of these two higher dimensional protocols
in the context of satellite-based secure quantum communication. To
analyze the effectiveness of these schemes for satellite-based quantum
communication, we employ a robust method known as the elliptic beam
approximation \cite{VSV16}. By employing a generalized
model using this approach, we assess the performance of the HD-Ext-B92
and HD-BB84 protocols. The key rate per pulse and QBER are plotted
against the noise parameter. Notably, our findings reveal that, in
higher dimensions, HD-BB84 outperforms HD-Ext-B92 in terms of both
key rate and noise tolerance. However, HD-BB84 experiences a more
pronounced saturation of QBER in high dimensions. We deduce the key
rate of the HD-Ext-B92 scheme without introducing any additional free
parameters, as opposed to the approach discussed in Ref. \cite{IK21},
and is elaborated in Appendix B. Our analysis \cite{DMB+24} comprehensively demonstrates
the impact of link transmittance on the weighted sum of key rate under
nominal noise levels for both the schemes (HD-Ext-B92 and HD-BB84)
under uplink and downlink configurations. Moreover, we delve into
the analysis of PDR across different values of noise parameter (at
the zenith position) and zenith angle (with nominal noise). Remarkably, the PDR
exhibits consistent shapes across all scenarios. It is noteworthy
that the graphical points are denser for HD-BB84; as anticipated this
is because the HD-BB84 protocol makes use of two complete bases. Additionally,
the probability tends to be higher for lower key rate values compared
to higher ones. In summary, our investigations into the performance of higher-dimensional QKD protocols over satellite-based systems may have a substantial impact
on both theoretical and experimental aspects of satellite-based quantum
communication.



\chapter{Finite and Asymptotic Key Analysis for CubeSat QKD}\label{chap2:Background}
\section{Introduction}

The performance of satellite-based QKD critically depends on accurate estimation of secure key rates under realistic operational conditions. Asymptotic key rate analysis remains a fundamental tool for understanding the ultimate performance limits of QKD systems. For instance, \cite{ABN21} demonstrates how optimizing signal and decoy intensities can significantly enhance secure key generation, while \cite{GBL+23} shows that limiting an eavesdropper's accessible quantum modes strengthens achievable key rates, particularly in high-loss scenarios. Although asymptotic analysis provides valuable theoretical insight, practical QKD systems operate with finite resources. Consequently, precise evaluation of finite-size effects has become essential. Recent progress in finite-key analysis incorporates statistical fluctuations and optimized security parameters \cite{PKH23}. Early numerical methods directly integrated statistical uncertainties into key rate calculations \cite{BGK+20}, where SDP frameworks later enabled generalized finite-key security proofs applicable even to device-independent QKD \cite{GLL21}. These techniques have since been adapted to satellite-based links, integrating orbital dynamics, loss variations, and statistical corrections \cite{SBM+22, ELJ+21, DTR+21}. More recent studies have further included practical constraints such as hardware limitations, link efficiency, and environmental conditions \cite{SBM+23}. In this context, Chapter~4 presents a detailed analysis of secure key generation in a CubeSat-based QKD downlink, combining both finite-key and asymptotic frameworks with realistic optical channel modeling. Tight statistical techniques for parameter estimation and error correction are utilized following \cite{LCW+14, SBM+22} to evaluate finite-key block sizes. For the asymptotic regime, key rates are computed for WCP implementations of both efficient and standard BB84 protocols. CubeSats have emerged as practical and cost-effective platforms for space-based quantum communication. Motivated by successful missions such as Jinan-1 \cite{LCJ+24}, this chapter considers a CubeSat LEO at an altitude of 400\,km, which provides reduced free-space loss and improved photon detection probability compared to MEO and GEO configurations \cite{ZSH+24, LVM+23, SLM+22, MTP+24}. To model beam propagation accurately, the elliptical beam approximation \cite{LKB19} is adopted. This model captures beam wandering, elliptical deformation, and orientation fluctuations, and additionally accounts for atmospheric absorption and backscattering. Such effects are crucial under varying weather conditions, for which turbulence and scattering processes play a significant role \cite{VSV16}. In contrast, earlier CubeSat QKD studies such as \cite{ZSH+24} primarily consider diffraction and background noise, without detailed turbulence modeling. This chapter focuses on a downlink configuration, as it suffers less turbulence-induced distortion than uplink transmission \cite{LKB19}. This makes downlink QKD particularly suitable for BB84 implementations \cite{GN22}, which continue to be widely adopted due to their simplicity, robustness, and proven security \cite{SP00}. Both efficient and standard BB84 protocols are analyzed using a two-decoy-state method to ensure strong protection against PNS attacks. The BB84 protocol has seen extensive experimental validation \cite{SP00, OKM+24} and has recently been deployed on CubeSat platforms \cite{ZSH+24, LSL22}. Unlike prior CubeSat analyses such as \cite{MTP+24}, which employ a single decoy state and do not quantify key rate variations under different weather conditions, this chapter incorporates two decoy states, evaluates PDR, and examines QKD performance across different zenith angles \cite{MDB25}. The efficient BB84 protocol, with its biased basis choice, provides a higher sifting ratio, making it advantageous for CubeSat missions where link duration is limited and channel loss is significant \cite{SBM+22, MTP+24}. Standard BB84 is also included as a benchmark due to its symmetric basis structure and strong theoretical guarantees. Overall, this chapter offers a comprehensive assessment of CubeSat-based QKD performance by integrating finite-key and asymptotic security analyses with detailed atmospheric and beam propagation modeling \cite{MDB25}. The results highlight the practical considerations and performance-security trade-offs essential for future satellite QKD missions.\\
The remainder of this chapter is organized as follows: Section \ref{Preliminaries} presents an explanation of the decoy-state-based efficient BB84 and standard BB84 protocols. Additionally, it explores the finite key rate analysis and the asymptotic key rate analysis for these protocols. Moreover, we explore the impact of atmospheric conditions on CubeSat communication links and analyze the elliptical beam deformation approximation at the receiver. Section \ref{Performance Analysis of QKD Implementation in CubeSat System} provides a detailed assessment of the performance of the efficient and standard BB84 protocols, supplemented with illustrative results obtained from simulations. Finally, Section \ref{Conclusion} concludes the paper by summarizing the key findings and discussing their implications.

\section{Protocol Description and Finite–Asymptotic Key Rate Analysis}
\label{Preliminaries}
\subsection {Protocol Description: efficient BB84 and standard BB84 (using two decoy settings)}
The BB84 QKD protocol \cite{BB14} has gained extensive adoption due to its straightforward design, robust performance, and theoretically sound security guarantees. Despite this, practical deployments of BB84 typically deviate from the idealized single-photon sources; instead, weak coherent laser pulses are favored for their widespread availability and implementation feasibility. Although such laser sources enhance repetition rates compared to current single-photon emitters, they also render BB84 susceptible to PNS attacks that exploit the multiphoton components of emitted pulses. Decoy state protocols effectively mitigate PNS vulnerabilities and increase resilience to substantial channel losses, requiring only minimal alterations to  BB84 implementations. To mitigate the challenges associated with multiphoton components and channel losses, we utilize both standard and efficient BB84 protocols augmented with decoy-state settings. The standard BB84 protocol with decoy states provides a robust framework for secure key generation, while the efficient version improves key rates by maximizing the utilization of transmitted quantum states. In the following subsections, we systematically analyze the performance of these protocols under practical conditions.
\subsubsection{ Efficient BB84 and standard BB84 protocol (using two decoy settings)}
In the efficient BB84 protocol \cite{LCA05}, Alice and Bob select between the Z basis (\(|0\rangle, |1\rangle\)) and the X basis (\(|+\rangle, |-\rangle\)) with asymmetric probabilities \( c_X \) and \( 1 - c_X \). The Z basis is used for parameter estimation, while the X basis is used for key generation. The protocol employs phase-randomized laser pulses and a decoy-state method with three intensity levels \( \mu_1, \mu_2, \mu_3 \) satisfying \( \mu_1 > \mu_2 + \mu_3 \) and \( \mu_2 > \mu_3 \geq 0 \) \cite{LCW+14}.\\  
 {\textit{Preparation and measurement}}:   {Alice randomly selects a bit \( m_i \), a basis \( A_i \in \{X, Z\} \) with probabilities \( c_X, 1 - c_X \), and an intensity \( q_i \in \mathcal{Q} = \{\mu_1, \mu_2, \mu_3\} \) with probabilities \( p_{\mu_1}, p_{\mu_2}, p_{\mu_3} \). She then transmits a weak laser pulse to Bob, who selects a basis \( B_i \) with the same probabilities and records the measurement result \( m_i' \). Bob’s outcomes include \( \{0, 1, \emptyset, \bot\} \), where \( \emptyset \) indicates no detection, and \( \bot \) (double detection) results in a random bit assignment.} \\ 
 {\textit{Basis reconciliation and raw key generation}}:   {Alice and Bob publicly announce bases and intensities, defining sets \( \chi_q, \mathscr{Z}_q = \{ i : A_i = B_i \in \{X, Z\}, q_i = q, m_i' \neq \emptyset \} \). If \( |\chi_q| \geq n_{X,q} \) and \( |\mathscr{Z}_q| \geq n_{Z,q} \) for all \( q \), they generate a raw key pair \( (X_{\text{Alice}}, X_{\text{Bob}}) \) by sampling \( w_X = \sum_{q \in \mathcal{Q}} w_{X,q} \) from \( \bigcup_{q \in \mathcal{Q}} \chi_q \).} \\ 
 {\textit{Error estimation and post-processing}}:  
 {Alice and Bob estimate errors using \( \mathscr{Z}_q \), determining bit errors \( m_{Z,q} \), vacuum (\( y_{X,0} \)) events,  single-photon (\( y_{X,1} \)) events and phase errors \( f_{X,1} \). If the phase error rate \( \varphi_X = f_{X,1}/y_{X,1} \) exceeds the threshold \( \varphi_{\text{tol}} \), they abort. Otherwise, they proceed with error correction (leaking at most \( \lambda_{\text{ec}} \) bits), error verification (leaking \( \log_2(1/\varepsilon_{\text{hash}}) \) bits), and privacy amplification to extract a final secret key \( (S_A, S_B) \) of length \( l \).}  \\
 {In the standard BB84 protocol with two decoy states \cite{LMC05}, Alice and Bob choose bases randomly, uniformly, and independently. Unlike the efficient BB84 protocol, it employs an unbiased basis choice and uses both bases for key generation and parameter estimation \cite{SBM+22}.}\\
The calculation cost of the efficient BB84 and standard BB84 protocols differs due to the asymmetric basis selection in the efficient variant, which optimizes key generation and error estimation, making it more efficient than standard BB84 \cite{LCA05}. Efficient BB84 reduces data loss through biased basis selection, lowering sifting and error correction costs, while the information leakage, dependent on the data block size \cite{SBM+22}, slightly increases due to a higher raw key rate. Additionally, its phase error estimation is simpler, requiring fewer computational resources than standard BB84. The table \ref{tcal_cost} presents a comparative analysis of the calculation costs for both protocols, outlining the key differences across various computational steps.\\
\begin{table}
    \centering
     \caption{Calculation cost comparison of Efficient BB84 and Standard BB84 protocols.}
     \label{tcal_cost}
    \begin{tabularx}{\textwidth}{|X|X|X|}
        \hline
        \textbf{Computation Step} & \textbf{Efficient BB84} & \textbf{Standard BB84} \\
        \hline
        \textbf{Basis Reconciliation}  & The efficiency of this scheme is asymptotically twice that of the standard BB84 protocol due to its biased basis selection, reducing the computational cost associated with sifting by minimizing data rejection. & Approximately 50\% of the data is discarded due to Bob's random basis selection, which results in incorrect measurements half of the time.
 \\
        \hline
        \textbf{Error Estimation} & In our analysis, the QBER for the efficient BB84 protocol is determined to be approximately 5\%, requiring a lower error correction cost. & Likewise, the QBER for the standard BB84 protocol is observed to be around 10\%,  leading to a higher computational cost for error correction. \\   
        \hline
      \textbf{Error Correction} & In the finite key regime, the information leakage during error correction is approximately 0.0004375 bits per pulse, depending on the data block size. Higher raw key rate in efficient BB84 leads to increased leakage. & Here, the leakage is around 0.00034 bits per pulse, as more data is discarded during sifting, resulting in a lower raw key rate and reduced leakage. \\  
        \hline
        \textbf{Phase Error Estimation} & Involves calculating phase error rate only from the X basis, reducing computational complexity. & Requires phase error estimation from both X and Z bases, increasing computational burden. \\
        \hline
    \end{tabularx}
    \label{tab:placeholder_label}
\end{table}
\subsection{Finite key rate and asymptotic analysis} 
\label{Finite key rate and Asymptotic analysis}
This work focuses on examining the finite key rate and asymptotic key rate for both efficient and standard BB84 protocols for two decoy states in CubeSats. CubeSats, with their ability to operate in LEO, play a crucial role in enabling QKD by providing a practical platform for secure communication. Their compact design and suitability for downlink scenarios make them an ideal platform for analyzing the performance of key rates under practical constraints. Detailed insights into this analysis are provided in the subsequent subsections.
\subsubsection{Finite key rate analysis for decoy-state BB84 in CubeSat systems}  
 The security of decoy-state QKD was initially developed under the assumption of the asymptotic-key regime \cite{W05, LMC05}. However, for practical implementations with finite data sizes, uncertainties in the channel parameters must be taken into account \cite{MQZ+05, HHH+07, CS09}. Early methods addressing finite-key effects relied on Gaussian approximations to quantify the discrepancy between asymptotic and finite-key results \cite{ZZR+17}. These approaches, however, limited the security analysis to collective and coherent attacks.  Later advancements extended the finite-key security analysis to include more general attack strategies \cite{HN14}, utilizing bounds such as the multiplicative Chernoff bound \cite{CXC+14, ZZR+17} and Hoeffding’s Inequality \cite{LCW+14} to quantify statistical fluctuations. A composable finite-key analysis for decoy-state efficient BB84 using the multiplicative Chernoff bound was introduced in \cite{YZG+20}, providing tighter security bounds and improving the estimation of key parameters.\\
 {In this work, we extend these finite-key rate analyses to CubeSat-based QKD}, where CubeSats in LEO enable secure key exchange over free-space optical links \cite{MDB25}. The overpass duration of a CubeSat, or time window, determines the total number of pulses transmitted during a single pass. Let \( N_{p} \) be the total number of pulses transmitted during a CubeSat pass, which depends on the source repetition rate and the duration of the CubeSat overpass, referred to as the time window. For a maximum zenith angle of \( 80^\circ \), the time window is typically limited to approximately \( 440~\mathrm{seconds} \) \cite{SBM+22}. The finite key rate
 {measured in bits per pulse} for a single pass, based on the efficient BB84 protocol, is then determined as \cite{SBM+22} \\
\begin{equation}
 R^{eff} = \frac{l}{N_p} =\left\lfloor y_{X,0}+y_{X,1}(1-h(\varphi_X))-\lambda_{ec}-6\log_2\frac{21}{\varepsilon_{sec}}-\log_2\frac{2}{\varepsilon_{corr}}\right\rfloor.
 \label{eq_reff}
\end{equation}
Here $y_{X,0}$, $y_{X,1}$, and $\varphi_X$ represent the vacuum yield, single-photon yield, and phase error rate in the $X$-basis, respectively  { and \( h(x) := -x \log_2 x - (1 - x) \log_2 (1 - x) \) is the binary entropy function.} The amount of information leakage is quantified by $\lambda_{ec}$, which is considered during privacy amplification. In the finite key regime, this leakage is fundamentally bounded by $\lambda_{\text{ec}} \leq \log|\mathcal{W}|$, where $\mathcal{W}$ denotes the set of syndromes involved in the information reconciliation process. We utilized an estimate of $\lambda_{ec}$ that varies with the block size, as described  below  \cite{TMP+17}
\begin{equation}
\begin{split}
\lambda_{ec} = & \, n_X h(Q) + n_X (1-Q) \log\left[\frac{(1-Q)}{Q}\right] \\
& - \left(F^{-1}(\varepsilon_{corr}; n_X, 1-Q) - 1\right) \log\left[\frac{(1-Q)}{Q}\right] \\
& - \frac{1}{2} \log(n_X) - \log\Big(\frac{1}{\varepsilon_{corr}}\Big).
\end{split}
\label{eq_lambda_ec}
\end{equation}
Here, $n_X$ represents the data block size, $Q$ denotes the QBER \cite{SBM+21}, and $F^{-1}$ refers to the inverse of the cumulative distribution function of the Binomial distribution. This definition is used to evaluate the quantity of information that is leaked during the error correction process in the finite key regime. The protocol's reliability and security are characterized by two parameters, $\varepsilon_{corr}$ and $\varepsilon_{sec}$. A protocol is considered $\varepsilon$-secure if it satisfies $\varepsilon = \varepsilon_{corr} +\varepsilon_{sec}$, where it is $\varepsilon_{corr}$-correct and $\varepsilon_{sec}$-secret.\\
We now analyze the standard BB84 protocol using WCPs with two-decoy-states. The finite key rate for single pass for standard BB84 protocol can be expressed as follows \cite{SBM+22}: 
\begin{equation}
\begin{split} 
R^s = \frac{l}{N_p} = \frac{1}{2} \left\lfloor y_{X,0} + y_{X,1} \left( 1 - h(\varphi_X) \right) - \lambda_{ecX} + y_{Z,0} + y_{Z,1} \left( 1 - h(\varphi_Z) \right) \right. \\
\left. - \lambda_{ecZ} - (12\log_2 \frac{21}{\varepsilon_{sec}} - 2\log_2 \frac{2}{\varepsilon_{corr}}\Big) \right\rfloor.
\end{split}
\label{eq_rs}
\end{equation}
Satellite-based quantum communication systems are significantly impacted by finite statistical effects due to the limited duration of transmission windows. We use improved analysis of \cite{SBM+22} in modelling statistical fluctuations arising from finite statistics. This enhances the robustness of the secret key rate and incorporates a finite-statistics correction term, denoted as \( \mathrm{\delta}^{\pm}_{X(Z),j} \). This correction term is defined using the inverse multiplicative Chernoff bound \cite{YZG+20, ZZR+17}. Specifically let Y denotes a sum of $\mathcal{T}$ independent Bernoulli samples, which need not be identical. Denote $y^\infty$ as the expectation value of Y, with $y$ the observed value for Y. The extent of the discrepancy between the observed and expected values is influenced by the available statistics. To quantify this deviation, the probability that \( y \leq y^\infty + \delta_Y^{+} \) is less than a fixed positive constant \( \epsilon > 0 \), and the probability that \( y \geq y^\infty - \delta_Y^{-} \) is less than \( \epsilon \). This is achieved through setting
\begin{equation}
    \mathrm{\delta}_Y^{+} = \beta + \sqrt{2\beta y + \beta^2}, \quad \mathrm{\delta}_Y^{-} = \frac{\beta}{2} + \sqrt{2\beta y + \frac{\beta^2}{4}},
    \label{eq_delta}
\end{equation}
where $\beta=\ln{\frac{1}{\epsilon}}$ \cite{YZG+20}. Hence, we define the following finite sample size data block size \cite{SBM+22, SBM+21}.
\begin{equation}
\begin{split}
    n^{\pm}_{X(Z),j} = \frac{e^j}{p_j}\Big[  n_{X(Z),j}\pm \mathrm{\delta}^{\pm}_{n_{X(Z),j}}\Big] ,\\
    m^{\pm}_{X(Z),j} = \frac{e^j}{p_j}\Big[  m_{X(Z),j}\pm \mathrm{\delta}^{\pm}_{m_{X(Z),j}}\Big],
\end{split}    
\label{eq_nene}
\end{equation}
for the number of events and errors respectively in the $X(Z)$ basis. From this the vaccum and single photon yields, and the phase error rate of single photon events are defined as given in \cite{LCW+14}. The number of vacuum events in $X_A$ satisfies
\begin{equation}
    y_{X,0}\geq \frac{\tau_0\mu_2 n^-_{X,\mu_3}-\mu_3 n^+_{X,\mu_2}}{\mu_2-\mu_3},
    \label{eq_yx0}
\end{equation}
where $\tau_n:=\sum_{j \in \kappa}e^{-j}j^np_j/n!$ is the probability that Alice sends n-photon state. The number of single photon events in $X_A$ is 
\begin{equation}
   y_{X,1} \geq \frac{\tau_1 \mu_1 \left[n_{X,\mu_2}^- - n_{X,\mu_3}^+ - \frac{\mu_2^2 - \mu_3^2}{\mu_1^2} \left(n_{X,\mu_1}^+ - \frac{y_{X,0}}{\tau_0}\right)\right]}{\mu_1 (\mu_2 - \mu_3) - (\mu_2^2 - \mu_3^2)}.
   \label{eq_yx1}
\end{equation}
 The number of vacuum events, \(y_{Z,0}\), and the number of single-photon events, \(y_{Z,1}\), using Eqs. (\ref{eq_yx0}) and (\ref{eq_yx1}) can also be defined. Additionally, the number of bit errors, \(v_{Z,1}\), associated with the single-photon events in the \(Z\)-basis is also computed. It is given by
\begin{equation}
    v_{Z,1}\leq \tau_1 \frac{m_{Z,\mu_2}^+-m_{Z,\mu_3}^-}{\mu_2-\mu_3}. 
    \label{eq_vz1}
\end{equation}
The formula for the phase error rate of the single-photon events in $X_A$ is \cite{LCW+14}
\begin{equation}
    \varphi_X:=\frac{c_{X,1}}{y_{X,1}}\leq \frac{v_{Z,1}}{y_{Z,1}}+\gamma\Big(\varepsilon_{sec},\frac{v_{Z,1}}{y_{Z,1}},y_{Z,1},y_{X,1}\Big),
    \label{eq_psix}
\end{equation}
where
\begin{equation}
    \gamma(a,b,c,d):=\sqrt{\frac{(c+d)(1-b)b}{cd\log2}}\log_2\Big(\frac{c+d}{cd(1-b)b}\frac{21^2}{a^2}\Big).
    \label{eq_gamma}
\end{equation}

\subsubsection{Asymptotic analysis of key rate per pass} 
The asymptotic key length is determined by increasing the number of CubeSat passes. Let $M$ denote the total number of CubeSat passes than the asymptotic secret key length is given by $l_\infty = \lim\limits_{M \to \infty} \frac{l_M}{M}$ \cite{SBM+22} where \( l_M \) represents the secret key length (SKL) achieved from \( M \) CubeSat passes. The quantity \( l_\infty \) is determined by analyzing the asymptotic scaling of the ratio \( \frac{l_M}{M} \).
The estimation of vacuum counts per pass is expressed as \cite{SBM+22}:
\begin{equation}
    \frac{y_{X(Z),0}}{M}= \frac{\tau_0}{\mu_2-\mu_3}\Big(\frac{\mu_2\Gamma_3 (n_{X(Z),3}-\mathrm{\delta}^-_{X(Z),3})-\mu_3\Gamma_2 (n_{X(Z),2}+\mathrm{\delta}^+_{X(Z),2})}{M}\Big),
    \label{eq_yx0/m}
\end{equation}
where $n_{X(Z),j}$ represents the number of sifted counts in the $X(Z)$ basis from pulses with intensity $j$. The term $\tau_0$ denotes the average probability that the laser transmits a vacuum state. Additionally, $\Gamma_j = \exp(\mu_j)/p_j$ and $\mathrm{\mathrm{\delta}}_{X(Z),j}^\pm$ are determined using the multiplicative Chernoff bound \cite{YZG+20}. The asymptotic behavior of these correction terms follows the scaling \( O\Big(\sqrt{n_{X(Z)}}\Big) \), which implies that the scaling with respect to the number of CubeSat passes is \( O(\sqrt{M}) \). As a result, the terms \( \frac{\mathrm{\delta}_{X(Z),j}^\pm}{M} \) scale as \( O\Big(\frac{1}{\sqrt{M}}\Big) \), and consequently, they approach zero as \( M \to \infty \).
 As expected, the finite statistical correction term diminish in this limit. Assuming each CubeSat pass follows the same orbit, the total number of counts \( n_{X(Z),j} \) can be expressed as \( M \) times the number of counts for a single pass, \( n^{(1)}_{X(Z),j} \).  From this we obtain \cite{SBM+22}.
\begin{equation}
 \lim\limits_{M \to \infty}\frac{y_{X(Z),0}}{M}= \frac{\tau_0}{\mu_2-\mu_3}\Big(\mu_2\Gamma_3 ( n^{(1)}_{X(Z),3} )-\mu_3\Gamma_2 ( n^{(1)}_{X(Z),2} )\Big)= y^\infty_{X(Z),0}, 
 \label{eq_yx0/m_inf}
\end{equation}
where $y_{X(Z),0}^\infty$ represents the asymptotic estimate of the vacuum counts. For a single transmission pass, which will be formally defined in the next paragraph, the key rate for the efficient BB84 protocol under asymptotic conditions can be determined. By applying a similar methodology to each term in $l_M/M$, the asymptotic key rate can be written as
\begin{equation}
R_\infty^{eff} = \frac{l_\infty }{N_p} =  \left\lfloor y_{X,0}^\infty+y_{X,1}^\infty(1-h(\varphi_X^\infty))-\lambda_{ec}^\infty\right\rfloor.
\label{eq_reff_inf}
\end{equation}
The phase error rate, denoted as \( \varphi_X^\infty \), is given by the ratio \( \frac{v_{Z,1}^\infty}{y_{Z,1}^\infty} \), where \( y_{X,1}^\infty \), \( y_{Z,1}^\infty \), and \( v_{Z,1}^\infty \) represent the asymptotic estimates for the single-photon counts in the \( X \)-basis, the \( Z \)-basis, and the number of single-photon errors in the \( Z \)-basis, respectively, for a single pass.
 These asymptotic quantities, including $v_{Z,1}^\infty$, $y_{X,0}^\infty$, and $y_{X,1}^\infty$, are determined by averaging the single-pass values over an infinite number of passes. A refined estimate of the error correction term, $\lambda_{ec}$, and its asymptotic upper bound is provided in \cite{TMP+17}, from which it follows that $\lambda_{ec}^\infty = n_X^{(1)} h(Q)$, where $Q$ is the QBER for a single pass. Similarly the asymptotic key rate for the standard BB84 protocol can be determined by utilizing both the $X$ and $Z$ bases for key generation and parameter estimation and can be written as-
\begin{equation}
R_\infty^s = \frac{l_\infty}{N_p} = \frac{1}{2} \left\lfloor y_{X,0}^\infty+(y_{X,1}^\infty(1-h(\varphi_X^\infty)))-\lambda_{ecX}^\infty+ y_{Z,0}^\infty+(y_{Z,1}^\infty(1-h(\varphi_Z^\infty)))-\lambda_{ecZ}^\infty\right\rfloor.
\label{eq_rs_inf}
\end{equation}
Our study focuses on evaluating the key rate performance for both the efficient BB84 and standard BB84 protocols using two decoy states under various weather conditions for CubeSat. Channel transmittance is modeled using the elliptical-beam approximation for atmospheric links proposed by Vasylyev et al. \cite{VSV16, VSV+17}, which has also been applied in the previous chapter (see Section~\ref{subsec:Elliptic_Beam_Model}). Additionally, we adopt a generalized approach alongside the varying weather conditions presented in \cite{LKB19}. This methodology significantly influences the transmittance values, as transmittance of the channel depends on the characteristics of the beam and the size of the receiving aperture.  In our study, we have chosen $\overline{L}$ to be $400$ $km$ for CubeSats \cite{LVM+23} and assumed that the zenith angle ranges from $[0^\circ,66^\circ]$ which is denoted by $\phi$ as shown in Fig. \ref{fig:Non-uniform_free_space_link}.  In the subsequent section, we will analyze the performance of the selected protocols for CubeSat-based links. This analysis requires the computation of average key rates over the PDT \cite{LKB19}, following the expression in Eq.~\ref{eq:Average key-rate} used in the previous chapter, which is evaluated for different link lengths and system configurations. The specific expressions for the different key rate implementations, $R(\eta)$, are described in Sec.~\ref{Finite key rate and Asymptotic analysis} (see Eqs.~\ref{eq_reff}, \ref{eq_rs}, \ref{eq_reff_inf}, and \ref{eq_rs_inf}).
\section{Results and Discussion}
\label{Performance Analysis of QKD Implementation in CubeSat System}
\begin{figure}[htbp!]
    \centering
        \hspace*{-1.2cm}
    \begin{subfigure}[b]{0.49\textwidth}
        \centering
        \includegraphics[width=1.2\linewidth]{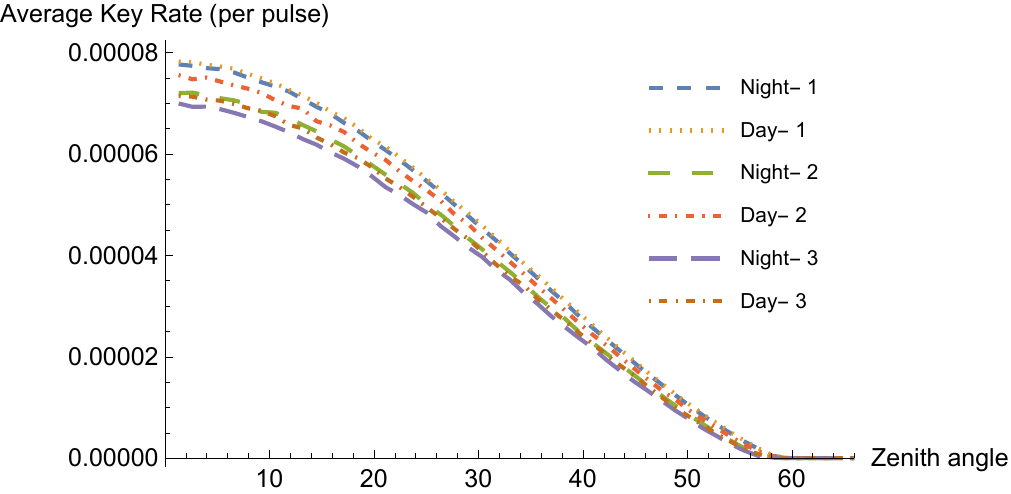} 
         \caption{}
       \label{fe2a}
    \end{subfigure}
    \hfill
    \begin{subfigure}[b]{0.49\textwidth}
        \centering
         \hspace*{0.8cm}
            
        \includegraphics[width=1.2\linewidth]{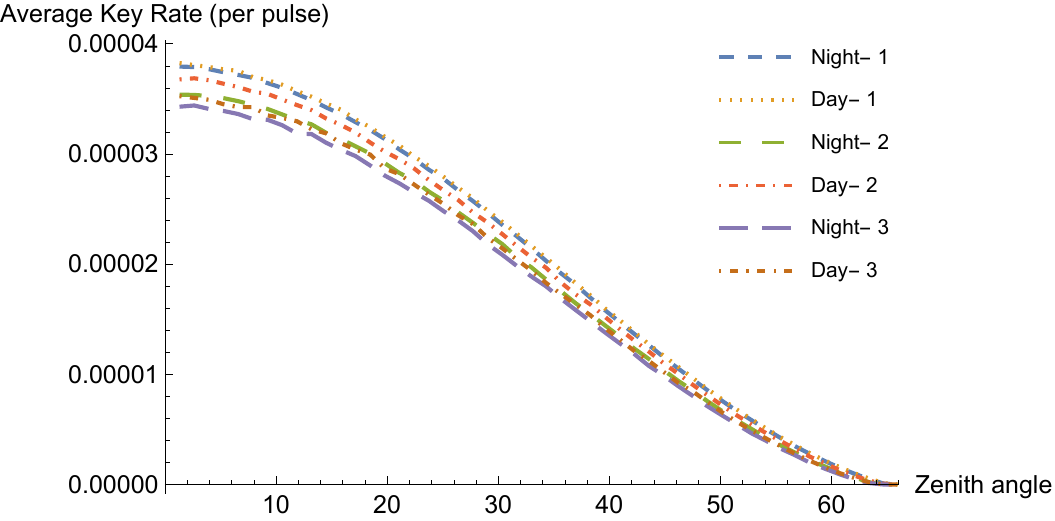} 
        \caption{}
        \label{fs2b}
    \end{subfigure}

    \vskip\baselineskip 

    \begin{subfigure}[b]{0.49\textwidth}
        \centering
        \hspace*{-1.2cm}
        \includegraphics[width=1.2\linewidth]{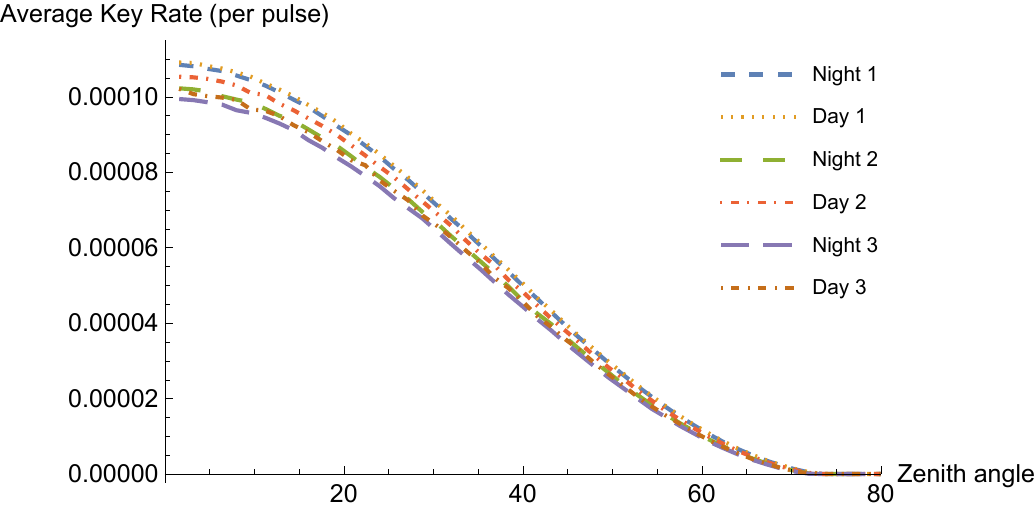} 
        \caption{}
        \label{ae2c}
    \end{subfigure}
    \hfill
    \begin{subfigure}[b]{0.49\textwidth}
        \centering
         \hspace*{-0.25cm}
        \includegraphics[width=1.2\linewidth]{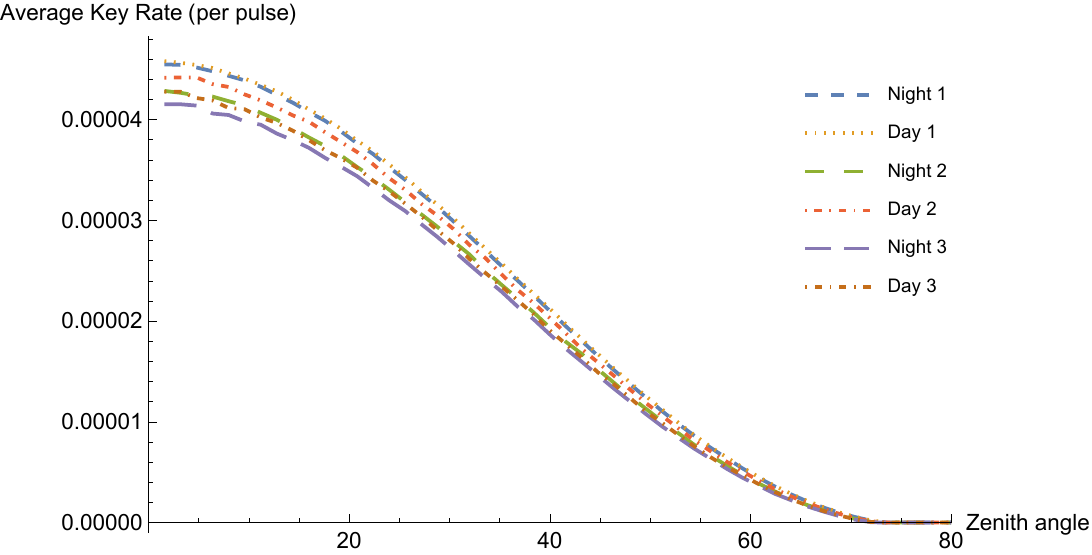} 
        \caption{}
        \label{as2d}
    \end{subfigure}
    \caption{The plot illustrates how the average key rate (per pulse) varies with the zenith angle in a downlink configuration across six distinct weather scenarios: clear (Day 1), moderate wind (Day 2), windy (Day 3), as well as clear (Night 1), slightly foggy (Night 2), and moderately foggy (Night 3). The upper row represents the finite key rate, and the lower row corresponds to asymptotic key rate. Figure 4.3.1(a) and 4.3.1(b) shows the average finite key rate for the efficient and standard BB84 protocols, respectively, while Figure 4.3.1(c) and 4.3.1(d) shows the average asymptotic key rate for the efficient and standard BB84 protocols, respectively under the weather conditions (Day 1/2/3 and Night 1/2/3).
}

    \label{fig2}
\end{figure}

 This section examines the average key rate as a function of the zenith angle using PDT, obtained after performing the weighted sum, for the efficient and standard BB84 protocols in CubeSat-based QKD under different weather conditions in a downlink scenario. To accurately model atmospheric losses, we employ the elliptical beam approximation, which effectively captures the impact of beam spreading and turbulence-induced distortions on photon transmission. Additionally, we analyze the PDR for different zenith angles, considering both protocols in the finite and asymptotic cases. CubeSats typically feature compact optics with apertures of $\leq 10\,\text{cm}$ \cite{LKB19}. We present the results of numerical simulations for CubeSat-based implementations of the efficient and standard BB84 protocols, evaluated under finite key and asymptotic key analyses \cite{MDB25}. The simulations incorporate the experimental parameters specified in Table \ref{table} \cite{SBM+22, LKB19}. These analyses consider varying atmospheric conditions, including clear, slightly foggy, and moderately foggy nights, as well as non-windy, moderately windy, and windy days~\cite{LKB19}. In this scenario, the critical factors include both atmospheric effects, the transmitter and receiver telescope radii, and the signal wavelength for CubeSats in orbit. For the CubeSat, a radius of $r_{\text{sat}} = 5\,\text{cm} \, (W_0)$ is considered, whereas the ground station telescope has a radius of $r_{\text{grnd}} = 50\,\text{cm}$, and the signal wavelength is $\lambda = 785\,\text{nm}$. We have opted for a downlink configuration due to its lower transmission losses \cite{DMB+24}.\\
\begin{table}[h!]
\centering
\caption{Parameters related to the baseline SatQKD system, various atmospheric weather conditions, and the optical and technical characteristics of the link.}
\begin{tabularx}{\textwidth}{>{\raggedright\arraybackslash}X >{\raggedright\arraybackslash}X >{\raggedright\arraybackslash}X}
\hline
\textbf{Parameter} & \textbf{Value} & \textbf{Short Description} \\
\hline
$\mathcal{W}_0$ & 5 cm & CubeSats Down-link \\
$r_a$ & 50 cm & CubeSats Down-link \\
$\lambda$ & 785 nm & Wavelength of the signal light \\
$\beta$ & 0.7 & Parameter in $\chi_{\text{ext}}(\phi)$ \\
$p_e$ & $2 \times 10^{-6}$ rad & Pointing error \\
$\overline{h}$ & $20$ km & Atmosphere thickness \\
$\overline{L}$ & $400$ km & Minimum altitude (at zenith) \\
$n_0$ & 0.61 m$^{-3}$ & Night-~1 \\
$n_0$ & 0.01 m$^{-3}$ & Day-~1 \\
$n_0$ & 3.00 m$^{-3}$ & Night-~2 \\
$n_0$ & 0.05 m$^{-3}$ & Day-~2 \\
$n_0$ & 6.10 m$^{-3}$ & Night-~3 \\
$n_0$ & 0.10 m$^{-3}$ & Day-~3 \\
$C_n^2$ & $1.12 \times 10^{-16}$ m$^{-2/3}$ & Night-~1 \\
$C_n^2$ & $1.64 \times 10^{-16}$ m$^{-2/3}$ & Day-~1 \\
$C_n^2$ & $5.50 \times 10^{-16}$ m$^{-2/3}$ & Night-~2 \\
$C_n^2$ & $8.00 \times 10^{-16}$ m$^{-2/3}$ & Day-~2 \\
$C_n^2$ & $1.10 \times 10^{-15}$ m$^{-2/3}$ & Night-~3 \\
$C_n^2$ & $1.60 \times 10^{-15}$ m$^{-2/3}$ & Day-~3 \\
$QBER_I$ & $5 \times 10^{-3}$ & Intrinsic QBER \\
$p_{ap}$ & $1 \times 10^{-3}$ & Afterpulse probability \\
$p_{ec}$ & $5 \times 10^{-7}$ & Extraneous count probability/pulse \\
$f_s$ & $1 \times 10^{8}$  {Hz} & Source rate \\
$\varepsilon_{corr}$ & $ 10^{-15}$ & Correctness parameter \\
$\varepsilon_{sec}$ & $ 10^{-9}$ & Secrecy parameter \\
$\lambda_{ec}$ &  depends on block size (see text) &  Error correction efficiency   \\
\hline
\end{tabularx}
\label{table}
\end{table}
The Eqs. (\ref{eq_reff}) and (\ref{eq_rs}) represent the finite key rate expressions, while Eqs. (\ref{eq_reff_inf}) and (\ref{eq_rs_inf}) provide the asymptotic key rate formulations for the efficient and standard BB84 protocols respectively. By incorporating the PDT in CubeSat-based communication, these expressions enable the computation of the average key rate for CubeSat-based quantum communication systems. Here, Fig. \ref{fig2} depicts the dependence of the average key rate on the zenith angle, incorporating the PDT. The analysis is performed for a downlink scenario across various weather conditions, as outlined in Table \ref{table}. Each value on the plot is determined from 1,000 samples of the parameters, based on Eq. (\ref{eq:Vector of beam-parameters}) in Sec.\ref{subsec:Elliptic_Beam_Model} and computed using Eq. (\ref{eq:PDT Equation}). {Figure. \ref{fig2} represents the finite and asymptotic key analysis, demonstrating that the secure key rate for efficient BB84 is generally higher than that of standard BB84, particularly during daytime conditions compared to nighttime.
 In Fig. \ref{fe2a}, at the zenith position (i.e., \(0^\circ\) zenith angle), the efficient BB84 protocol achieves a key rate of approximately \(8 \times 10^{-5}\) per pulse under clear daytime conditions (Day 1), whereas under clear nighttime conditions (Night 1), the key rate slightly decreases to \(7.7 \times 10^{-5}\) per pulse. This reduction is primarily due to increased aerosol scattering and the potential formation of haze at night due to low temperature. In Fig. \ref{fs2b}, the standard BB84 protocol follows a similar trend, with key rates of approximately \(4 \times 10^{-5}\) per pulse in Day 1 and \(3.8 \times 10^{-5}\) per pulse in Night 1, yielding approximately half the key rate of the efficient BB84 protocol in both cases. The disparity in key rates becomes more pronounced at higher zenith angles, such as \(50^\circ\). At a zenith angle of \(50^\circ\), the reduction in key rate compared to the zenith position is significant for both the efficient and standard BB84 protocols. This decline can be theoretically attributed to the increased path length at larger zenith angles. As the zenith angle increases, the signal traverses a longer atmospheric path, leading to greater attenuation due to absorption and scattering. Furthermore, enhanced beam divergence reduces the overlap with the receiving telescope aperture, further lowering the detected photon count. In Fig. \ref{fe2a}, for the efficient BB84 protocol, the key rate decreases from \(8 \times 10^{-5}\) per pulse at zenith position to \(1.09 \times 10^{-5}\) per pulse at \(50^\circ\) during Day 1, representing approximately a 7.34-fold reduction. Similarly, during Night 1, the key rate decreases from \(7.7 \times 10^{-5}\) per pulse at zenith position to \(1.06 \times 10^{-5}\) per pulse at \(50^\circ\), yielding a 7.26-fold reduction. In Fig. \ref{fs2b}, for the standard BB84 protocol, the key rate drops from \(4 \times 10^{-5}\) per pulse at \(0^\circ\) to \(7.8 \times 10^{-6}\) per pulse at \(50^\circ\) during Day 1, corresponding to a 5.13-fold reduction. In Night 1, the key rate decreases from \(3.8 \times 10^{-5}\) per pulse at zenith position to \(7.6 \times 10^{-6}\) per pulse at \(50^\circ\), leading to a 5-fold reduction. These reductions highlight the significant impact of increasing zenith angle on the key rate, primarily due to enhanced atmospheric losses and longer path lengths through the atmosphere at higher zenith angles. Furthermore, the results consistently demonstrate the superior performance of the efficient BB84 protocol over the standard BB84 protocol, as it achieves significantly higher key rates under both conditions.}  Across all the plots in Fig. \ref{fig2}, the pattern of the plots, corresponding to the sequential arrangement of different weather conditions, remains consistent for both protocols.  The sequence of weather conditions yielding higher key rate values follows the order: day-~1, night-~1, day-~2, night-~2, day-~3, and night-~3. A key aspect of interest is the comparison of system performance between daytime and nighttime operations.
 
\begin{figure}[htbp!]
    \centering
    \begin{subfigure}[b]{0.493\textwidth}
        \centering
         \hspace*{-0.9cm}
        \includegraphics[width=1.3\linewidth]{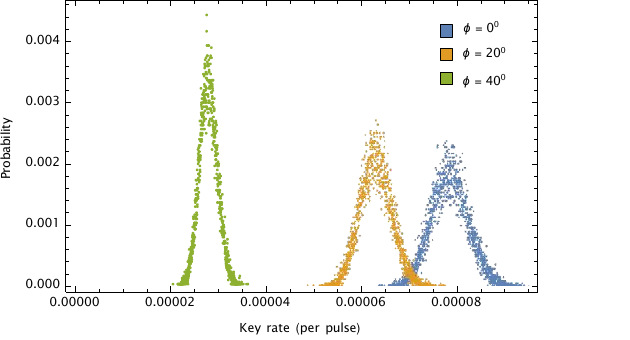} 
         \caption{}
       \label{pdrfe3a}
    \end{subfigure}
    \hfill
    \begin{subfigure}[b]{0.493\textwidth}
        \centering
          \hspace*{0.3cm}
        \includegraphics[width=1.3\linewidth]{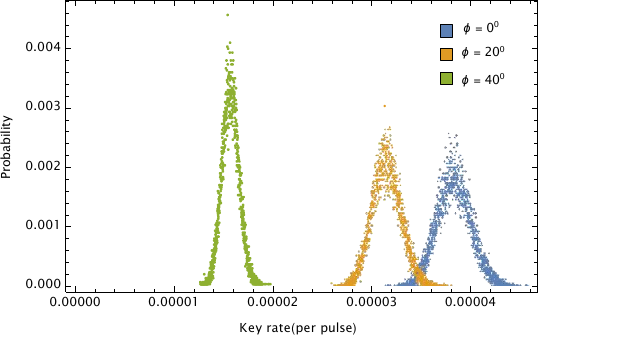} 
        \caption{}
         \label{pdrfs3b}
    \end{subfigure}

    \vskip\baselineskip 

    \begin{subfigure}[b]{0.493\textwidth}
        \centering
         \hspace*{-0.9cm}
        \includegraphics[width=1.3\linewidth]{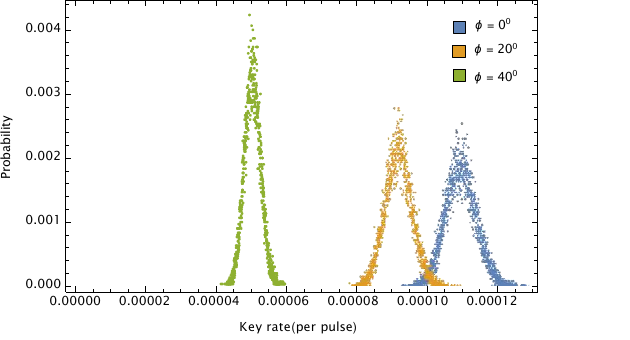} 
        \caption{}
        \label{pdrae3c}
    \end{subfigure}
    \hfill
    \begin{subfigure}[b]{0.493\textwidth}
        \centering
          \hspace*{0.3cm}
        \includegraphics[width=1.3\linewidth]{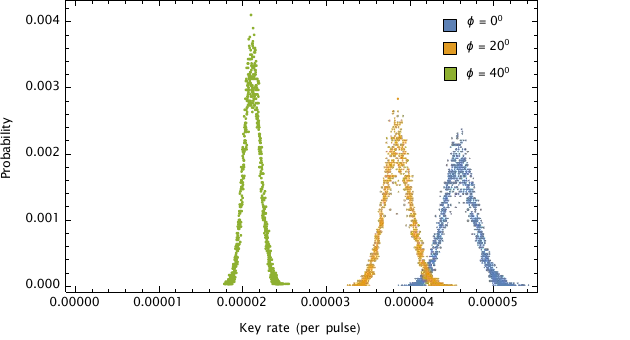} 
        \caption{}
        \label{pdras3d}
    \end{subfigure}

    \caption{Plot depicting the variation in key rate distribution across different zenith angles $(\phi)$ under condition Day-1 a) PDR for Finite efficient BB84 protocol, b) PDR for Finite standard BB84 protocol, c) PDR for Asymptotic efficient BB84 protocol, d) PDR for Asymptotic standard BB84 protocol.}
    \label{pdr3}
\end{figure}
 
In the daytime, elevated temperatures lead to more intense winds and enhanced mixing among various atmospheric layers, leading to more significant turbulence effects. However, on average, clear daytime conditions exhibit less moisture in the lower part of the atmosphere than nighttime conditions which results in reduced beam spreading due to scattering particles. In contrast, cooler nighttime temperatures lead to a less turbulent atmosphere, coupled with formation of haze and mist. As a result, scattering has a more pronounced impact at night than turbulence does during the day.  {The secure key rate exhibits a progressive decline with increasing atmospheric turbulence and fog. Under moderate wind conditions (Day 2) and slightly foggy conditions (Night 2), the efficient BB84 protocol achieves a key rate of \(7.5 \times 10^{-5}\) per pulse during Day 2 and \(7.2 \times 10^{-5}\) per pulse during Night 2 at the zenith position. At a zenith angle of \(50^\circ\), the key rate decreases to the order of \(10^{-6}\), reflecting the impact of increased optical path length and atmospheric attenuation. The standard BB84 protocol follows a similar trend but exhibits a lower key rate, reducing to \(3.6 \times 10^{-5}\) per pulse during Day 2 and \(3.5 \times 10^{-5}\) per pulse during Night 2 at the zenith position. At \(50^\circ\) zenith angle, the standard BB84 protocol experiences a more pronounced decline, highlighting its greater susceptibility to turbulence-induced fluctuations and channel losses compared to the efficient BB84 protocol.  With further intensification of atmospheric disturbances in windy (Day 3) and moderately foggy (Night 3) conditions, the degradation in key rate becomes more severe. The efficient BB84 protocol remains operational across all conditions, sustaining key generation even at higher zenith angles, whereas the standard BB84 protocol approaches the threshold beyond which key generation becomes impractical. The increased optical path length at larger zenith angles exacerbates scattering and absorption effects, leading to heightened attenuation. This degradation is particularly detrimental to the standard BB84 protocol, which exhibits greater sensitivity to statistical fluctuations and environmental losses.  These results reinforce that clear daytime conditions (Day 1) are optimal for CubeSat-based QKD due to minimal scattering losses, while nighttime conditions (Night 1) introduce slightly higher losses due to increased aerosol scattering. However, severe turbulence and fog (Day 3 and Night 3) significantly impair performance, particularly for the standard BB84 protocol. The superior performance of the efficient BB84 protocol aligns with theoretical expectations, as its biased basis selection enhances the sifting ratio, thereby improving parameter estimation accuracy and increasing overall key generation efficiency.} Additionally, the efficient BB84 protocol uses one basis for key generation and the other for parameter estimation, thereby maximizing the utilization of measurement outcomes and minimizing unused data. In contrast, the standard BB84 protocol employs both bases with equal probability and utilizes both for key generation, necessitating parameter estimation for each basis. To estimate the signal parameters, the protocol reveals a random sample of results from each measurement basis separately. Consequently, only half of the revealed results from each basis are used for parameter estimation, introducing greater statistical uncertainty compared to the efficient BB84 protocol.  {In Fig.~\ref{ae2c} and Fig.~\ref{as2d}, the asymptotic regime follows a similar trend to the finite-key regime, with key rates increasing by approximately 1.2 to 1.5 times. However, the relative advantage of the efficient BB84 protocol remains consistent across all conditions, resulting in a higher asymptotic average key rate compared to the standard BB84 protocol. This trend is evident in Fig.~\ref{ae2c} and Fig.~\ref{as2d}, where the asymptotic key rates exceed those in the finite-key regime, as shown in Fig.~\ref{fe2a} and Fig.~\ref{fs2b}, further reinforcing the performance superiority of the efficient BB84 protocol.}  Notably, in the finite-key scenario, the key rate drops to zero earlier at a zenith angle of $62^\circ$, whereas in the asymptotic case, it extends up to $75.2^\circ$. This difference arises because the asymptotic analysis assumes an infinite number of satellite passes, thereby eliminating statistical fluctuations and finite-size effects. As a result, the key rate remains higher and persists to a greater zenith angle compared to the finite-key case.\\
 In Fig. \ref{pdr3}, we present the PDR at different zenith positions under downlink configuration. In this scenario, we consider the optimal performance under daytime condition 1. A dataset comprising $3 \times 10^4$ beam parameters served to simulate the average key rate, with outcomes rounded to five decimal places for generating the PDR graphs for efficient and standard BB84 protocols in both finite and asymptotic cases. A comparison between $\phi=0^\circ$ and $\phi=40^\circ$ shows that a higher key rate is observed at $\phi=0^\circ$, whereas the maximum probability of the key rate is higher at $\phi=40^\circ$. Notably, a larger key rate correlates with a decreased probability of occurrence. At lower zenith angles (e.g., $\phi = 0^\circ$), the key rate distribution is broader, indicating greater variation in key rates. Shorter optical path lengths result in stronger signal transmission, leading to higher key rates. However, atmospheric effects such as turbulence still introduce fluctuations, causing a wider distribution and reducing the probability of key rate at lower zenith angles. At higher zenith angles (e.g., $\phi = 20^\circ$ and $\phi = 40^\circ$), the probability of the key rate increases, but the key rate distribution becomes narrower, and the overall key rate decreases. The longer atmospheric path length results in higher attenuation, beam spreading, and turbulence effects, reducing transmittance and increasing photon losses. Consequently, the probability distribution of the key rate becomes more concentrated at lower values, reflecting a reduced spread of key rates. The higher uncertainity of keyrates at lower zenith angles (e.g., $\phi = 0^\circ$) in PDR is primarily due to the combined effect of high transmittance and residual atmospheric fluctuation. At small zenith angles, the shorter propagation path results in strong signal transmission and higher average keyrates. However, in this regime, the key rate becomes more sensitive to fluctuations caused by turbulence, beam wandering and pointing errors. Because the signal strength is relatively high in this regime, even small variations in atmospheric conditions produce noticeable variations in the received photon statistics, which directly impacts the estimated keyrate. Consequently, this results in more uncertainity at low zenith angles. 
In contrast, at higher zenith angles (e.g., $\phi = 20^\circ$ and $\phi = 40^\circ$ ), the optical path length increases significantly, leading to stronger attenuation that results reduced transmittance which decreases the effect of turbulence . In this regime, the received signal is consistently weak, and the keyrate is predominantly limited by channel loss rather than fluctuations. As a result, the key rate values tend to cluster around lower values, producing a narrower and more concentrated distribution with reduced apparent uncertainity. Therefore, the increased uncertainity at lower zenith angles is not due to poorer channel conditions, but rather due to the greater sensitivity of higher keyrates to stochastic channel variations. The overall key rate declines due to increased channel losses and lower detection efficiency.\\
 In both finite and asymptotic cases, the efficient BB84 protocol consistently achieves a higher key rate than the standard protocol. This is evident in the PDR, where the efficient BB84 protocol exhibits a broader distribution at lower zenith angles and maintains higher key rates compared to the standard BB84 protocol. Additionally, as the zenith angle increases, the PDR of the efficient BB84 protocol narrows more gradually compared to the standard BB84 protocol, which experiences a sharper reduction in key rate probability. This trend is consistently observed across all subplots in Fig.~\ref{pdr3}, highlighting the superior performance of the efficient BB84 protocol in CubeSat-based QKD implementations. 
To highlight the advantages of the efficient BB84 protocol over the standard BB84 protocol, we provide a comparative functional table. The below table \ref{func_comp} outlines key performance metrics, emphasizing improvements in key rate, error resilience, and protocol efficiency \cite{SBM+22}. 

\begin{table}[htbp]
    \centering
    \caption{Functional Comparison Between Standard and Efficient BB84 Protocols}
    \label{func_comp}
    \renewcommand{\arraystretch}{1.3} 
    \begin{tabular}{|p{4.5cm}|p{5.5cm}|p{5.5cm}|}
        \hline
        \textbf{Feature} & \textbf{Standard BB84 Protocol} & \textbf{Efficient BB84 Protocol} \\
        \hline
        \textbf{Basis Selection and Sifting Ratio} & Symmetric basis choice (X and Z); lower sifting efficiency. & Asymmetric basis choice; improved sifting efficiency. \\
        \hline
        \textbf{Key Generation and Parameter Estimation} & Both bases used for key generation and parameter estimation. & One basis used for key generation; the other for parameter estimation. \\
        \hline
        \textbf{Classical Communication} & Higher communication overhead for separate QBER estimation. & Reduced overhead; QBER inferred from basis choice during sifting. \\
        \hline
        \textbf{Finite Key Rate} & Lower average key rate due to inefficient sifting. & Higher average key rate due to increased raw key length from efficient sifting. \\
        \hline
        \textbf{Asymptotic Key Rate} & Lower key rate across weather conditions. & Higher rate due to improved protocol efficiency and channel loss resilience. \\
        \hline
        \textbf{Finite Statistical Uncertainty} & Estimation from two bases increases uncertainty. & Reduced uncertainty from single-basis parameter estimation. \\
        \hline
        \textbf{Key Rate Distribution (PDR)} & Narrower distribution at lower zenith angles with generally lower rates. & Broader distribution at low zenith angles with higher key rates. \\
        \hline
    \end{tabular}
    \label{tab:BB84Comparison}
\end{table}

\section{Conclusion}
\label{Conclusion}

In this chapter, we investigated the performance of two QKD protocols-the efficient BB84 and the standard BB84-under a two-decoy state setting in a CubeSat-based free-space communication downlink scenario \cite{MDB25}. A key aspect of this study is the application of the elliptical beam approximation to model transmittance more accurately compared to the conventional circular beam approach. This refinement provides a more realistic characterization of beam divergence and atmospheric effects, thereby improving the reliability of key rate estimation for CubeSat-based QKD.  

A comprehensive finite and asymptotic key rate analysis has been conducted to evaluate and compare the performance of the efficient and standard BB84 protocols under varying atmospheric conditions in a CubeSat-based QKD system. The analysis utilized the elliptical beam approximation model to accurately characterize channel transmittance and its impact on key rate generation. Our results demonstrate that the efficient BB84 protocol consistently achieves higher key rates across different transmission conditions, particularly in low-transmittance regimes, making it a more suitable candidate for CubeSat-based quantum communication. The dependence of the average key rate per pulse on the \textquotedbl Zenith angle\textquotedbl  under different atmospheric conditions has been analyzed, along with the PDR. The results indicate that the PDR exhibits a consistent shape across all analyzed scenarios, revealing critical insights into the performance variations of the protocols.  

While the present study utilizes uniform and normal distributions for beam parameter modeling, other transmittance models-such as log-normal, Gamma-Gamma, and Double Weibull distributions-may further enhance accuracy under varying atmospheric conditions. The selection of an appropriate distribution is influenced by factors like the  intensity of turbulence, link distance, and the optical system configuration. Future work will focus on incorporating these alternative models and optimizing finite key generation techniques for higher-dimensional QKD protocols in CubeSat-based quantum communication. Additionally, real-time adaptive beam correction methods could be explored to mitigate the impact of atmospheric fluctuations.  

Our study \cite{MDB25} integrates finite and asymptotic key analysis with elliptical beam modeling across various atmospheric conditions, along with PDR, to establish a more precise analytical framework for key rate estimation in CubeSat-based QKD. By enhancing the understanding of CubeSat-based quantum key distribution, it contributes to the advancement of satellite-based quantum communication networks. The findings presented provide a practical foundation for bridging the gap between theoretical modeling and real-world implementation, supporting the future development of secure satellite-based quantum communication systems.


\chapter{Conclusion and Future Outlook}\label{conclusion}
Satellite-based quantum key distribution has emerged as a crucial pathway for enabling secure global quantum communication beyond the limits of terrestrial networks. This thesis presents a comprehensive investigation of satellite-based quantum key distribution by progressing from foundational protocol analysis to advanced high-dimensional schemes and finally to realistic finite keyrate analysis. The study begins by evaluating several widely used prepare-and-measure and entanglement-based QKD protocols under asymptotic assumptions, using circular-beam modeling to understand how atmospheric attenuation, pointing errors, and orbital geometry influence secure key generation in satellite links. Motivated by the inherent rate and noise-tolerance limitations of qubit-level encoding, the work then extends the analysis to HD-QKD, where additional photonic degrees of freedom and an elliptical-beam turbulence model enable a more robust characterization of key rates under realistic atmospheric conditions. Building further toward practical implementation, the thesis incorporates finite-size statistical effects and decoy-state analysis within a CubeSat downlink scenario, integrating turbulence, link dynamics, and background noise into a unified framework for evaluating efficient and standard BB84 protocols under mission-relevant constraints. Overall, the work follows a natural progression-from understanding baseline protocol behavior, to exploring performance enhancements through higher-dimensional encoding, and finally to quantifying realistic finite-key performance-thus forming a coherent pathway toward designing secure and efficient satellite-based QKD systems. The remainder of this chapter presents a concise summary of the key findings of this thesis and outlines potential directions for future research.
\section{Summary of the thesis}
We now provide a concise summary of the chapters in this thesis and discuss the key findings of our research.\\
\textbf{Chapter 1:} This chapter provides an introductory foundation for the thesis, which focuses on the broad overview of quantum physics and its impact on modern technologies, emphasizing the growing importance of quantum communication in the era of quantum computing. As classical cryptographic schemes such as RSA \cite{RSA78} and Diffie-Hellman \cite{DH22} face potential threats from quantum computers-particularly due to Shor’s algorithm \cite{S94}- QKD emerges as a fundamentally secure alternative based on quantum mechanical principles. The chapter introduces key concepts essential for understanding quantum communication, including qubits, quantum states, density matrices \cite{NC10}, composite quantum systems, entanglement \cite{P13}, and quantum entropy. These tools establish the theoretical groundwork for analyzing QKD protocols. Next, the chapter provides an overview of QKD, explaining its goal of enabling information-theoretic security by exploiting principles such as the no-cloning theorem \cite{WZ82, SIG+05} and the Heisenberg uncertainty principle. These principles ensure that any eavesdropping attempt introduces detectable disturbances. Motivated by limitations in optical-fiber-based QKD \cite{MPR+19, BBR+18}, the chapter highlights satellite-based QKD as a promising solution for long-distance secure communication \cite{BBM+15}. Finally, the chapter outlines the structure of the thesis and briefly introduces the motivations and objectives of subsequent chapters, which focus on protocol performance comparison, HD enhancements, and finite-key security analysis.\\
\textbf{Chapter 2:} This chapter presented a comprehensive analysis of the performance of four QKD protocols-BB84, B92, BBM92, and E91-within the context of satellite-based quantum communication. Motivated by the need for long-distance, high-security communication, the study focused on LEO satellite links, which offer reduced channel loss and improved photon detection probability compared to MEO and GEO systems.
The chapter implements the circular beam propagation model as the core framework for computing channel transmittance in both uplink and downlink configurations. Key atmospheric effects, including diffraction, turbulence, attenuation, and pointing errors, were incorporated to accurately characterize photon propagation through the free-space channel. Uplink modeling accounted for turbulence-induced beam broadening using the Hufnagel-Valley profile, whereas downlink modeling included Gaussian-distributed pointing errors. Atmospheric transmittance values derived from MODTRAN~6 \cite{BCK+14} further enabled realistic simulation of channel losses. Environmental noise contributions, such as stray photons during day and night conditions, were evaluated to determine their impact on detector performance and QBER. The chapter then formulated QBER expressions and asymptotic key rate equations for each of the considered protocols, followed by numerical evaluation under practical LEO mission parameters. Results demonstrated that protocol performance is strongly influenced by channel asymmetries, beam propagation characteristics, environmental noise, and inherent protocol resource requirements. The comparative study highlights the conditions under which each protocol is best suited for satellite-based QKD. The result in this chapter is presented in \cite{MRS26}.\\ \textbf{Chapter 3:} This chapter presented a detailed performance analysis of two high-dimensional QKD protocols-HD-Ext-B92 and HD-BB84-within the framework of satellite-based quantum communication. Motivated by the need for secure long-distance quantum links, the study examined the suitability of these protocols for LEO scenarios, where atmospheric turbulence, beam wandering, and diffraction significantly influence channel transmittance. The atmospheric channel was modeled using the elliptic-beam approximation introduced by Vasylyev \cite{VSV+17}, which provides realistic characterization of turbulence effects in free-space optical links. The chapter began by outlining the theoretical foundations of HD-QKD and summarizing the modifications made to the HD-Ext-B92 key-rate formulation to eliminate additional free parameters, as described in Appendix B. Both HD-Ext-B92 and HD-BB84 were analyzed under depolarizing noise using WCP sources, with the decoy-state method incorporated to mitigate PNS attacks. A comprehensive numerical study was conducted to investigate the behavior of the key rate, QBER, and the PDR across different system dimensions, weather conditions, noise levels, and zenith angles. Using the elliptic-beam-based channel model, the simulations demonstrated that the HD-BB84 protocol consistently achieves higher key rates and better noise tolerance than the HD-Ext-B92 protocol for both uplink and downlink configurations. The PDR analysis further revealed that while both protocols exhibit similar distribution shapes, HD-BB84 produces more densely clustered and higher-probability regions around favorable key-rate values, indicating stronger stability and performance across channel fluctuations. The chapter concluded that HD-BB84 outperforms HD-Ext-B92 in terms of key-rate magnitude, noise resilience, and probability distribution characteristics. Overall, the findings confirm that HD encoding can significantly enhance the performance of satellite-based QKD systems.  The work presented in this chapter is published in Ref. \cite{DMB+24}\\
\textbf{Chapter 4:} This chapter presented a comprehensive finite-key and asymptotic-key performance analysis of the efficient and standard BB84 QKD protocols for CubeSat-based downlink communication. Motivated by the growing feasibility of CubeSats as low-cost quantum communication platforms, the study focused on realistic modeling of atmospheric effects, link geometry, and statistical constraints that govern secure key generation in short-duration satellite passes. To model photon propagation through the turbulent atmosphere, the chapter employed the elliptical beam approximation which is also implemented in Chapter 3. This model provides a more realistic characterization of channel transmittance than the conventional circular beam approximation, particularly in downlink configurations where atmospheric distortions near the ground station dominate. Weather-dependent turbulence and scattering effects were incorporated to evaluate performance under clear, hazy, foggy, and windy conditions. Both protocols were analyzed under two-decoy-state settings using weak coherent pulses. The chapter detailed the finite-key framework, incorporating statistical fluctuations through multiplicative Chernoff bounds and refined error-correction leakage models. The asymptotic analysis established the fundamental key-rate limits by removing finite-size corrections and averaging performance over an infinite number of CubeSat passes. Key rate expressions for both regimes were implemented, and the probability distribution of transmittance was used to compute average key rates across varying zenith angles. Numerical simulations demonstrated that the efficient BB84 protocol consistently outperforms the standard BB84 protocol in both finite and asymptotic regimes. The efficient protocol achieves higher key rates due to its biased-basis strategy, which improves sifting efficiency, reduces statistical uncertainty, and enhances resilience to atmospheric loss. In contrast, the standard BB84 protocol exhibits lower performance, especially under severe turbulence and fog, where its symmetric basis structure increases data rejection and error sensitivity. Probability distribution of key rates analysis further confirmed that efficient BB84 maintains broader and higher-rate distributions across zenith angles, reflecting greater robustness in CubeSat channels. Overall, this chapter provides a complete evaluation of CubeSat-based BB84 QKD, integrating atmospheric modeling, statistical corrections, PDT-based averaging, and weather-dependent link analysis. The results show that efficient BB84 is the preferred protocol for CubeSat missions, offering superior key-rate performance, improved stability, and greater operational range. The results reported in this chapter are published in Ref. \cite{MDB25}.
\section{Limitations and Future Scope}
The limitations of any research work serve as a motivation for future advancements toward more feasible and effective results. In the present study, we have analyzed both prepare-and-measure and entanglement-based QKD protocols using the circular beam approximation to evaluate the secure key rate. However, our analysis does not encompass all the protocols or alternative beam propagation models. Future research could focus on a comprehensive comparative by incorporating additional QKD protocols, including CV-QKD schemes. Moreover, the analysis can be extended by employing different beam models beyond the circular approximation, which may provide deeper insights into system performance under diverse atmospheric and optical conditions. In this thesis, we have carried out a comprehensive numerical study of high-dimensional QKD protocols, specifically HD-BB84 and HD-Extended B92, to examine the behavior of the key rate, QBER, and probability distribution of keyrate using elliptical beam approximation model under depolarizing noise. While this analysis provides valuable insights into system performance, it is limited to a single noise model and a specific attack strategy. Future research can extend this work by analyzing these HD-protocols under alternative noise models, such as amplitude-damping and phase-damping noise, to better capture diverse physical scenarios. Additionally, exploring other HD-protocols and evaluating their robustness against various eavesdropping strategies beyond the PNS attack would further strengthen the understanding of their practical security.  And also the work in chapter 3 definitely establishes the advantages
of using higher-dimensional states in satellite-based quantum communication; but there are challenges associated with the experimental generation and maintenance of the qudits. In the near future, we would like to
address this technical issue and also to find the optimal choice of dimension that can provide the desired key rate.
Finally, we explored the finite key rate analysis for two variants of the BB84 protocol. Since our study focused on two variants of BB84 protocol, future work could extend finite-key analysis to other QKD schemes, including HD and CV QKD protocols. Moreover, while our analysis is limited to LEO scenarios, it can be further expanded to MEO and GEO configurations to provide a broader understanding of satellite-based QKD performance. In this work, statistical fluctuations were addressed using the Chernoff bound. To obtain tighter bounds on parameter estimation, future research may consider employing alternative inequalities such as Azuma’s or Kato’s inequalities, which could potentially yield more accurate finite-key performance estimates.








\newpage
\appendix


\addcontentsline{toc}{chapter}{REFERENCES
}
\printbibliography

@article{WZ82,
  title={A single quantum cannot be cloned},
  author={Wootters, William K and Zurek, Wojciech H},
  journal={Nature},
  volume={299},
  number={5886},
  pages={802--803},
  year={1982},
  publisher={Nature Publishing Group UK London}
}

@article{SIG+05,
  title={Quantum cloning},
  author={Scarani, Valerio and Iblisdir, Sofyan and Gisin, Nicolas and Ac{\'\i}n, Antonio},
  journal={Reviews of Modern Physics},
  volume={77},
  number={4},
  pages={1225--1256},
  year={2005},
  publisher={APS}
}

@book{NC10,
  title={Quantum computation and quantum information},
  author={Nielsen, Michael A and Chuang, Isaac L},
  year={2010},
  publisher={Cambridge university press}
}

@article{H27,
  title={Heisenberg uncertainty principle},
  author={Heisenberg, Werner},
  journal={Zeitschrift f{\"u}r Physik},
  volume={43},
  pages={172},
  year={1927}
}

@article{W83,
  title={Conjugate coding},
  author={Wiesner, Stephen},
  journal={ACM Sigact News},
  volume={15},
  number={1},
  pages={78--88},
  year={1983},
  publisher={ACM New York, NY, USA}
}

@inproceedings{NS18,
  title={Quantum key distribution (QKD) protocols: A survey},
  author={Nurhadi, Ali Ibnun and Syambas, Nana Rachmana},
  booktitle={2018 4th International Conference on Wireless and Telematics (ICWT)},
  pages={1--5},
  year={2018},
  organization={IEEE}
}

@article{PAB+20,
  title={Advances in quantum cryptography},
  author={Pirandola, Stefano and Andersen, Ulrik L and Banchi, Leonardo and Berta, Mario and Bunandar, Darius and Colbeck, Roger and Englund, Dirk and Gehring, Tobias and Lupo, Cosmo and Ottaviani, Carlo and others},
  journal={Advances in optics and photonics},
  volume={12},
  number={4},
  pages={1012--1236},
  year={2020},
  publisher={Optical Society of America}
}

@article{B92,
  title = {Quantum cryptography using any two nonorthogonal states},
  author = {Bennett, Charles H.},
  journal = {Phys. Rev. Lett.},
  volume = {68},
 year = {1992}
 }

@article{BBM92,
  title={Quantum cryptography without Bell’s theorem},
  author={Bennett, Charles H and Brassard, Gilles and Mermin, N David},
  journal={Physical Review Letters},
  volume={68},
  number={5},
  pages={557},
  year={1992},
  publisher={APS}
}

@article{LCW+14,
  title={Concise security bounds for practical decoy-state quantum key distribution},
  author={Lim, Charles Ci Wen and Curty, Marcos and Walenta, Nino and Xu, Feihu and Zbinden, Hugo},
  journal={Physical Review A},
  volume={89},
  number={2},
  pages={022307},
  year={2014},
  publisher={APS}
}

@article{SBM+22,
  title={Finite key effects in satellite quantum key distribution},
  author={Sidhu, Jasminder S and Brougham, Thomas and McArthur, Duncan and Pousa, Roberto G and Oi, Daniel KL},
  journal={npj Quantum Information},
  volume={8},
  number={1},
  pages={18},
  year={2022},
  publisher={Nature Publishing Group UK London}
}

@article{TMP+17,
  title={Fundamental finite key limits for one-way information reconciliation in quantum key distribution},
  author={Tomamichel, Marco and Martinez-Mateo, Jesus and Pacher, Christoph and Elkouss, David},
  journal={Quantum Information Processing},
  volume={16},
  pages={1--23},
  year={2017},
  publisher={Springer}
}

@article{W05,
  title={Beating the photon-number-splitting attack in practical quantum cryptography},
  author={Wang, Xiang-Bin},
  journal={Physical Review Letters},
  volume={94},
  number={23},
  pages={230503},
  year={2005},
  publisher={APS}
}

@article{MQZ+05,
  title={Practical decoy state for quantum key distribution},
  author={Ma, Xiongfeng and Qi, Bing and Zhao, Yi and Lo, Hoi-Kwong},
  journal={Physical Review A—Atomic, Molecular, and Optical Physics},
  volume={72},
  number={1},
  pages={012326},
  year={2005},
  publisher={APS}
}

@article{HHH+07,
  title={Security analysis of decoy state quantum key distribution incorporating finite statistics},
  author={Hasegawa, Jun and Hayashi, Masahito and Hiroshima, Tohya and Tomita, Akihisa},
  journal={arXiv preprint arXiv:0707.3541},
  year={2007}
}

@article{CS09,
  title={Finite-key analysis for practical implementations of quantum key distribution},
  author={Cai, Raymond YQ and Scarani, Valerio},
  journal={New Journal of Physics},
  volume={11},
  number={4},
  pages={045024},
  year={2009},
  publisher={IOP Publishing}
}

@article{ZZR+17,
  title={Improved key-rate bounds for practical decoy-state quantum-key-distribution systems},
  author={Zhang, Zhen and Zhao, Qi and Razavi, Mohsen and Ma, Xiongfeng},
  journal={Physical Review A},
  volume={95},
  number={1},
  pages={012333},
  year={2017},
  publisher={APS}
}

@article{HN14,
  title={Security analysis of the decoy method with the {B}ennett--{B}rassard 1984 protocol for finite key lengths},
  author={Hayashi, Masahito and Nakayama, Ryota},
  journal={New Journal of Physics},
  volume={16},
  number={6},
  pages={063009},
  year={2014},
  publisher={IOP Publishing}
}

@article{CXC+14,
  title={Finite-key analysis for measurement-device-independent quantum key distribution},
  author={Curty, Marcos and Xu, Feihu and Cui, Wei and Lim, Charles Ci Wen and Tamaki, Kiyoshi and Lo, Hoi-Kwong},
  journal={Nature Communications},
  volume={5},
  number={1},
  pages={3732},
  year={2014},
  publisher={Nature Publishing Group UK London}
}

@article{YZG+20,
  title={Tight security bounds for decoy-state quantum key distribution},
  author={Yin, Hua-Lei and Zhou, Min-Gang and Gu, Jie and Xie, Yuan-Mei and Lu, Yu-Shuo and Chen, Zeng-Bing},
  journal={Scientific Reports},
  volume={10},
  number={1},
  pages={1--10},
  year={2020},
  publisher={Nature Publishing Group}
}

@article{SBM+21,
  title={Satellite quantum modelling \& analysis software version 1.1: documentation},
  author={Sidhu, Jasminder S and Brougham, Thomas and McArthur, Duncan and Pousa, Roberto G and Oi, Daniel KL},
  journal={arXiv preprint arXiv:2109.01686},
  year={2021}
}

@article{VSV16,
  title={Atmospheric quantum channels with weak and strong turbulence},
  author={Vasylyev, D and Semenov, AA and Vogel, W},
  journal={Physical Review Letters},
  volume={117},
  number={9},
  pages={090501},
  year={2016},
  publisher={APS}
}

@article{VSV+17,
  title={Free-space quantum links under diverse weather conditions},
  author={Vasylyev, D and Semenov, AA and Vogel, W and G{\"u}nthner, K and Thurn, A and Bayraktar, {\"O} and Marquardt, Ch},
  journal={Physical Review A},
  volume={96},
  number={4},
  pages={043856},
  year={2017},
  publisher={APS}
}

@article{LKB19,
  title={Satellite-based links for quantum key distribution: beam effects and weather dependence},
  author={Liorni, Carlo and Kampermann, Hermann and Bru{\ss}, Dagmar},
  journal={New Journal of Physics},
  volume={21},
  number={9},
  pages={093055},
  year={2019},
  publisher={IOP Publishing}
}

@article{TP88,
  title={Vertical distribution features of atmospheric water vapour in the Po Valley area},
  author={Tomasi, Claudio and Paccagnella, Tiziana},
  journal={Pure and Applied Geophysics},
  volume={127},
  pages={93--115},
  year={1988},
  publisher={Springer}
}

@article{T84,
  title={Vertical distribution features of atmospheric water vapor in the Mediterranean, Red Sea, and Indian Ocean},
  author={Tomasi, Claudio},
  journal={Journal of Geophysical Research: Atmospheres},
  volume={89},
  number={D2},
  pages={2563--2566},
  year={1984},
  publisher={Wiley Online Library}
}

@article{WHW+18,
  title={Atmospheric effects on continuous-variable quantum key distribution},
  author={Wang, Shiyu and Huang, Peng and Wang, Tao and Zeng, Guihua},
  journal={New Journal of Physics},
  volume={20},
  number={8},
  pages={083037},
  year={2018},
  publisher={IOP Publishing}
}

@article{DMB+24,
  title={Analysis for Satellite-Based High-Dimensional Extended {B92} and High-Dimensional {BB84} Quantum Key Distribution},
  author={Dutta, Arindam and Muskan and Banerjee, Subhashish and Pathak, Anirban},
  journal={Advanced Quantum Technologies},
  volume={7},
  number={11},
  pages={2400149},
  year={2024},
  publisher={Wiley Online Library}
}

@inproceedings{BB84,
  title={Quantum key distribution and coin tossing},
  author={Bennett, CH and Brassard, G},
  booktitle={Proc. of IEEE Int. Conf. on Computers, Systems, and Signal Processing (Bangalore, India, 1984)},
  pages={175--179},
  year={1984}
}

@article{GRT+02,
  title={Quantum cryptography},
  author={Gisin, Nicolas and Ribordy, Gr{\'e}goire and Tittel, Wolfgang and Zbinden, Hugo},
  journal={Reviews of Modern Physics},
  volume={74},
  number={1},
  pages={145},
  year={2002},
  publisher={APS}
}

@article{H03,
  title={Quantum key distribution with high loss: toward global secure communication},
  author={Hwang, Won-Young},
  journal={Physical Review Letters},
  volume={91},
  number={5},
  pages={057901},
  year={2003},
  publisher={APS}
}

@article{LMC05,
  title={Decoy state quantum key distribution},
  author={Lo, Hoi-Kwong and Ma, Xiongfeng and Chen, Kai},
  journal={Physical Review Letters},
  volume={94},
  number={23},
  pages={230504},
  year={2005},
  publisher={APS}
}

@article{LCL+17,
  title={Satellite-to-ground quantum key distribution},
  author={Liao, Sheng-Kai and Cai, Wen-Qi and Liu, Wei-Yue and Zhang, Liang and Li, Yang and Ren, Ji-Gang and Yin, Juan and Shen, Qi and Cao, Yuan and Li, Zheng-Ping and others},
  journal={Nature},
  volume={549},
  number={7670},
  pages={43--47},
  year={2017},
  publisher={Nature Publishing Group UK London}
}

@article{ZSH+24,
  title={End-to-end demonstration for CubeSatellite quantum key distribution},
  author={Zhang, Peide and Sagar, Jaya and Hastings, Elliott and Stefko, Milan and Joshi, Siddarth and Rarity, John},
  journal={IET Quantum Communication},
  volume = {5},
  pages = {291--302},
  year={2024},
  publisher={Wiley Online Library}
}

@article{MPR+19,
  title={Experimental quantum key distribution beyond the repeaterless secret key capacity},
  author={Minder, Mariella and Pittaluga, Mirko and Roberts, George Lloyd and Lucamarini, Marco and Dynes, James F and Yuan, ZL and Shields, Andrew J},
  journal={Nature Photonics},
  volume={13},
  number={5},
  pages={334--338},
  year={2019},
  publisher={Nature Publishing Group UK London}
}

@article{BBR+18,
  title={Secure quantum key distribution over 421 km of optical fiber},
  author={Boaron, Alberto and Boso, Gianluca and Rusca, Davide and Vulliez, C{\'e}dric and Autebert, Claire and Caloz, Misael and Perrenoud, Matthieu and Gras, Ga{\"e}tan and Bussi{\`e}res, F{\'e}lix and Li, Ming-Jun and others},
  journal={Physical Review Letters},
  volume={121},
  number={19},
  pages={190502},
  year={2018},
  publisher={APS}
}

@article{ATL15,
  title={All-photonic quantum repeaters},
  author={Azuma, Koji and Tamaki, Kiyoshi and Lo, Hoi-Kwong},
  journal={Nature Communications},
  volume={6},
  number={1},
  pages={1--7},
  year={2015},
  publisher={Nature Publishing Group}
}

@article{ZPD+18,
  title={Long-range big quantum-data transmission},
  author={Zwerger, Michael and Pirker, Alexander and Dunjko, Vedran and Briegel, Hans J and D{\"u}r, Wolfgang},
  journal={Physical Review Letters},
  volume={120},
  number={3},
  pages={030503},
  year={2018},
  publisher={APS}
}

@article{SGL18,
  title={Efficient quantum repeater with respect to both entanglement-concentration rate and complexity of local operations and classical communication},
  author={Su, Zhaofeng and Guan, Ji and Li, Lvzhou},
  journal={Physical Review A},
  volume={97},
  number={1},
  pages={012325},
  year={2018},
  publisher={APS}
}

@article{SSD+11,
  title={Quantum repeaters based on atomic ensembles and linear optics},
  author={Sangouard, Nicolas and Simon, Christoph and De Riedmatten, Hugues and Gisin, Nicolas},
  journal={Reviews of Modern Physics},
  volume={83},
  number={1},
  pages={33--80},
  year={2011},
  publisher={APS}
}

@article{BBM+15,
  title={Entanglement over global distances via quantum repeaters with satellite links},
  author={Boone, K and Bourgoin, J-P and Meyer-Scott, E and Heshami, K and Jennewein, T and Simon, C},
  journal={Physical Review A},
  volume={91},
  number={5},
  pages={052325},
  year={2015},
  publisher={APS}
}

@article{BAL17,
  title={Progress in satellite quantum key distribution},
  author={Bedington, Robert and Arrazola, Juan Miguel and Ling, Alexander},
  journal={npj Quantum Information},
  volume={3},
  number={1},
  pages={30},
  year={2017},
  publisher={Nature Publishing Group UK London}
}

@inproceedings{LVM+23,
  title={A Study on Quantum Key Distribution Satellite Communications},
  author={Lauterbach, Filip and Van{\"e}k, Michal and Mehic, Miralem and Voznak, Miroslav},
  booktitle={2023 15th International Congress on Ultra Modern Telecommunications and Control Systems and Workshops (ICUMT)},
  pages={128--133},
  year={2023},
  organization={IEEE}
}

@article{LCA05,
  title={Efficient quantum key distribution scheme and a proof of its unconditional security},
  author={Lo, Hoi-Kwong and Chau, Hoi Fung and Ardehali, Mohammed},
  journal={Journal of Cryptology},
  volume={18},
  pages={133--165},
  year={2005},
  publisher={Springer}
}

@article{DP22,
  title={A short review on quantum identity authentication protocols: how would {B}ob know that he is talking with {A}lice?},
  author={Dutta, Arindam and Pathak, Anirban},
  journal={Quantum Information Processing},
  volume={21},
  number={11},
  pages={369},
  year={2022},
  publisher={Springer}
}

@INPROCEEDINGS{SB18,
  author={Sharma, Vishal and Banerjee, Subhashish},
  booktitle={2018 9th International Conference on Computing, Communication and Networking Technologies (ICCCNT)}, 
  title={Analysis of Quantum Key Distribution Based Satellite Communication}, 
  pages={1--5},
  year={2018},
  publisher={IEEE}
}

@article{ABN21,
  title={Optimizing the decoy-state {BB}84 {QKD} protocol parameters},
  author={Attema, Thomas and Bosman, Joost W and Neumann, Niels MP},
  journal={Quantum Information Processing},
  volume={20},
  number={4},
  pages={154},
  year={2021},
  publisher={Springer}
}

@article{GBL+23,
  title={Satellite-based quantum key distribution in the presence of bypass channels},
  author={Ghalaii, Masoud and Bahrani, Sima and Liorni, Carlo and Grasselli, Federico and Kampermann, Hermann and Wooltorton, Lewis and Kumar, Rupesh and Pirandola, Stefano and Spiller, Timothy P and Ling, Alexander and others},
  journal={PRX Quantum},
  volume={4},
  number={4},
  pages={040320},
  year={2023},
  publisher={APS}
}

@article{BGK+20,
  title={Numerical finite-key analysis of quantum key distribution},
  author={Bunandar, Darius and Govia, Luke CG and Krovi, Hari and Englund, Dirk},
  journal={npj Quantum Information},
  volume={6},
  number={1},
  pages={104},
  year={2020},
  publisher={Nature Publishing Group UK London}
}

@article{GLL21,
  title={Numerical calculations of the finite key rate for general quantum key distribution protocols},
  author={George, Ian and Lin, Jie and L{\"u}tkenhaus, Norbert},
  journal={Physical Review Research},
  volume={3},
  number={1},
  pages={013274},
  year={2021},
  publisher={APS}
}

@article{SBM+23,
  title={Finite key performance of satellite quantum key distribution under practical constraints},
  author={Sidhu, Jasminder S and Brougham, Thomas and McArthur, Duncan and Pousa, Roberto G and Oi, Daniel KL},
  journal={Communications Physics},
  volume={6},
  number={1},
  pages={210},
  year={2023},
  publisher={Nature Publishing Group UK London}
}

@article{MTP+24,
  title={Optical payload design for downlink quantum key distribution and keyless communication using CubeSats},
  author={Mendes, Pedro Neto and Teixeira, Gon{\c{c}}alo Lobato and Pinho, David and Rocha, Rui and Andr{\'e}, Paulo and Niehus, Manfred and Faleiro, Ricardo and Rusca, Davide and Cruzeiro, Emmanuel Zambrini},
  journal={EPJ Quantum Technology},
  volume={11},
  number={1},
  pages={48},
  year={2024},
  publisher={Springer Berlin Heidelberg}
}

@article{PKH23,
  title={Statistical fluctuation analysis for decoy-state quantum secure direct communication},
  author={Park, Jooyoun and Kim, Bumil and Heo, Jun},
  journal={Quantum Information Processing},
  volume={22},
  number={2},
  pages={112},
  year={2023},
  publisher={Springer}
}

@article{ELJ+21,
  title={Strategies for achieving high key rates in satellite-based QKD},
  author={Ecker, Sebastian and Liu, Bo and Handsteiner, Johannes and Fink, Matthias and Rauch, Dominik and Steinlechner, Fabian and Scheidl, Thomas and Zeilinger, Anton and Ursin, Rupert},
  journal={npj Quantum Information},
  volume={7},
  number={1},
  pages={5},
  year={2021},
  publisher={Nature Publishing Group UK London}
}

@article{DTR+21,
  title={Feasibility of satellite-to-ground continuous-variable quantum key distribution},
  author={Dequal, Daniele and Trigo Vidarte, Luis and Roman Rodriguez, Victor and Vallone, Giuseppe and Villoresi, Paolo and Leverrier, Anthony and Diamanti, Eleni},
  journal={npj Quantum Information},
  volume={7},
  number={1},
  pages={3},
  year={2021},
  publisher={Nature Publishing Group UK London}
}

@inproceedings{SLM+22,
  title={A CubeSat platform for space based quantum key distribution},
  author={Sivasankaran, Srihari and Liu, Clarence and Mihm, Moritz and Ling, Alexander},
  booktitle={2022 IEEE international conference on space optical systems and applications (ICSOS)},
  pages={51--56},
  year={2022},
  organization={IEEE}
}

@article{GN22,
  title={LEO small satellite {QKD} downlink performance: QuantSat-PT case study},
  author={Galetsky, Vladlen and Niehus, Manfred},
  journal={arXiv preprint arXiv:2209.10293},
  year={2022}
}

@article{SP00,
  title={Simple proof of security of the {BB84} quantum key distribution protocol},
  author={Shor, Peter W and Preskill, John},
  journal={Physical review letters},
  volume={85},
  number={2},
  pages={441},
  year={2000},
  publisher={APS}
}

@article{LSL22,
  title={Eavesdropping detection in {BB84} quantum key distribution protocols},
  author={Lee, Chankyun and Sohn, Ilkwon and Lee, Wonhyuk},
  journal={IEEE Transactions on Network and Service Management},
  volume={19},
  number={3},
  pages={2689--2701},
  year={2022},
  publisher={IEEE}
}

@article{OKM+24,
  title={Assessment of Practical Satellite Quantum Key Distribution Architectures for Current and Near-Future Missions},
  author={Orsucci, Davide and Kleinpa{\ss}, Philipp and Meister, Jaspar and De Marco, Innocenzo and H{\"a}usler, Stefanie and Strang, Thomas and Walenta, Nino and Moll, Florian},
  journal={International Journal of Satellite Communications and Networking},
  year={2024},
  publisher={Wiley Online Library}
}

@article{LCJ+24,
  title={Microsatellite-based real-time quantum key distribution},
  author={Li, Yang and Cai, Wen-Qi and Ren, Ji-Gang and Wang, Chao-Ze and Yang, Meng and Zhang, Liang and Wu, Hui-Ying and Chang, Liang and Wu, Jin-Cai and Jin, Biao and others},
  journal={arXiv preprint arXiv:2408.10994},
  year={2024}
}

@article{BT00,
  title={Quantum cryptography using larger alphabets},
  author={Bechmann-Pasquinucci, Helle and Tittel, Wolfgang},
  journal={Physical Review A},
  volume={61},
  number={6},
  pages={062308},
  year={2000},
  publisher={APS}
}

@article{CBK+02,
  title={Security of quantum key distribution using d-level systems},
  author={Cerf, Nicolas J and Bourennane, Mohamed and Karlsson, Anders and Gisin, Nicolas},
  journal={Physical review letters},
  volume={88},
  number={12},
  pages={127902},
  year={2002},
  publisher={APS}
}

@article{CDB+19,
  title={High-dimensional quantum communication: benefits, progress, and future challenges},
  author={Cozzolino, Daniele and Da Lio, Beatrice and Bacco, Davide and Oxenl{\o}we, Leif Katsuo},
  journal={Advanced Quantum Technologies},
  volume={2},
  number={12},
  pages={1900038},
  year={2019},
  publisher={Wiley Online Library}
}

@article{SPD+25,
  title={High-dimensional coherent one-way quantum key distribution},
  author={Sulimany, Kfir and Pelc, Guy and Dudkiewicz, Rom and Korenblit, Simcha and Eisenberg, Hagai S and Bromberg, Yaron and Ben-Or, Michael},
  journal={NPJ Quantum Information},
  volume={11},
  number={1},
  pages={16},
  year={2025},
  publisher={Nature Publishing Group UK London}
}

@article{HLS+24,
  title={High-dimensional quantum key distribution using orbital angular momentum of single photons<? pag$\backslash$break?> from a colloidal quantum dot at room temperature},
  author={Halevi, Dotan and Lubotzky, Boaz and Sulimany, Kfir and Bowes, Eric G and Hollingsworth, Jennifer A and Bromberg, Yaron and Rapaport, Ronen},
  journal={Optica Quantum},
  volume={2},
  number={5},
  pages={351--357},
  year={2024},
  publisher={Optica Publishing Group}
}

@article{LZL+22,
  title={High dimensional quantum key distribution with temporal and polarization hybrid encoding},
  author={Li, Dong-Dong and Zhao, Mei-Sheng and Li, Zhi and Tang, Yan-Lin and Dai, Yun-Qi and Tang, Shi-Biao and Zhao, Yong},
  journal={Optical Fiber Technology},
  volume={68},
  pages={102828},
  year={2022},
  publisher={Elsevier}
}

@article{HK05,
  author = {H.-K. Lo, X. Ma, and K. Chen},
  title = {Decoy State Quantum Key Distribution},
  journal = {Physical Review Letters},
  year = {2005},
  volume = {94},
  pages = {230504}
}

@article{X05,
  author = {X. Ma et al.},
  title = {Practical Decoy State for Quantum Key Distribution},
  journal = {Physical Review A},
  year = {2005},
  volume = {72},
  pages = {012326}
}

@article{DW17,
  author = {D. Vasylyev, A. Semenov, and W. Vogel},
  title = {Atmospheric Quantum Channels with Weak and Strong Turbulence},
  journal = {Physical Review Letters},
  year = {2017},
  volume = {117},
  pages = {090501}
}

@article{IK21,
  title={Analysis of a high-dimensional extended {B}92 protocol},
  author={Iqbal, Hasan and Krawec, Walter O},
  journal={Quantum Information Processing},
  volume={20},
  number={10},
  pages={344},
  year={2021},
  publisher={Springer}
}

@book{W21,
  title={Quantum key distribution},
  author={Wolf, Ramona},
  volume={988},
  year={2021},
  publisher={Springer}
}

@book{P13,
  title={Elements of quantum computation and quantum communication},
  author={Pathak, Anirban},
  year={2013},
  publisher={CRC Press Boca Raton}
}

@book{M08,
  title={Quantum computing explained},
  author={McMahon, David},
  year={2008},
  publisher={John Wiley \& Sons}
}

@article{MMF+13,
  title={On quantum R{\'e}nyi entropies: A new generalization and some properties},
  author={M{\"u}ller-Lennert, Martin and Dupuis, Fr{\'e}d{\'e}ric and Szehr, Oleg and Fehr, Serge and Tomamichel, Marco},
  journal={Journal of Mathematical Physics},
  volume={54},
  number={12},
  year={2013},
  publisher={AIP Publishing}
}

@article{E91,
  title={Quantum cryptography based on Bell’s theorem},
  author={Ekert, Artur K},
  journal={Physical Review Letters},
  volume={67},
  number={6},
  pages={661},
  year={1991},
  publisher={APS}
}

@article{L00,
  title={Security against individual attacks for realistic quantum key distribution},
  author={L{\"u}tkenhaus, Norbert},
  journal={Physical Review A},
  volume={61},
  number={5},
  pages={052304},
  year={2000},
  publisher={APS}
}

@article{SBC+09,
  title={The security of practical quantum key distribution},
  author={Scarani, Valerio and Bechmann-Pasquinucci, Helle and Cerf, Nicolas J and Du{\v{s}}ek, Miloslav and L{\"u}tkenhaus, Norbert and Peev, Momtchil},
  journal={Reviews of modern physics},
  volume={81},
  number={3},
  pages={1301--1350},
  year={2009},
  publisher={APS}
}

@article{SPR17,
  title={Quantum cryptography: Key distribution and beyond},
  author={Shenoy-Hejamadi, Akshata and Pathak, Anirban and Radhakrishna, Srikanth},
  journal={Quanta},
  volume={6},
  number={1},
  pages={1--47},
  year={2017}
}

@article{V80,
  title={Isoplanatic degradation of tilt correction and short-term imaging systems},
  author={Valley, George C},
  journal={Applied Optics},
  volume={19},
  number={4},
  pages={574--577},
  year={1980},
  publisher={Optical Society of America}
}

@article{HS64,
  title={Modulation transfer function associated with image transmission through turbulent media},
  author={Hufnagel, RE and Stanley, NR},
  journal={Journal of the Optical Society of America},
  volume={54},
  number={1},
  pages={52--61},
  year={1964},
  publisher={Optical Society of America}
}

@inproceedings{BCK+14,
  title={MODTRAN{\textregistered} 6: A major upgrade of the MODTRAN{\textregistered} radiative transfer code},
  author={Berk, Alexander and Conforti, Patrick and Kennett, Rosemary and Perkins, Timothy and Hawes, Fred and Van Den Bosch, Jeannette},
  booktitle={2014 6th Workshop on Hyperspectral Image and Signal Processing: Evolution in Remote Sensing (WHISPERS)},
  pages={1--4},
  year={2014},
  organization={IEEE}
}

@article{MDB25,
  title={Finite and asymptotic key analysis for CubeSat-based {BB}84 {QKD} with elliptical beam approximation},
  author={Muskan and Dutta, Arindam and Banerjee, Subhashish},
  journal={Physica Scripta},
  volume={100},
  number={5},
  pages={055117},
  year={2025},
  publisher={IOP Publishing}
}

@incollection{DH22,
  title={New directions in cryptography},
  author={Diffie, Whitfield and Hellman, Martin E},
  booktitle={Democratizing cryptography: the work of Whitfield Diffie and Martin Hellman},
  pages={365--390},
  year={2022}
}

@article{RSA78,
  title={A method for obtaining digital signatures and public-key cryptosystems},
  author={Rivest, Ronald L and Shamir, Adi and Adleman, Leonard},
  journal={Communications of the ACM},
  volume={21},
  number={2},
  pages={120--126},
  year={1978},
  publisher={ACM New York, NY, USA}
}

@article{S02,
  title={Moore's law: past, present and future},
  author={Schaller, Robert R},
  journal={IEEE spectrum},
  volume={34},
  number={6},
  pages={52--59},
  year={2002},
  publisher={IEEE}
}

@article{GT07,
  title={Quantum communication},
  author={Gisin, Nicolas and Thew, Rob},
  journal={Nature photonics},
  volume={1},
  number={3},
  pages={165--171},
  year={2007},
  publisher={Nature Publishing Group UK London}
}

@article{R99,
  title={Continuous variable quantum cryptography},
  author={Ralph, Timothy C},
  journal={Physical Review A},
  volume={61},
  number={1},
  pages={010303},
  year={1999},
  publisher={APS}
}

@article{GG02,
  title={Continuous variable quantum cryptography using coherent states},
  author={Grosshans, Fr{\'e}d{\'e}ric and Grangier, Philippe},
  journal={Physical review letters},
  volume={88},
  number={5},
  pages={057902},
  year={2002},
  publisher={APS}
}

@article{WLB04,
  title={Quantum cryptography without switching},
  author={Weedbrook, Christian and Lance, Andrew M and Bowen, Warwick P and Symul, Thomas and Ralph, <? format?> Timothy C and Lam, Ping Koy},
  journal={Physical review letters},
  volume={93},
  number={17},
  pages={170504},
  year={2004},
  publisher={APS}
}

@article{LG09,
  title={Unconditional security proof of long-distance continuous-variable quantum key distribution with discrete modulation},
  author={Leverrier, Anthony and Grangier, Philippe},
  journal={Physical review letters},
  volume={102},
  number={18},
  pages={180504},
  year={2009},
  publisher={APS}
}

@article{UG15,
  title={Unidimensional continuous-variable quantum key distribution},
  author={Usenko, Vladyslav C and Grosshans, Fr{\'e}d{\'e}ric},
  journal={Physical Review A},
  volume={92},
  number={6},
  pages={062337},
  year={2015},
  publisher={APS}
}

@article{GVK+22,
  title={The rationale for the optimal continuous-variable quantum key distribution protocol},
  author={Goncharov, Roman and Vorontsova, Irina and Kirichenko, Daniil and Filipov, Ilya and Adam, Iurii and Chistiakov, Vladimir and Smirnov, Semyon and Nasedkin, Boris and Pervushin, Boris and Kargina, Daria and others},
  journal={Optics},
  volume={3},
  number={4},
  pages={338--351},
  year={2022},
  publisher={MDPI}
}

@inproceedings{SAR03,
  title={Quantum cryptography protocols robust against photon number splitting attacks},
  author={Scarani, Valerio and Acin, A and Ribordy, Gr{\'e}goire and Gisin, Nicolas},
  booktitle={ERATO Conference on Quantum Information Science},
  pages={4--6},
  year={2003}
}

@article{IWY02,
  title={Differential phase shift quantum key distribution},
  author={Inoue, Kyo and Waks, Edo and Yamamoto, Yoshihisa},
  journal={Physical review letters},
  volume={89},
  number={3},
  pages={037902},
  year={2002},
  publisher={APS}
}

@article{SBG+05,
  title={Fast and simple one-way quantum key distribution},
  author={Stucki, Damien and Brunner, Nicolas and Gisin, Nicolas and Scarani, Valerio and Zbinden, Hugo},
  journal={Applied Physics Letters},
  volume={87},
  number={19},
  year={2005},
  publisher={AIP Publishing}
}

@inproceedings{S94,
  title={Algorithms for quantum computation: discrete logarithms and factoring},
  author={Shor, Peter W},
  booktitle={Proceedings 35th annual symposium on foundations of computer science},
  pages={124--134},
  year={1994},
  organization={Ieee}
}

@article{LYD+18,
  title={Overcoming the rate--distance limit of quantum key distribution without quantum repeaters},
  author={Lucamarini, Marco and Yuan, Zhiliang L and Dynes, James F and Shields, Andrew J},
  journal={Nature},
  volume={557},
  number={7705},
  pages={400--403},
  year={2018},
  publisher={Nature Publishing Group UK London}
}

@article{LCQ12,
  title={Measurement-device-independent quantum key distribution},
  author={Lo, Hoi-Kwong and Curty, Marcos and Qi, Bing},
  journal={Physical review letters},
  volume={108},
  number={13},
  pages={130503},
  year={2012},
  publisher={APS}
}

@article{BF02,
  title={Deterministic secure direct communication using entanglement},
  author={Bostr{\"o}m, Kim and Felbinger, Timo},
  journal={Physical Review Letters},
  volume={89},
  number={18},
  pages={187902},
  year={2002},
  publisher={APS}
}

@article{MAK+15,
  title={Inside quantum repeaters},
  author={Munro, William J and Azuma, Koji and Tamaki, Kiyoshi and Nemoto, Kae},
  journal={IEEE Journal of Selected topics in quantum electronics},
  volume={21},
  number={3},
  pages={78--90},
  year={2015},
  publisher={IEEE}
}

@article{BMH+13,
  title={A comprehensive design and performance analysis of low Earth orbit satellite quantum communication},
  author={Bourgoin, Jean-Philippe and Meyer-Scott, Evan and Higgins, Brendon L and Helou, B and Erven, Chris and Huebel, Hannes and Kumar, B and Hudson, D and D'Souza, Ian and Girard, Ralph and others},
  journal={New Journal of Physics},
  volume={15},
  number={2},
  pages={023006},
  year={2013},
  publisher={IOP Publishing}
}

@article{BTD+09,
  title={Feasibility of satellite quantum key distribution},
  author={Bonato, Cristian and Tomaello, Andrea and Da Deppo, Vania and Naletto, Giampiero and Villoresi, Paolo},
  journal={New Journal of Physics},
  volume={11},
  number={4},
  pages={045017},
  year={2009},
  publisher={IOP Publishing}
}

@article{YCL+17,
  title={Satellite-to-ground entanglement-based quantum key distribution},
  author={Yin, Juan and Cao, Yuan and Li, Yu-Huai and Ren, Ji-Gang and Liao, Sheng-Kai and Zhang, Liang and Cai, Wen-Qi and Liu, Wei-Yue and Li, Bo and Dai, Hui and others},
  journal={Physical review letters},
  volume={119},
  number={20},
  pages={200501},
  year={2017},
  publisher={APS}
}

@article{RXY+17,
  title={Ground-to-satellite quantum teleportation},
  author={Ren, Ji-Gang and Xu, Ping and Yong, Hai-Lin and Zhang, Liang and Liao, Sheng-Kai and Yin, Juan and Liu, Wei-Yue and Cai, Wen-Qi and Yang, Meng and Li, Li and others},
  journal={Nature},
  volume={549},
  number={7670},
  pages={70--73},
  year={2017},
  publisher={Nature Publishing Group UK London}
}

@article{MRS26,
  title={Performance Analysis of Satellite-Based QKD Protocols},
  author={Muskan and Meena, Ramniwas and Banerjee, Subhashish},
  journal={Modern Physics Letters A},
  year={2026},
  publisher={World Scientific}
}

@article{MB25,
  title={Continuous variable-based quantum communication in the ocean: R. Meena, S. Banerjee},
  author={Meena, Ramniwas and Banerjee, Subhashish},
  journal={Quantum Information Processing},
  volume={24},
  number={2},
  pages={46},
  year={2025},
  publisher={Springer}
}

@article{MB23,
  title={Characterization of quantumness of non-Gaussian states under the influence of Gaussian channel: R. Meena, S. Banerjee},
  author={Meena, Ramniwas and Banerjee, Subhashish},
  journal={Quantum Information Processing},
  volume={22},
  number={8},
  pages={298},
  year={2023},
  publisher={Springer}
}

@article{MKB25,
  title={Husimi phase distribution in non-Gaussian operations},
  author={Meena, Ramniwas and Kumar, Chandan and Banerjee, Subhashish},
  journal={Physical Review A},
  volume={112},
  number={2},
  pages={023702},
  year={2025},
  publisher={APS}
}

@article{PMB26,
  title={Dynamics of quantum coherence and non-classical correlations in open quantum system coupled to a squeezed thermal bath},
  author={Pathania, Neha and Meena, Ramniwas and Banerjee, Subhashish},
  journal={Physica A: Statistical Mechanics and its Applications},
  pages={131405},
  year={2026},
  publisher={Elsevier}
}

@article{NMA+26,
  title={A Tutorial on Quantum Key Distribution Protocols With Interactive Simulation of QBER and SKR Performance Metrics},
  author={Nath, Ushnik and Meena, Ramniwas and Ali, Shahnoor and Kundu, Neel Kanth},
  journal={IET Quantum Communication},
  volume={7},
  number={1},
  pages={e70033},
  year={2026},
  publisher={Wiley Online Library}
}

@article{EK91,
  title={Quantum cryptography based on Bell’s theorem},
  author={Ekert, Artur K},
  journal={Physical Review Letters},
  volume={67},
  number={6},
  pages={661},
  year={1991},
  publisher={APS}
}

@article{WZY02,
  title = {Security of quantum key distribution with entangled photons against individual attacks},
  author = {Waks, Edo and Zeevi, Assaf and Yamamoto, Yoshihisa},
  journal = {Phys. Rev. A},
  volume = {65},
  issue = {5},
  pages = {052310},
  numpages = {16},
  year = {2002},
  }

@article{RFH+11,
title = {Performance Comparison of {B}{B}84 and {B}92 Satellite-Based Free Space Quantum Optical Communication Systems in the Presence of Channel Effects},
author = {R. Etengu and F. M. Abbou and H. Y. Wong and A. Abid and N. Nortiza and A. Setharaman},
pages = {37--47},
volume = {32},
number = {1},
journal={Journal of Optical Communications},
year = {2011},
}

@article{SBC09,
  title={The security of practical quantum key distribution},
  author={Scarani, Valerio and Bechmann-Pasquinucci, Helle and Cerf, Nicolas J and Du{\v{s}}ek, Miloslav and L{\"u}tkenhaus, Norbert and Peev, Momtchil},
  journal={Reviews of modern physics},
  volume={81},
  number={3},
  pages={1301--1350},
  year={2009},
  publisher={APS}
}

@article{B09,
  title={Operational and convolution properties of two-dimensional Fourier transforms in polar coordinates},
  author={Baddour, Natalie},
  journal={Journal of the Optical Society of America A},
  volume={26},
  number={8},
  pages={1767--1777},
  year={2009},
  publisher={Optical Society of America}
}

@article{AS93,
  title={The infrared and electro-optical systems handbook},
  author={Accetta, Joseph S and Shumaker, David L},
  year={1993}
}

@book{G96,
  author    = {Goodman, Joseph W.},
  title     = {Introduction to Fourier Optics},
  edition   = {2nd},
  year      = {1996},
  publisher = {McGraw-Hill},
  address   = {New York}
}

@inproceedings{AOW+12,
  title={Double blinding-attack on entanglement-based quantum key distribution protocols},
  author={Adenier, Guillaume and Ohya, Masanori and Watanabe, Noboru and Basieva, Irina and Khrennikov, Andrei Yu},
  booktitle={AIP Conference Proceedings},
  volume={1424},
  number={1},
  pages={9--16},
  year={2012},
  organization={American Institute of Physics}
}

@article{RGK05,
  title={Information-theoretic security proof for quantum-key-distribution protocols},
  author={Renner, Renato and Gisin, Nicolas and Kraus, Barbara},
  journal={Physical Review A},
  volume={72},
  number={1},
  pages={012332},
  year={2005},
  publisher={APS}
}

@article{CRE_04,
  title={A generic security proof for quantum key distribution},
  author={Christandl, Matthias and Renner, Renato and Ekert, Artur},
  journal={arXiv preprint quant-ph/0402131},
  year={2004}
}

@article{C00,
  title={Pauli cloning of a quantum bit},
  author={Cerf, Nicolas J},
  journal={Physical Review Letters},
  volume={84},
  number={19},
  pages={4497},
  year={2000},
  publisher={APS}
}

@article{LP21,
  title={Quantum-dot single-photon sources for the quantum internet},
  author={Lu, Chao-Yang and Pan, Jian-Wei},
  journal={Nature Nanotechnology},
  volume={16},
  number={12},
  pages={1294--1296},
  year={2021},
  publisher={Nature Publishing Group}
}

@article{TS21,
  title={The race for the ideal single-photon source is on},
  author={Thomas, Sarah and Senellart, Pascale},
  journal={Nature Nanotechnology},
  volume={16},
  number={4},
  pages={367--368},
  year={2021},
  publisher={Nature Publishing Group}
}

@article{DW05,
  title={Distillation of secret key and entanglement from quantum states},
  author={Devetak, Igor and Winter, Andreas},
  journal={Proceedings of The Royal Society A: Mathematical, Physical and Engineering Sciences},
  volume={461},
  number={2053},
  pages={207--235},
  year={2005},
  publisher={The Royal Society}
}

@article{K16,
  title={Quantum key distribution with mismatched measurements over arbitrary channels},
  author={Krawec, Walter O},
  journal={arXiv preprint arXiv:1608.07728},
  year={2016}
}

@article{BCC+10,
  title={The uncertainty principle in the presence of quantum memory},
  author={Berta, Mario and Christandl, Matthias and Colbeck, Roger and Renes, Joseph M and Renner, Renato},
  journal={Nature Physics},
  volume={6},
  number={9},
  pages={659--662},
  year={2010},
  publisher={Nature Publishing Group UK London}
}

@article{K87,
  title={Complementary observables and uncertainty relations},
  author={Kraus, Karl},
  journal={Physical Review D},
  volume={35},
  number={10},
  pages={3070},
  year={1987},
  publisher={APS}
}

@article{MU88,
  title={Generalized entropic uncertainty relations},
  author={Maassen, Hans and Uffink, Jos BM},
  journal={Physical Review Letters},
  volume={60},
  number={12},
  pages={1103},
  year={1988},
  publisher={APS}
}

@article{BMA+09,
  title={Analysis of subcarrier multiplexed quantum key distribution systems: signal, intermodulation, and quantum bit error rate},
  author={Capmany, Jos{\'E} and Ortigosa-Blanch, Arturo and Mora, Jos{\'e} and Ruiz-Alba, Antonio and Amaya, Waldimar and Martinez, Alfonso},
  journal={IEEE Journal of Selected Topics in Quantum Electronics},
  volume={15},
  number={6},
  pages={1607--1621},
  year={2009},
  publisher={IEEE}
}

@article{VSV12,
  title={Toward global quantum communication: beam wandering preserves nonclassicality},
  author={Vasylyev, D Yu and Semenov, AA and Vogel, W},
  journal={Physical Review Letters},
  volume={108},
  number={22},
  pages={220501},
  year={2012},
  publisher={APS}
}

@inproceedings{LC06,
  title={Using historic models of $C_{n}^{2}$ to predict $r_{0}$ and regimes affected by atmospheric turbulence for horizontal, slant, and topological paths},
  author={Lawson, Janice K and Carrano, Carmen J},
  booktitle={Atmospheric Optical Modeling, Measurement, and Simulation II},
  volume={6303},
  pages={38--49},
  year={2006},
  organization={SPIE}
}

@article{FSV+10,
  title={Estimates of $C_{n}^{2}$ from numerical weather prediction model output and comparison with thermosonde data},
  author={Frehlich, Rod and Sharman, Robert and Vandenberghe, Francois and Yu, Wei and Liu, Yubao and Knievel, Jason and Jumper, George},
  journal={Journal of Applied Meteorology and Climatology},
  volume={49},
  number={8},
  pages={1742--1755},
  year={2010},
  publisher={American Meteorological Society}
}

@article{BSH+13,
  title={A comprehensive design and performance analysis of low Earth orbit satellite quantum communication},
  author={Bourgoin, JP and Meyer-Scott, Evan and Higgins, Brendon L and Helou, B and Erven, Chris and Huebel, Hannes and Kumar, B and Hudson, D and D'Souza, Ian and Girard, Ralph and others},
  journal={New Journal of Physics},
  volume={15},
  number={2},
  pages={023006},
  year={2013},
  publisher={IOP Publishing}
}

@article{BHP93,
  title={Eavesdropping strategies and rejected-data protocols in quantum cryptography},
  author={Barnett, Stephen M and Huttner, Bruno and Phoenix, Simon JD},
  journal={Journal of Modern Optics},
  volume={40},
  number={12},
  pages={2501--2513},
  year={1993},
  publisher={Taylor \& Francis}
}

@article{WMU08,
  title={Tomography increases key rates of quantum-key-distribution protocols},
  author={Watanabe, Shun and Matsumoto, Ryutaroh and Uyematsu, Tomohiko},
  journal={Physical Review A},
  volume={78},
  number={4},
  pages={042316},
  year={2008},
  publisher={APS}
}

@article{MW08,
  title={Key rate available from mismatched measurements in the {BB84} protocol and the uncertainty principle},
  author={Matsumoto, Ryutaroh and Watanabe, Shun},
  journal={IEICE Transactions on Fundamentals of Electronics, Communications and Computer Sciences},
  volume={91},
  number={10},
  pages={2870--2873},
  year={2008},
  publisher={The Institute of Electronics, Information and Communication Engineers}
}

@article{TCK+14,
  title={Loss-tolerant quantum cryptography with imperfect sources},
  author={Tamaki, Kiyoshi and Curty, Marcos and Kato, Go and Lo, Hoi-Kwong and Azuma, Koji},
  journal={Physical Review A},
  volume={90},
  number={5},
  pages={052314},
  year={2014},
  publisher={APS}
}

@article{LGT09,
  title={Robust unconditionally secure quantum key distribution with two nonorthogonal and uninformative states},
  author={Lucamarini, Marco and Di Giuseppe, Giovanni and Tamaki, Kiyoshi},
  journal={Physical Review A},
  volume={80},
  number={3},
  pages={032327},
  year={2009},
  publisher={APS}
}

@article{MFR12,
  title={Statistical fluctuation analysis for measurement-device-independent quantum key distribution},
  author={Ma, Xiongfeng and Fung, Chi-Hang Fred and Razavi, Mohsen},
  journal={Physical Review A},
  volume={86},
  number={5},
  pages={052305},
  year={2012},
  publisher={APS}
}

@article{XXL14,
  title={Protocol choice and parameter optimization in decoy-state measurement-device-independent quantum key distribution},
  author={Xu, Feihu and Xu, He and Lo, Hoi-Kwong},
  journal={Physical Review A},
  volume={89},
  number={5},
  pages={052333},
  year={2014},
  publisher={APS}
}

@article{YCLL+17,
  title={Satellite-based entanglement distribution over 1200 kilometers},
  author={Yin, Juan and Cao, Yuan and Li, Yu-Huai and Liao, Sheng-Kai and Zhang, Liang and Ren, Ji-Gang and Cai, Wen-Qi and Liu, Wei-Yue and Li, Bo and Dai, Hui and others},
  journal={Science},
  volume={356},
  number={6343},
  pages={1140--1144},
  year={2017},
  publisher={American Association for the Advancement of Science}
}

@article{TKI03,
  title={Unconditionally secure key distribution based on two nonorthogonal states},
  author={Tamaki, Kiyoshi and Koashi, Masato and Imoto, Nobuyuki},
  journal={Physical Review Letters},
  volume={90},
  number={16},
  pages={167904},
  year={2003},
  publisher={APS}
}

@inproceedings{M13,
  title={Improved asymptotic key rate of the {B92} protocol},
  author={Matsumoto, Ryutaroh},
  booktitle={2013 IEEE International Symposium on Information Theory},
  pages={351--353},
  year={2013},
  organization={IEEE}
}

@article{VBB2000,
  title={Atmospheric extinction of stellar radiation in the optical domain},
  author={Vargas, M Jurado and Ben{\'\i}tez, P Merch{\'a}n and Bajo, F S{\'a}nchez},
  journal={European Journal of Physics},
  volume={21},
  number={3},
  pages={245},
  year={2000},
  publisher={IOP Publishing}
}

@inproceedings{AK20,
  title={Finite key analysis of the extended {B}92 protocol},
  author={Amer, Omar and Krawec, Walter O},
  booktitle={2020 IEEE International Symposium on Information Theory (ISIT)},
  pages={1944--1948},
  year={2020},
  organization={IEEE}
}

@article{BPM+97,
  title={Experimental quantum teleportation},
  author={Bouwmeester, Dik and Pan, Jian-Wei and Mattle, Klaus and Eibl, Manfred and Weinfurter, Harald and Zeilinger, Anton},
  journal={Nature},
  volume={390},
  number={6660},
  pages={575--579},
  year={1997},
  publisher={Nature Publishing Group UK London}
}

@article{FDI+04,
  title={Quantum teleportation with a quantum dot single photon source},
  author={Fattal, David and Diamanti, Eleni and Inoue, Kyo and Yamamoto, Yoshihisa},
  journal={Physical Review Letters},
  volume={92},
  number={3},
  pages={037904},
  year={2004},
  publisher={APS}
}

@article{OMM+09,
  title={Quantum teleportation between distant matter qubits},
  author={Olmschenk, Steven and Matsukevich, DN and Maunz, P and Hayes, D and Duan, L-M and Monroe, C},
  journal={Science},
  volume={323},
  number={5913},
  pages={486--489},
  year={2009},
  publisher={American Association for the Advancement of Science}
}

@article{FSB+98,
  title={Unconditional quantum teleportation},
  author={Furusawa, Akira and S{\o}rensen, Jens Lykke and Braunstein, Samuel L and Fuchs, Christopher A and Kimble, H Jeff and Polzik, Eugene S},
  journal={science},
  volume={282},
  number={5389},
  pages={706--709},
  year={1998},
  publisher={American Association for the Advancement of Science}
}

@article{YBF07,
  title={Experimental demonstration of quantum teleportation of broadband squeezing},
  author={Yonezawa, Hidehiro and Braunstein, Samuel L and Furusawa, Akira},
  journal={Physical Review Letters},
  volume={99},
  number={11},
  pages={110503},
  year={2007},
  publisher={APS}
}

@article{LBT+11,
  title={Teleportation of nonclassical wave packets of light},
  author={Lee, Noriyuki and Benichi, Hugo and Takeno, Yuishi and Takeda, Shuntaro and Webb, James and Huntington, Elanor and Furusawa, Akira},
  journal={Science},
  volume={332},
  number={6027},
  pages={330--333},
  year={2011},
  publisher={American Association for the Advancement of Science}
}

@article{HZL+20,
  title={Experimental high-dimensional quantum teleportation},
  author={Hu, Xiao-Min and Zhang, Chao and Liu, Bi-Heng and Cai, Yu and Ye, Xiang-Jun and Guo, Yu and Xing, Wen-Bo and Huang, Cen-Xiao and Huang, Yun-Feng and Li, Chuan-Feng and others},
  journal={Physical Review Letters},
  volume={125},
  number={23},
  pages={230501},
  year={2020},
  publisher={APS}
}

@article{DLB+11,
  title={Experimental high-dimensional two-photon entanglement and violations of generalized Bell inequalities},
  author={Dada, Adetunmise C and Leach, Jonathan and Buller, Gerald S and Padgett, Miles J and Andersson, Erika},
  journal={Nature Physics},
  volume={7},
  number={9},
  pages={677--680},
  year={2011},
  publisher={Nature Publishing Group UK London}
}

@article{MGT+17,
  title={Quantifying photonic high-dimensional entanglement},
  author={Martin, Anthony and Guerreiro, Thiago and Tiranov, Alexey and Designolle, S{\'e}bastien and Fr{\"o}wis, Florian and Brunner, Nicolas and Huber, Marcus and Gisin, Nicolas},
  journal={Physical Review Letters},
  volume={118},
  number={11},
  pages={110501},
  year={2017},
  publisher={APS}
}

@article{KRR+17,
  title={On-chip generation of high-dimensional entangled quantum states and their coherent control},
  author={Kues, Michael and Reimer, Christian and Roztocki, Piotr and Cort{\'e}s, Luis Romero and Sciara, Stefania and Wetzel, Benjamin and Zhang, Yanbing and Cino, Alfonso and Chu, Sai T and Little, Brent E and others},
  journal={Nature},
  volume={546},
  number={7660},
  pages={622--626},
  year={2017},
  publisher={Nature Publishing Group UK London}
}

@article{HXL+20,
  title={Efficient generation of high-dimensional entanglement through multipath down-conversion},
  author={Hu, Xiao-Min and Xing, Wen-Bo and Liu, Bi-Heng and Huang, Yun-Feng and Li, Chuan-Feng and Guo, Guang-Can and Erker, Paul and Huber, Marcus},
  journal={Physical Review Letters},
  volume={125},
  number={9},
  pages={090503},
  year={2020},
  publisher={APS}
}

@article{VSP+20,
  title={High-dimensional pixel entanglement: efficient generation and certification},
  author={Valencia, Natalia Herrera and Srivastav, Vatshal and Pivoluska, Matej and Huber, Marcus and Friis, Nicolai and McCutcheon, Will and Malik, Mehul},
  journal={Quantum},
  volume={4},
  pages={376},
  year={2020},
  publisher={Verein zur F{\"o}rderung des Open Access Publizierens in den Quantenwissenschaften}
}

@article{ZHG18,
  title={Long-distance continuous-variable quantum key distribution using separable Gaussian states},
  author={Zhou, Jian and Huang, Duan and Guo, Ying},
  journal={Physical Review A},
  volume={98},
  number={4},
  pages={042303},
  year={2018},
  publisher={APS}
}

@article{ZWF+23,
  title={Four-state continuous-variable quantum key distribution with a hybrid linear amplifier},
  author={Zhou, Jian and Wu, Leixin and Feng, Yanyan and Li, Hui and Shi, Jinjing and Shi, Ronghua},
  journal={Quantum Information Processing},
  volume={22},
  number={9},
  pages={356},
  year={2023},
  publisher={Springer}
}

@article{CDG+21,
  title={A secure blockchain-based group key agreement protocol for IoT},
  author={Chen, Chien-Ming and Deng, Xiaoting and Gan, Wensheng and Chen, Jiahui and Islam, SK Hafizul},
  journal={The Journal of Supercomputing},
  volume={77},
  pages={9046--9068},
  year={2021},
  publisher={Springer}
}

@inproceedings{FLL+21,
  title={A continuous variable quantum key distribution protocol based on multi-dimensiondata reconciliation with Polar code},
  author={Feng, Bao and Lv, Chao and Liu, Jinsuo and Zhang, Tianbing},
  booktitle={Journal of Physics: Conference Series},
  volume={1757},
  number={1},
  pages={012111},
  year={2021},
  organization={IOP Publishing}
}

@inproceedings{BZJ+21,
  title={Cascade protocol with parameters and backtracking optimization},
  author={Bai, Enjian and Zhang, Yun and Jiang, Xueqin and Wu, Yun and Shi, Zhen and Lin, Xiaodong and Pei, Zelin},
  booktitle={2021 IEEE International Conference on Power Electronics, Computer Applications (ICPECA)},
  pages={571--574},
  year={2021},
  organization={IEEE}
}

@inproceedings{ZM20,
  title={Security Analysis and Optimization of BB84 QKD System Post-Processing},
  author={Zhang, Pin and Mao, Xiangjie},
  booktitle={Journal of Physics: Conference Series},
  volume={1621},
  number={1},
  pages={012017},
  year={2020},
  organization={IOP Publishing}
}

@article{CLB+19,
  title={High-dimensional quantum communication: benefits, progress, and future challenges},
  author={Cozzolino, Daniele and Da Lio, Beatrice and Bacco, Davide and Oxenl{\o}we, Leif Katsuo},
  journal={Advanced Quantum Technologies},
  volume={2},
  number={12},
  pages={1900038},
  year={2019},
  publisher={Wiley Online Library}
}

@article{BB14,
  title={Quantum cryptography: Public key distribution and coin tossing},
  author={Bennett, Charles H and Brassard, Gilles},
  journal={Theoretical Computer Science},
  volume={560},
  pages={7--11},
  year={2014},
  publisher={Elsevier}
}
\chapter*{Appendices}
\addcontentsline{toc}{chapter}{Appendices} 
 \refstepcounter{section}
\section*{Appendix \Alph{section}: Double-Blinding Attack }
\addcontentsline{toc}{section}{Appendix \Alph{section}:  }\label{sec:Appendix-A}

\renewcommand{\theequation}{\Alph{section}\arabic{equation}}

\setcounter{equation}{0}

In the context of the BBM92 protocol, the existing blinding attack are of the intercept and resend type. In this type of attack, a malicious entity, often referred to as Eve, intercepts the signal that was originally intended for Bob. Eve then proceeds to perform measurements using random bases in order to obtain the raw key, just as Bob would have done in the intended communication process.
To conceal her presence, Eve forwards a signal to Bob whenever she successfully obtains a measurement result. This signal ensures that Bob receives an identical outcome, while in the case of diagonal alignment, no detection occurs at all.
 In practical implementation using QKD devices \cite{AOW+12}, Eve employs techniques to blind Bob's detectors to single-photon detection. She achieves this by manipulating the detectors to shift from Geiger mode to linear mode, where a detector only registers a click if the incoming signal intensity exceeds a preset discriminator threshold, denoted as $I_{th}$. After each detection, Eve sends a bright pulse with linear polarization aligned to her own measurement result. When Eve and Bob randomly select identical measurement bases, the pulse deterministically generates a click in one of Bob's detectors. This ensures that Bob's measurement outcomes match those of Eve because the pulse is either fully reflected or transmitted at Bob's polarizing beamsplitter. However, to prevent double counting and incorrect results when Eve and Bob randomly select bases that are diagonal to each other, Eve adjusts the intensity of the pulses to be lower than twice the threshold intensity of the detectors. Consequently, the pulse is split in half at Bob's polarizing beamsplitter, resulting in an output that is insufficient to surpass the threshold and produce a click in either of Bob's detectors.
The objective of the attack is for Eve to obtain an exact replica of Bob's key at the conclusion of the raw key distribution process. If Alice and Bob are sufficiently satisfied with the measured QBER on a subset of the key, Eve can eavesdrop on the error correction protocol that Alice and Bob employ. By performing the same operations as Bob during the error correction phase, Eve can successfully acquire an exact copy of the sifted key in the end.
One limitation of single-blinding attacks is that, on average, Bob's resulting key size is reduced by half compared to what he would have obtained without the attack. This reduction occurs because approximately half of the time, the randomly chosen measurement bases of Eve and Bob turn out to be diagonally opposite to each other. Consequently, Bob's detectors do not register any clicks in such cases. Therefore, the efficiency of this attack, by design, is fundamentally limited to $50\%$ on Bob's side.\par
Here the proposed double-blinding attack involves a similar implementation to the single-blinding attack, but with the key difference that Eve blinds all detectors on both sides instead of just Bob's detectors. Due to the double-blinding attack, Alice and Bob are unable to detect the presence of Eve, resulting in a complete elimination of information leakage. In other words, the measure of information leakage, denoted as  $\tau$ becomes zero in this scenario.

\refstepcounter{section}{}
\section*{Appendix \Alph{section}: Detailed Security Analysis and Key Rate Derivation for the HD-Ext-B92 Protocol }
\addcontentsline{toc}{section}{Appendix \Alph{section}:  }\label{Appendix B}

\renewcommand{\theequation}{\Alph{section}\arabic{equation}}

\setcounter{equation}{0}

We recap the security analysis proposed in Ref. \cite{IK21} and show
our important modification in the investigation of the minimum value
key rate (per pulse) for HD-Ext-B92 protocol. We elaborate the theorem
\cite{K16} which provides the lower bound of the conditional von
Neumann entropy of classical-quantum state $\rho_{aE}$ in Hilbert
space\footnote{Alice's register and Eve's quantum memory are represented in Hilbert
space $\mathcal{H}_{a}$ and $\mathcal{H}_{E}$, respectively.} $\mathcal{H}_{a}\otimes\mathcal{H}_{E}$.

\emph{Theorem: }Let $\mathcal{H}_{a}$ and $\mathcal{H}_{E}$ are finite-dimensional
Hilbert space and consider the following state of Alice and Eve in
the form of density matrix,

\begin{equation}
\rho_{aE}=\frac{1}{M}\left(|0\rangle\langle0|_{a}\otimes\left[\sum_{x=1}^{{\rm d}}|E_{x}^{0}\rangle\langle E_{x}^{0}|\right]+|1\rangle\langle1|_{a}\otimes\left[\sum_{x=1}^{{\rm d}}|E_{x}^{1}\rangle\langle E_{x}^{1}|\right]\right),\label{eq:Classical-Quantum state of ALice and Bob}
\end{equation}
where $M(>0)$ is normalization factor, ${\rm d}$ has finite value,
and each state\footnote{Eve's states are not necessarily normalized, nor orthogonal; it might
be that $|E_{x}^{y}\rangle\equiv0$ also.} $|E_{x}^{y}\rangle\in\mathcal{H}_{E}$. Assuming $K_{x}^{y}=\langle E_{x}^{y}|E_{x}^{y}\rangle\geq0$,
then,

\[
S\left(a|E\right)_{\rho_{aE}}\geq\sum_{x=1}^{{\rm d}}\left(\frac{K_{x}^{0}+K_{x}^{1}}{M}\right)S_{x},
\]
where
\[
S_{x}=\begin{cases}
h\left(\frac{K_{x}^{0}}{K_{x}^{0}\,+\,K_{x}^{1}}\right)-h\left(\delta_{x}\right) & {\rm if}\,\,K_{x}^{0}\wedge K_{x}^{1}\geq0,\\
0 & {\rm otherwise},
\end{cases}
\]
and
\[
\delta_{x}=\frac{1}{2}+\frac{\sqrt{\left(K_{x}^{0}-K_{x}^{1}\right)^{2}+4{\rm Re^{2}}\langle E_{x}^{0}|E_{x}^{1}\rangle}}{2\left(K_{x}^{0}+K_{x}^{1}\right)}.
\]
This \emph{Theorem}{}
serves as the foundation for our analysis, facilitating the derivation
of the key rate equation for the protocol discussed throughout the
remainder of this appendix.

The action of Eve's unitary operation $\mathcal{E}_{TE}$ on Alice's
transmitted state $|\Upsilon\rangle_{T}$ and Eve's ancilla state
$|\chi\rangle_{E}$ is described in the following,

\[
\mathcal{E}_{TE}|\Upsilon\rangle_{T}\otimes|\chi\rangle_{E}=\sum_{c=1}^{d}|c,E_{c}^{\Upsilon}\rangle_{TE},
\]
and

\[
\begin{array}{lcl}
\mathcal{E}_{TE}|\psi\rangle_{T}\otimes|\chi\rangle_{E} & = & \mathcal{E}_{TE}\frac{1}{\sqrt{2}}\left(|m\rangle+|n\rangle\right)_{T}\otimes|\chi\rangle_{E}\\
 & = & \frac{1}{\sqrt{2}}\sum_{c=1}^{d}|c\rangle_{T}\otimes|F_{c}\rangle_{E},
\end{array}
\]
where $|F_{c}\rangle_{E}:=|E_{c}^{m}\rangle_{E}+|E_{c}^{n}\rangle_{E}$,
and $E_{c}^{\Upsilon}$ is an arbitrary state in Eve's ancillary basis
when Alice's transmitted state before and after Eve's operation are
$|\Upsilon\rangle_{T}$ and $|c\rangle_{T}$, respectively. As $\mathcal{E}_{TE}$
is a unitary operation the relation holds as $\sum_{c=1}^{d}\langle E_{c}^{\Upsilon}|E_{c}^{\Upsilon}\rangle=1$.
After Eve's unitary operation on the classical-quantum state, $\rho_{aT}=\frac{1}{2}\left(|0\rangle\langle0|_{a}\otimes|m\rangle\langle m|_{T}+|1\rangle\langle1|_{a}\otimes|\psi\rangle\langle\psi|_{T}\right)$
is as following,

\begin{equation}
\begin{array}{lcl}
\rho_{aTE} & = & \mathcal{E}_{TE}\left(\rho_{aT}\right)\\
 & = & \frac{1}{2}\left[|0\rangle\langle0|_{a}\otimes P\left(\sum_{c=1}^{d}|c,E_{c}^{m}\rangle_{TE}\right)+|1\rangle\langle1|_{a}\otimes P\left(\frac{1}{\sqrt{2}}\sum_{c=1}^{d}|c,F_{c}\rangle_{TE}\right)\right],
\end{array}\label{eq:Alice-Transmitted-Eve's state}
\end{equation}
where $P\left(|\upsilon\rangle\right)=|\upsilon\rangle\langle\upsilon|$
is projection operator. After receiving the transmitted register $T$
Bob will apply the measurement operators $M_{0}=I_{a}\otimes\left(I-|\psi\rangle\langle\psi|\right)_{T}\otimes I_{E}$
and $M_{1}=I_{a}\otimes\left(I-|m\rangle\langle m|\right)_{T}\otimes I_{E}$
on $T$. Using Eq. (\ref{eq:Alice-Transmitted-Eve's state}) we can
write density state after Bob's operations,

\begin{equation}
\begin{array}{lcl}
\rho_{aTE}^{0} & = & M_{0}\left(\rho_{aTE}\right)M_{0}^{\dagger}\\
 & = & \frac{1}{2}\left[|0\rangle\langle0|\otimes P\left\{ \underset{c\ne m,c\ne n}{\sum}|c,E_{c}^{m}\rangle+\frac{1}{2}\,|m,E_{m}^{m}-E_{n}^{m}\rangle-\frac{1}{2}\,|n,E_{m}^{m}-E_{n}^{m}\rangle\right\} \right.\\
 & + & \left.|1\rangle\langle1|\otimes P\left\{ \frac{1}{\sqrt{2}}\left(\underset{c\ne m,c\ne n}{\sum}|c,F_{c}\rangle+\frac{1}{2}\,|m,F_{m}-F_{n}\rangle-\frac{1}{2}\,|n,F_{m}-F_{n}\rangle\right)\right\} \right]_{aTE},
\end{array}\label{eq:Alice-Transmit-Eve when Bob gets 0}
\end{equation}
and
\begin{equation}
\begin{array}{lcl}
\rho_{aTE}^{1} & = & M_{1}\left(\rho_{aTE}\right)M_{1}^{\dagger}\\
 & = & \frac{1}{2}\left[|0\rangle\langle0|\otimes P\left(\underset{c\ne m}{\sum}|c,E_{c}^{m}\rangle\right)+|1\rangle\langle1|\otimes P\left(\frac{1}{\sqrt{2}}\underset{c\ne m}{\sum}|c,F_{c}\rangle\right)\right]_{aTE}.
\end{array}\label{eq:Alice-Transmit-Eve when Bob gets 1}
\end{equation}
After Bob gets his outcomes Eqs. (\ref{eq:Alice-Transmit-Eve when Bob gets 0})
and (\ref{eq:Alice-Transmit-Eve when Bob gets 1}) may be traced out
the transit register $T$ and include Bob's classical register $b$
to keep his measurement result. Now the Eqs. (\ref{eq:Alice-Transmit-Eve when Bob gets 0})
and (\ref{eq:Alice-Transmit-Eve when Bob gets 1}) can be written
like,

\begin{equation}
\begin{array}{lcl}
\rho_{aEb}^{0} & = & \frac{1}{2}\left[|0\rangle\langle0|_{a}\otimes\left\{ \underset{c\ne m,c\ne n}{\sum}|E_{c}^{m}\rangle\langle E_{c}^{m}|+\frac{1}{2}\,|\left(E_{m}^{m}-E_{n}^{m}\right)\rangle\langle\left(E_{m}^{m}-E_{n}^{m}\right)|\right\} _{E}\otimes|0\rangle\langle0|_{b}\right.\\
 & + & \left.|1\rangle\langle1|_{a}\otimes\frac{1}{2}\left\{ \underset{c\ne m,c\ne n}{\sum}|F_{c}\rangle\langle F_{c}|+\frac{1}{2}\,|\left(F_{m}-F_{n}\right)\rangle\langle\left(F_{m}-F_{n}\right)|\right\} _{E}\otimes|0\rangle\langle0|_{b}\right],
\end{array}\label{eq:Alice-Eve-Bob-bit_0}
\end{equation}
and
\begin{equation}
\rho_{aEb}^{1}=\frac{1}{2}\left[|0\rangle\langle0|_{a}\otimes\sum_{c\ne m}|E_{c}^{m}\rangle\langle E_{c}^{m}|\otimes|1\rangle\langle1|_{b}+|1\rangle\langle1|_{a}\otimes\frac{1}{2}\sum_{c\ne m}|F_{c}\rangle\langle F_{c}|\otimes|1\rangle\langle1|_{b}\right].\label{eq:Alice-Eve-Bob-bit_1}
\end{equation}
Adding up Eqs. (\ref{eq:Alice-Eve-Bob-bit_0}) and (\ref{eq:Alice-Eve-Bob-bit_1}),
the non-normalized density operator which represents in one key-bit
generation round is,
{\small
\begin{align}
\rho_{aEb} & =  \rho_{aEb}^{1}+\rho_{aEb}^{0}\nonumber\\
 & = \frac{1}{2}\,|0\rangle\langle0|_{a}\otimes\left[\left\{ \underset{c\ne m,c\ne n}{\sum}|E_{c}^{m}\rangle\langle E_{c}^{m}|+\frac{1}{2}\,|\left(E_{m}^{m}-E_{n}^{m}\right)\rangle\langle\left(E_{m}^{m}-E_{n}^{m}\right)|\right\} \otimes|0\rangle\langle0|_{b}+\underset{c\ne m}{\sum}|E_{c}^{m}\rangle\langle E_{c}^{m}|\otimes|1\rangle\langle1|_{b}\right]\nonumber\\
 & +  \frac{1}{2}\,|1\rangle\langle1|_{a}\otimes\left[\frac{1}{2}\left\{ \underset{c\ne m,c\ne n}{\sum}|F_{c}\rangle\langle F_{c}|+\frac{1}{2}\,|\left(F_{m}-F_{n}\right)\rangle\langle\left(F_{m}-F_{n}\right)|\right\} \otimes|0\rangle\langle0|_{b}+\frac{1}{2}\underset{c\ne m}{\sum}|F_{c}\rangle\langle F_{c}|\otimes|1\rangle\langle1|_{b}\right].\label{eq:Non-normalized density operator}
\end{align}
}

For computing the conditional entropy $H\left(a|b\right)$, we will
show how the Eq. (\ref{eq:Non-normalized density operator}) is utilized
to get the statistics for all combinations of Alice's and Bob's sifted
key. Now, trace out Bob's register from Eq. (\ref{eq:Non-normalized density operator})
to keep the composite state of Alice's register and Eve's memory which
is important to calculate $S(a|E)$. The final expression of the required
density matrix is,

\begin{align}
\rho_{aE} & =  \frac{1}{M}\left[|0\rangle\langle0|_{a}\otimes\left(\underset{c\ne m,c\ne n}{\sum}|E_{c}^{m}\rangle\langle E_{c}^{m}|+\frac{1}{4}\,|E_{m}^{m}\rangle\langle E_{m}^{m}|-\frac{1}{4}\,|E_{m}^{m}\rangle\langle E_{n}^{m}|-\frac{1}{4}\,|E_{n}^{m}\rangle\langle E_{m}^{m}|+\frac{3}{4}\,|E_{n}^{m}\rangle\langle E_{n}^{m}|\right)_{E}\right.\nonumber\\
 & + \left.|1\rangle\langle1|_{a}\otimes\left(\underset{c\ne m,c\ne n}{\sum}\frac{1}{2}\,|F_{c}\rangle\langle F_{c}|+\frac{1}{8}\,|F_{m}\rangle\langle F_{m}|-\frac{1}{8}\,|F_{m}\rangle\langle F_{n}|-\frac{1}{8}\,|F_{n}\rangle\langle F_{m}|+\frac{3}{8}\,|F_{n}\rangle\langle F_{n}|\right)_{E}\right],\label{eq:Alice-Eve required final expression}
\end{align}

where $M$ is the normalization factor that can be calculated as,

\begin{align*}
M & =  \underset{c\ne m,c\ne n}{\sum}\langle E_{c}^{m}|E_{c}^{m}\rangle+\frac{1}{2}\,\langle E_{n}^{m}|E_{n}^{m}\rangle+\frac{1}{4}\,\langle\left(E_{m}^{m}-E_{n}^{m}\right)|\left(E_{m}^{m}-E_{n}^{m}\right)\rangle\\
 & +  \frac{1}{2}\underset{c\ne m,c\ne n}{\sum}\langle F_{c}|F_{c}\rangle+\frac{1}{4}\,\langle F_{n}|F_{n}\rangle+\frac{1}{8}\,\langle\left(F_{m}-F_{n}\right)|\left(F_{m}-F_{n}\right)\rangle .
\end{align*}
We modify the derivation for $S(a|E)$ using $\rho_{aE}$ in comparison
with the seminal work \cite{IK21}. In our modified calculation, we
express the terms of $\rho_{aE}$ in Eve's two bases states, i.e.,
$\{E_{c}^{m}\}$ and $\{F_{c}\}$ which correspond to the bit values
(i.e., $0$ and $1$) in Alice's register\footnote{The above \emph{Theorem }allows our expression of Eq. (\ref{eq:Alice-Eve required final expression})
unlike the Eq. (5) in Ref. \cite{IK21}.}.

Applying this above \emph{theorem} we calculate the conditional von
Neumann entropy,

\begin{equation}
S(a|E)\ge\underset{c\ne m,c\ne n}{\sum}\left(\frac{K_{c}^{0}+K_{c}^{1}}{M}\right)S_{c}+\left(\frac{K_{m}^{0}+K_{m}^{1}}{M}\right)S_{m}+\left(\frac{K_{n}^{0}+K_{n}^{1}}{M}\right)S_{n},\label{eq:Value of S(A|E)}
\end{equation}
where

\[
\begin{array}{lclcclcl}
K_{c}^{0} & := & \langle E_{c}^{m}|E_{c}^{m}\rangle, &  &  & K_{c}^{1} & := & \frac{1}{2}\,\langle F_{c}|F_{c}\rangle,\forall\,c\ne m,n\\
\\
K_{m}^{0} & := & \frac{1}{4}\,\langle E_{m}^{m}|E_{m}^{m}\rangle, &  &  & K_{m}^{0} & := & \frac{1}{8}\,\langle F_{m}|F_{m}\rangle,\\
\\
K_{n}^{0} & := & \frac{3}{4}\,\langle E_{n}^{m}|E_{n}^{m}\rangle, &  &  & K_{n}^{1} & := & \frac{3}{8}\,\langle F_{n}|F_{n}\rangle.
\end{array}
\]
And
\[
\begin{array}{lcl}
S_{c} & = & h\left(\frac{K_{c}^{0}}{K_{c}^{0}\,+\,K_{c}^{1}}\right)-h\left(\frac{1}{2}+\frac{\sqrt{\left(K_{c}^{0}\,-\,K_{c}^{1}\right)^{2}+4\,{\rm Re^{2}}\langle E_{c}^{m}|\frac{1}{\sqrt{2}}F_{c}\rangle}}{2\,\left(K_{c}^{0}\,+\,K_{c}^{1}\right)}\right),\\
S_{m} & = & h\left(\frac{K_{m}^{0}}{K_{m}^{0}\,+\,K_{m}^{1}}\right)-h\left(\frac{1}{2}+\frac{\sqrt{\left(K_{m}^{0}\,-\,K_{m}^{1}\right)^{2}+4\,{\rm Re^{2}}\langle\frac{1}{2}E_{m}^{m}|\frac{1}{2\sqrt{2}}F_{m}\rangle}}{2\,\left(K_{m}^{0}\,+\,K_{m}^{1}\right)}\right),\\
S_{n} & = & h\left(\frac{K_{n}^{0}}{K_{n}^{0}\,+\,K_{n}^{1}}\right)-h\left(\frac{1}{2}+\frac{\sqrt{\left(K_{n}^{0}\,-\,K_{n}^{1}\right)^{2}+4\,{\rm Re^{2}}\frac{3}{4\sqrt{2}}\langle E_{n}^{m}|F_{n}\rangle}}{2\,\left(K_{n}^{0}\,+\,K_{n}^{1}\right)}\right).
\end{array}
\]
Here, we briefly describe the \emph{parameter estimation} for the
required statistics to get the values in the above equations. Let
$p_{\upsilon c}\,(p_{\upsilon\psi})$ be the observable parameter
when Bob's measurement outcome is $|c\rangle\,(|\psi\rangle)$ using
the $Z\,(X)$ basis when Alice sends state\footnote{Here, the generalized state is $|\upsilon\rangle\in\{|m\rangle,|n\rangle,|\psi\rangle\}$,
these statistics $\left(p_{\upsilon c(\psi)}\right)$ come from the
rounds where Alice and Bob do the same or different basis measurement
(see Table $1$ in Ref. \cite{IK21}).} $|\upsilon\rangle$. We may write in the form of the observable parameters
$K_{c}^{0}=p_{mc},$ $K_{c}^{1}=p_{\psi c},$ $K_{m}^{0}=\frac{1}{4}\,p_{mm},$
$K_{m}^{1}=\frac{1}{4}\,p_{\psi m},$ $K_{n}^{0}=\frac{3}{4}\,p_{mn},$
and $K_{n}^{1}=\frac{3}{4}\,p_{\psi n}.$

\[
\begin{array}{lcl}
{\rm Re\,}\langle E_{c}^{m}|\frac{1}{\sqrt{2}}F_{c}\rangle & = & \frac{1}{\sqrt{2}}\left(\frac{p_{mc}}{2}+p_{\psi c}-\frac{p_{nc}}{2}\right),\\
\\
{\rm Re\,}\langle\frac{1}{2}E_{m}^{m}|\frac{1}{2\sqrt{2}}F_{m}\rangle & = & \frac{1}{4\sqrt{2}}\left(\frac{p_{mm}}{2}+p_{\psi m}-\frac{p_{nm}}{2}\right),\\
\\
{\rm Re}\,\frac{3}{4\sqrt{2}}\langle E_{n}^{m}|F_{n}\rangle & = & \frac{3}{4\sqrt{2}}\left(\frac{p_{mn}}{2}+p_{\psi n}-\frac{p_{nn}}{2}\right).
\end{array}
\]

In our study, we take only the depolarizing channel to evaluate the
satellite-based effect of the HD-Ext-B92 protocol. Suppose the depolarizing
channel $\mathcal{D}_{q}(\rho)$ with parameter ${\rm q}$ acting
on a density operator $\rho$ on a Hilbert space of dimension ${\rm d}$.
$\mathcal{D_{{\rm q}}}(\rho)$ acts as follows,

\[
\mathcal{D}_{{\rm q}}(\rho)=\left(1-\frac{{\rm d}}{{\rm d}-1}{\rm q}\right)\rho+\frac{{\rm q}}{{\rm d}-1}I.
\]
In the above, we have already mentioned the required parameter to
calculate the key rate in terms of observable statistics. The observable
statistics may be written in the effect of depolarizing channel scenario,

\[
\begin{array}{lccclclcl}
 &  & p_{mm} & = & p_{nn} & = & p_{\psi\psi} & = & 1-{\rm q},\\
 &  & p_{mc} & = & p_{nc} & = & p_{\psi c} & = & \frac{{\rm q}}{{\rm d}-1},\\
p_{m\psi} & = & p_{n\psi} & = & p_{\psi m} & = & p_{\psi n} & = & \frac{1}{2}\left(1-\frac{{\rm q}\,{\rm d}}{{\rm d}-1}\right)+\frac{{\rm q}}{{\rm d}-1}.
\end{array}
\]
The above analysis is sufficient to evaluate $S(a|E)$ using Eq. (\ref{eq:Value of S(A|E)}),
and to get the key rate we need the value of $H(a|b)$ which is analyzed\footnote{Assuming $p_{ij}$ is the joint probability when Alice's and Bob's
raw bit are ``$i$'' and ``$j$'' given that not eliminating that
iteration \cite{DP22}.} in the following,

\begin{equation}
\begin{array}{lcl}
H\left(a|b\right) & = & H\left(p_{00},\,p_{01},\,p_{10},\,p_{11}\right)-h\left(p_{00}+p_{10}\right).\end{array}\label{eq:Value of H(a|b)}
\end{equation}
To compute Eq. (\ref{eq:Value of H(a|b)}), Alice and Bob use classical
sampling i.e., the values of observable probabilities under the simulated
channel. Using Eq. (\ref{eq:Non-normalized density operator}) with
normalization term $M$,

\[
\begin{array}{lcl}
p_{00} & = & \frac{1}{2M}\left(1-p_{m\psi}\right),\\
p_{01} & = & \frac{1}{2M}\left(1-p_{mm}\right),\\
p_{10} & = & \frac{1}{2M}\left(1-p_{\psi\psi}\right),\\
p_{11} & = & \frac{1}{2M}\left(1-p_{\psi i}\right).
\end{array}
\]
These are the needful analyses that we recap above for estimating
the minimum value of the key rate in Eq. (\ref{eq:Key-rate equantion}).

\setcounter{section}{3}
\section*{Appendix \Alph{section}: First and Second Moments of Elliptical Beam Parameters for Uplink and Downlink }
\addcontentsline{toc}{section}{Appendix \Alph{section}: }

\makeatletter
\def\@currentlabel{\Alph{section}} 
\label{sec:Appendix-B}
\makeatother
\renewcommand{\theequation}{\Alph{section}\arabic{equation}}
\setcounter{equation}{0}

We may write the first and second moments of the beam parameters in
Eq. (\ref{eq:Vector of beam-parameters}) concerning the connection
detailed in Eq. (\ref{eq:Down-Link and Up-Link condition}). The angle
of orientation of the elliptical profile $\varphi$ is presumed to
have a uniform distribution within the interval $[0,\frac{\pi}{2}]$.
The mean value and the variance in the centroid position of the beam,
in the case of up-links, are consistent for both the $x$ and $y$
directions, and they are equal to\footnote{See for details Appendix C in Ref. \cite{VSV+17}.},

\[
\begin{array}{lclcl}
\left\langle x_{0}\right\rangle  & = & \left\langle y_{0}\right\rangle  & = & 0,\\
\left\langle x_{0}^{2}\right\rangle  & = & \left\langle y_{0}^{2}\right\rangle  & = & 0.419\,\sigma_{R}^{2}\mathcal{W}_{0}^{2}\Omega^{-\frac{7}{6}}{\rm \frac{h}{L}},
\end{array}
\]
in this context, the term \textquotedbl Rytov parameter\textquotedbl{}
represents the quantity $\sigma_{R}=1.23\,C_{n}^{2}k^{\frac{7}{6}}{\rm L}^{\frac{11}{6}}$,
while $\Omega=\frac{k\mathcal{W}_{0}^{2}}{2\,{\rm L}}$ stands for
the Fresnel number, $k$ is the optical wave number. The selected
reference frame is such that $\left\langle x_{0}\right\rangle =\left\langle y_{0}\right\rangle =0$.
The mean and covariance of $\mathcal{W}_{i}^{2}$ can be written
as,

\[
\begin{array}{lcl}
\langle\mathcal{W}_{i}^{2}\rangle & = & \frac{\mathcal{W}_{0}^{2}}{\Omega^{2}}\left(1+\frac{\pi}{8}\,{\rm L}n_{0}\mathcal{W}_{0}^{2}{\rm \frac{h}{L}}+2.6\,\sigma_{R}^{2}\Omega^{\frac{5}{6}}{\rm \frac{h}{L}}\right),\\
\langle\Delta\mathcal{W}_{i}^{2}\Delta\mathcal{W}_{j}^{2}\rangle & = & \left(2\delta_{ij}-0.8\right)\frac{\mathcal{W}_{0}^{4}}{\Omega^{\frac{19}{6}}}\left(1+\frac{\pi}{8}\,{\rm L}n_{0}\mathcal{W}_{0}^{2}{\rm \frac{h}{L}}\right)\sigma_{R}^{2}{\rm \frac{h}{L}}.
\end{array}
\]
The same type of expressions also applies to downlinks when considering
the position of the beam centroid,

\[
\begin{array}{lclcl}
\left\langle x_{0}\right\rangle  & = & \left\langle y_{0}\right\rangle  & = & 0,\\
\left\langle x_{0}^{2}\right\rangle  & = & \left\langle y_{0}^{2}\right\rangle  & = & \alpha\,{\rm L}.
\end{array}
\]
Additionally, for the semi-major and semi-minor axes of the elliptical
beam profile,

\[
\begin{array}{lcl}
\langle\mathcal{W}_{i}^{2}\rangle & = & \frac{\mathcal{W}_{0}^{2}}{\Omega^{2}}\left(1+\frac{\pi}{24}\,{\rm L}n_{0}\mathcal{W}_{0}^{2}\left({\rm \frac{h}{L}}\right)^{3}+1.6\,\sigma_{R}^{2}\Omega^{\frac{5}{6}}\left({\rm \frac{h}{L}}\right)^{\frac{8}{3}}\right),\\
\langle\Delta\mathcal{W}_{i}^{2}\Delta\mathcal{W}_{j}^{2}\rangle & = & \left(2\delta_{ij}-0.8\right)\,\frac{3}{8}\,\frac{\mathcal{W}_{0}^{4}}{\Omega^{\frac{19}{6}}}\left(1+\frac{\pi}{24}\,{\rm L}n_{0}\mathcal{W}_{0}^{2}\left({\rm \frac{h}{L}}\right)^{3}\right)\sigma_{R}^{2}\left({\rm \frac{h}{L}}\right)^{\frac{8}{3}},
\end{array}
\]
here, $\alpha\approx2$ $\mu{\rm rad}$ refers to the angular pointing
error. Subsequently, the understanding of the probability distribution
concerning the elliptic beam parameters (Eq. (\ref{eq:Vector of beam-parameters}))
is applied to calculate the PDT utilizing Eq. (\ref{eq:PDT Equation})
and a process of random sampling which is mentioned in Sec. \ref{subsec:Elliptic_Beam_Model}.

\end{document}